\documentclass[apj,twocolumn]{emulateapj}
\usepackage{apjfonts}
\usepackage{epsfig}
\usepackage{amsmath, amsthm, amssymb}

\slugcomment{Draft version - \today}
\begin{document}

\title{Aspherical Core-Collapse Supernovae in Red Supergiants Powered by Nonrelativistic Jets }
\author{Sean M. Couch, J. Craig Wheeler, and Milo\v s Milosavljevi\'c}
\affil{Department of Astronomy, The University of Texas, Austin, TX, 78712, smc@astro.as.utexas.edu}
\righthead{JET-DRIVEN SUPERNOVAE}
\lefthead{COUCH, WHEELER, \& MILOSAVLJEVI\'C}

\begin{abstract}

We explore the observational characteristics of jet-driven supernovae by simulating bipolar-jet-driven explosions in a red supergiant progenitor.  We present results of four models in which we hold the injected kinetic energy at a constant $10^{51}$ ergs across all jet models but vary the specific characteristics of the jets to explore the influence of the nature of jets on the structure of the supernova ejecta.  We evolve the explosions past shock-breakout and into quasi-homologous expansion of the supernova envelope into a red supergiant wind.  The simulations have sufficient numerical resolution to study the stability of the flow.  Our simulations show the development of fluid instabilities that produce pristine helium clumps in the hydrogen envelope.  The oppositely-directed, nickel-rich jets give a large-scale asymmetry that may account for the non-spherical excitation and substructure of spectral lines such as H$\alpha$ and He I 10830\AA.  Jets with a large fraction of kinetic to thermal energy punch through the progenitor envelope and give rise to explosions that would be observed to be asymmetric from the earliest epochs, inconsistent with spectropolarimetric measurements of Type II supernovae.  Jets with higher thermal energy fractions result in explosions that are roughly spherical at large radii but are significantly elongated at smaller radii, deep inside the ejecta, in agreement with the polarimetric observations.  We present shock breakout light curves that indicate that strongly aspherical shock breakouts are incompatible with recent {\it GALEX} observations of shock breakout from red supergiant stars.  Comparison with observations indicates that jets must deposit their kinetic energy efficiently throughout the ejecta while in the hydrogen envelope.  Thermal energy-dominated jets satisfy this criterion and yield many of the observational characteristics of Type II supernovae.

\end{abstract}

\section{Introduction}

The observations of core-collapse supernovae are replete with evidence of asphericity.  Hints of aspherical structure have been found in many supernovae (SNe) of various types, including both Type Ib/c  \citep{maund07c, maund07b, Modjaz:08, WW08} and Type II events \citep{Fassia:98, Leonard:01, Elmhamdi:03, Chugai:05, Leonard:06}.  Challenges to the classic spherical model of supernovae (SNe) have been building since the observation of Type II SN 1987A, wherein asphericity was inferred from net polarization measurements \citep{Jeffery:91} and from substructure in spectral lines \citep[i.e., the `Bochum' event,][]{Hanuschik:88, Lucy:88, Phillips:89}.  Mixing one or two nickel clumps outward in the ejecta has been shown to reproduce the substructure of the H$\alpha$ line in SN 1987A \citep{Spyromilio:90, Utrobin:95}.  With the Hubble Space Telescope (HST), we have gained late-time, resolved images of the structure of the nascent SN remnant that dramatically reveal aspherical, elongated ejecta \citep{wang02}.

Supernova 1987A, while especially observable, was a single event, and with an atypical blue supergiant progenitor; however, typical Type II-P supernovae (SNeII-P), thought to arise from a red supergiant progenitor, show similar signs of asphericity and asymmetry.  \citet{Fassia:98} present infrared spectra of Type II-P SN 1995V and discuss implications of nonspherical elemental mixing.  They find that in order to fit the He I 10830-\AA\ emission, significant amounts of $^{56}$Ni must be ``dredged-up'' to high velocities in order to excite the He I line through radioactive decay.  Fassia et al. also conclude that about 10\% of the He mass must be contained in pure He clumps that are not microscopically mixed with hydrogen.  When mixed with hydrogen, collisional ionization of hydrogen \citep[Penning ionization,][]{Bell:70, Chugai:91} is the dominant de-excitation mechanism for He in the hydrogen envelope.  This analysis suggests the need for a mechanism that produces significant outward propagation of nickel and the presence of fluid instabilities that lead to helium clumping.  

The detailed shape of spectral lines provides indirect evidence of asymmetry and asphericity of supernova ejecta.  We can gain a direct measure of the shape of the supernova photosphere and line-forming regions from continuum and line polarization \citep[see][]{WW08}.   While the sample size of Type II-P SNe for which polarization measurements have been made is modest, the data to-date suggest asphericity which depends on the depth of the photosphere.  Leonard et al. (2001) made spectropolarimetric observations of SN 1999em and find a net continuum polarization of about 0.2\% on day 7 after maximum light, increasing to 0.5\% at day 165.  Citing the calculations of polarization from electron scattering spheroids by \citet{hoflich91}, \citet{Leonard:01} estimate the asphericity of SN 1999em to be $\sim$7\% on day 7 and $\sim$10\% on day 165.  As they discuss, the late time polarization may be reduced due to the decreased optical depth to electron scattering in the SN atmosphere, implying the intrinsic asphericity of SN 1999em may be greater than their estimates based on the electron scattering calculations of \citet{hoflich91}.  Further hints at the asphericity of the ejecta in SN 1999em are discussed by \citet{Elmhamdi:03}.  They report substructure in the H$\alpha$ and He I lines that is reminiscent of the `Bochum' event in SN 1987A.  The spectral modeling of \citet{Elmhamdi:03} shows that an asymmetric $^{56}$Ni distribution may account for the structure of these lines.

A striking, and provocative, example of time-dependent SNeII-P polarization is found in SN 2004dj.  In this event, the continuum polarization was practically zero, within the errors, during the plateau phase \citep{Leonard:06}, but during the transition from photospheric to nebular phases, as the photosphere receded to reveal the core of the SN, the polarization increased dramatically to 0.56\% on day 91, implying an asphericity of $\sim$30\%, and then slowly decreased until no polarization was detected on day 271.  \citet{Leonard:06} show that the decay of the polarization signal is roughly consistent with the expected decrease in electron scattering opacity of the receding photosphere.  This seems to imply that the core of the explosion was significantly aspherical but was masked by a roughly spherical envelope.  As discussed by \citet{Leonard:06}, and first argued by \citet{Wang:01}, when combined with data from core-collapse SNe of other types (i.e., Ib/c), this may indicate that the core-collapse SN mechanism is similar across all spectral types and is intrinsically aspherical.  Similar to SN 1999em, the spectral lines of SN 2004dj imply an asymmetric structure in the ejecta, inferred from the secondary peaks to the red and blue of the central peak.  Spectral modeling including the energy deposition from radioactive decay of $^{56}$Ni show that a {\it bipolar} nickel distribution in a spherical hydrogen envelope can reproduce the substructure of the H$\alpha$, [O I], and [Ca II] lines due to asymmetric line excitation from the decay of $^{56}$Ni \citep{Chugai:05}.    Furthermore, \citet{Chugai:06} shows that a bipolar nickel distribution in a spherical envelope can also approximately account for the time evolution of the decreasing continuum polarization during the nebular phase.

Based on the preceding discussion, explosion models for SNeII-P must account for ``dredge-up'' of nickel to higher expansion velocities than expected from spherically symmetric explosions, clumps of pristine helium in the hydrogen envelope, an asymmetric and possibly bipolar distribution of nickel, and roughly spherical electron-scattering photospheres during the plateau phase that become remarkably aspherical as the photosphere recedes below the hydrogen envelope.  The jet-driven model of core-collapse SNe \citep[e.g.,][]{kho99, hof01, maeda03} has features that may produce some of the necessary observed characteristics.  In this model, the explosion is driven by nonrelativistic bipolar jets that send strong bow shocks into the progenitor envelope.  These bows shocks intersect near the equatorial plane of the star and establish an equatorial outflow that complements the polar outflows and completes the explosion.  The material that comprises the jets is elongated along the jet axis and is expected to be nickel-rich owing to the jets' origin in the inner core of the explosion.  Fluid instabilities at the interface between the helium core and hydrogen envelope may produce clumps of helium in the hydrogen envelope, as seen in non-jet-driven supernova calculations  \citep[e.g.,][]{kifon03, kifon06}.  The character of such an unstable flow, however, has not been investigated for the case of jet-driven SNe.  It is also unclear that a jet-driven explosion in a red supergiant progenitor can produce a photosphere that is nearly spherical early-on and becomes aspherical later \citep[see][]{hof01}.

In this work, we present a numerical study of the structure of jet-driven SNe with adequate resolution to explore the nature of any unstable flows and time evolution sufficient to make a direct comparison with observations.   We assume that bipolar jets are formed as a result of the core-collapse process and do not self-consistantly simulate the jet production process.   Collapsing stellar cores could produce bipolar jets  in the presence of rotation and magnetic field amplification \citep{Wheeler:00, Wheeler:02, Akiyama:03}.  The process of magnetic jet formation in collapsing stars has been studied in previous simulations \citep{LeBlanc:70, Shibata:06, Obergaulinger:06b, Obergaulinger:06a,  Burrows:07, kom07}, however, the rotation rate and initial magnetic field structure of the progenitor stars, two very critical properties, are not well understood from first principles.  Therefore, we vary the parameters of the jets in four, physically-plausible jet models in order to explore the importance of the nature of the jets on the final structure of the SN.  The progenitor model in all cases is a red supergiant star of 15 $M_\odot$, roughly analogous to Betelgeuse, and the models are evolved to nearly six days after the initiation of the jets, several days after shock breakout.  

Many previous studies have focused on highly energetic jets in connection with Gamma-Ray Bursts (GRBs) \citep[e.g.,][]{MacFadyen:01, Zhang:03} and explosions of Pop III stars \citep[e.g.,][]{maeda03, Tominaga:07}.  We focus explicitly on explosions of typical supernova energies ($10^{51}$ erg) driven by {\it nonrelativistic} jets.  We use a progenitor star of solar metallicity that is not extremely massive.  The goal of this work is to explore the observational features of jet-driven supernovae in common Type II events.

This work is organized as follows.  In \S\ref{sec:sims} we describe our numerical approach and discuss the details of our jet models.   In \S\ref{sec:dynamics} we provide a narrative of the general evolution of the explosions, from shock propagation in the core to eruption from the progenitor's surface.  The transport of nickel to radii and velocities greater than those of lighter elements is a feature of a jet-driven explosion model in which the jets contain a significant amount of nickel.  In \S\ref{sec:clumps} we discuss the significance of this source of overturn, as well as the development of fluid instabilities that lead to the mixing of elements, such as nickel and helium.  In \S\ref{sec:geometry} we discuss the resultant shapes of the explosions and discuss the evolution of the surfaces of constant electron scattering optical depth.  In \S\ref{sec:breakout} we calculate the bolometric shock breakout light curves of our models and compare them to {\it GALEX} observations of supernova shock breakout from red supergiant stars.  We discuss the implications of our study and list our conclusions in \S\ref{sec:discussion}.

\section{The Simulations}
\label{sec:sims}

We use the adaptive-mesh refinement (AMR) hydrodynamics code, FLASH \citep{fryx00} in 2D spherical-polar coordinates.  FLASH is an Eulerian, higher-order Gudonov-type code that solves the hydrodynamic equations using a modified, piecewise-parabolic method (PPM) based on that of \citet{1984JCoPh..54..174C}.   We simulate a fluid with 8 atomic species: H, He, C, O, Ca, Mg, Si, and Ni.   Since in the present work we do not seek to precisely calculate resulting metal abundances, we do not simulate the transmutation of species via nuclear burning.  We use FLASH's Helmholtz equation of state (EoS) that includes contributions to the internal energy from ions (via the ideal gas law), electrons, positrons and radiation in thermal equilibrium.  We calculate the gravitational potential from an evolving, spherically-averaged density profile.  The refinement criterion was calculated based upon the second spatial derivatives of the density and pressure \citep[see \S2 of][]{fryx00}.

We have run four different models with varied parameters for the bipolar jets.  All models use a 15 $M_\odot$ red supergiant progenitor \citep[model ``s15s7b2" of][]{1995ApJS..101..181W} for initial conditions.  We map this model into 2D spherical geometry ($r$, $\theta$) and place the initial inner boundary just outside the iron core ($\sim 3.8\times10^8$ cm, or $\sim 1.6\ M_\odot$ in mass); we do not simulate the collapse of the iron core.  For all of our models, the angular extent of our 2D grid is a full 180\degr, corresponding to 4$\pi$ steradians in 3D.  We use seven levels of refinement.  The base-level number of angular zones is 16 throughout the simulation, whereas the base-level number of radial zones is changed as the simulation proceeds to optimize the resolution (see re-gridding discussion below).  The maximum angular resolution is 0.17\degr throughout the entirety of the simulations.  All spatial derivatives of physical variables vanish at the inner and outer radial boundaries.  Material is allowed to flow out of the computational domain through the inner and outer boundaries, but no material, other than that of the jets, is allowed to enter the domain through the inner boundary \citep[the so-called ``diode'' boundary condition of][]{Zingale:02}.  Mass that flows out of the domain through the inner boundary is added to the central point mass contribution to the gravitational potential.  The initial model is spherically symmetric and all velocities are initially set to zero.  In the region beyond the surface of the progenitor ($\sim 3\times10^{13}$ cm), we place a simple $1/r^2$ density profile with a constant outflow velocity of 10 km s$^{-1}$ that corresponds to a red supergiant wind with a mass-loss rate of $10^{-4}\ M_\odot$ yr$^{-1}$ \citep[see, e.g.,][]{2005ApJ...630..892D}.  Each model requires approximately 10,000 CPU hours to complete.

In all cases, the north and south jets are identical in injection method and initial parameters.  We keep the total injected energy ($10^{51}$ erg) and total time-integrated power ($5\times10^{50}$ erg s$^{-1}$) constant in all four models.  We set the jets to be composed entirely of nickel.  In reality, the composition of the jets may be time-dependent and not necessarily dominated by nickel.  Simulating the jet formation mechanism self-consistently would be required to determine the exact composition of the jets.  The jets are injected into the progenitor model along both poles with an opening half-angle of about 15\degr\ for a total of two seconds, roughly the free-fall time of the inner core.  During the first half-second, the injection velocity is kept fixed.  After that, the velocity is linearly reduced to zero after two seconds. The time-dependent power of all jet models may then be described by
\begin{equation}
\label{eq:jetp}
P(t) = 
\begin{cases}  
P_0, & \text{$t \leq 0.5\ {\rm s}$}, \\
\frac{2}{3} (2 - t) P_0, & \text{$0.5\ {\rm s} < t \leq 2\ {\rm s}$}, \\
0, & \text{$t > 2\ {\rm s}$} ,
\end{cases}
\end{equation}
where $P_0$ is the initial jet power.  The total energy (both kinetic and internal) deposited by the jets in all models is $10^{51}$ erg, split equally between the two jets in each model.  By setting the energy and power equal in all models, the jets are described by just two parameters:  maximum injection velocity and total injected mass.  This is the origin of our naming scheming for the models:  v{\it A}m{\it B}, where {\it A} is the velocity in units of $10^9$ cm s$^{-1}$ and {\it B} is the mass injected in units of hundredths of solar masses.  

Our method of jet injection is similar to that of \citet{kho99} and \citet{hof01} except that by fixing the power of the jets as well as the total energy, the physical parameters are further constrained.  We do not connect the accretion rate into the inner boundary to the jet power \citep[as in, e.g.,][]{MacFadyen:01, maeda03}, the nature of this accretion and the material that is accreted is lost in our calculations other than for its contribution to the gravitational potential.  Notably, most of the silicon that lies just inside our computational domain falls through the inner boundary and so the final ejecta in our calculations have very little silicon.  Since we are not conducting a careful study of nucleosynthesis and abundance yields this loss is acceptable.  

The parameters of the jets are given in Table \ref{tab:jets} along with ambient values from the progenitor for comparison.  Model v3m12 is very similar to the jets employed by \citet{kho99}, however, in that work the progenitor was a helium star.  The speed and density of v5m06 correspond approximately to those of the magnetically formed jets of \citet{Burrows:07}.  Models v1m12 and v6m04 are similar in injection velocity to models simulated in \citet{hof01}.  While models v3m12, v5m06, and v6m04 differ significantly in velocity and mass, the total energy in all three models is dominated by kinetic energy; hence, we shall refer to these models as the `kinetic energy' models.  This sets model v1m12 apart as the only model in which the thermal energy in the jet is greater than the kinetic energy.  The thermal to kinetic energy fractions for the four models, found by integrating equation (\ref{eq:jetp}), are 8 (v1m12), 0.05 (v3m12), 0.06 (v5m06), and 0.1 (v6m04).

As the jets and shock waves expand to larger radii, we use a data regridding method to expand the grid and optimize the resolution, allowing us to achieve high-resolution, long time-scale simulations in reasonable amounts of computer time.   While expanding the grid we also increase the radius of the inner radial boundary, excluding the small zones in the center of the grid where the Courant condition is very limiting and, thus, increasing the minimum time step.  At each re-gridding we decrease the radial resolution but keep the angular resolution fixed throughout the entire simulation.  We change the grid resolution parameters a total of ten times throughout the course of the simulations.  The details of our changes in grid parameters are given in Table \ref{tab:grid}.  All four models are evolved for a total 5.79 days, covering a spatial range from $\sim 4\times10^8$ cm to $\sim 10^{15}$ cm.   
 
\section{Dynamics and Evolution of Jet-Driven Explosions}
\label{sec:dynamics}

\begin{figure}
\centering
\plottwo{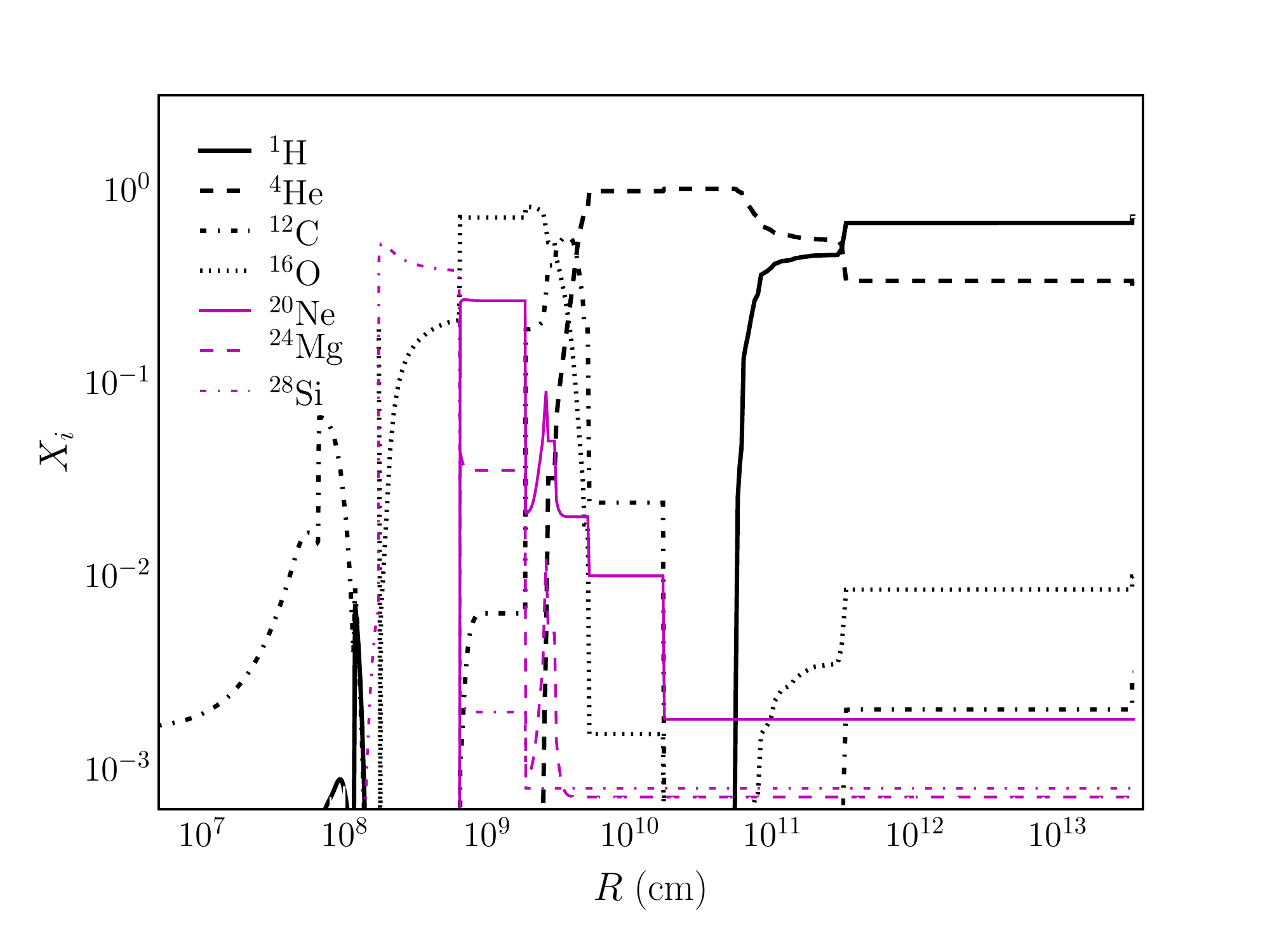}{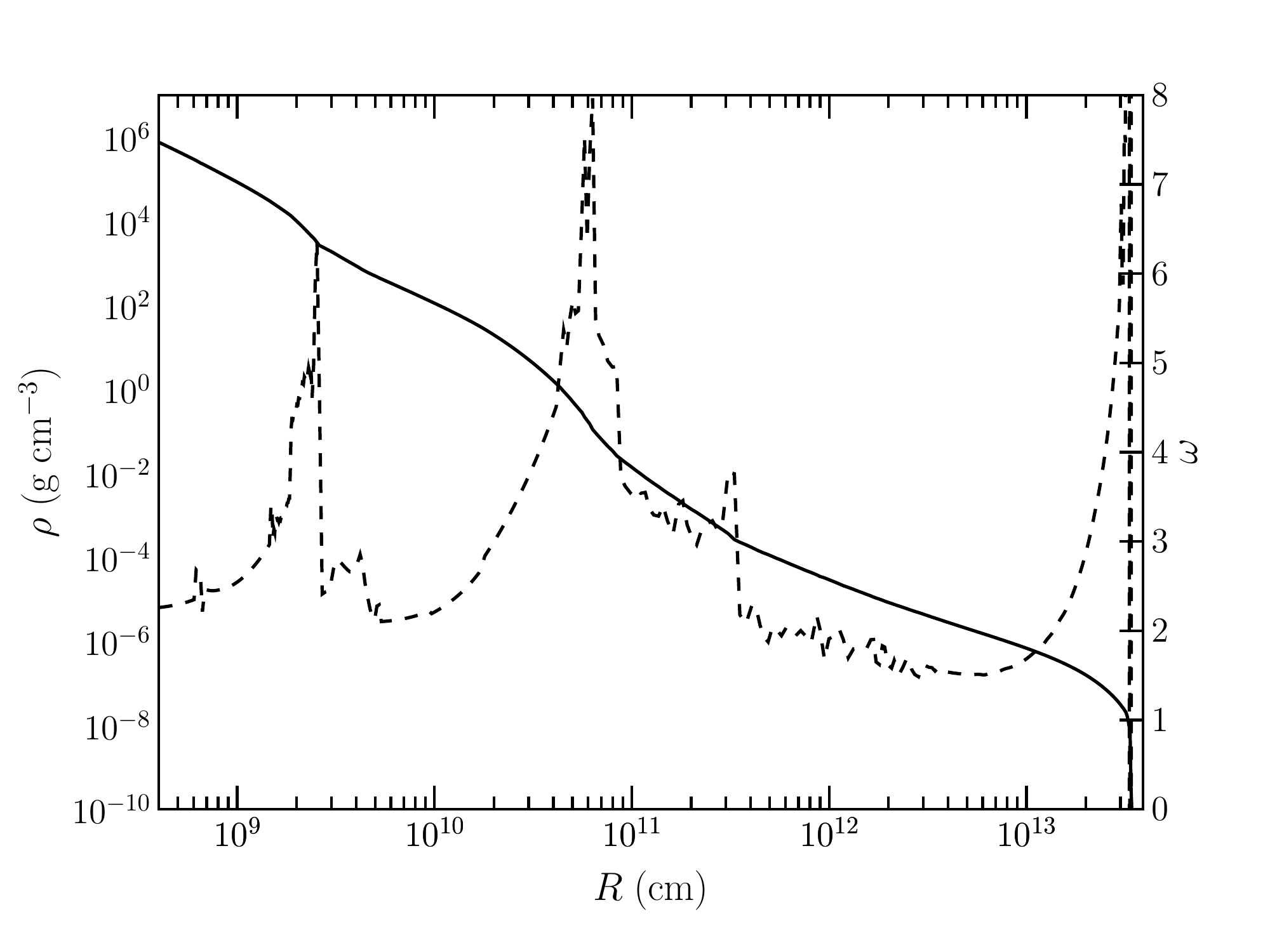}
\caption{Radial abundance (left) and density (right) profiles of the red supergiant progenitor model used as initial conditions for our simulations.  In the left panel, $X_i$ is the mass fraction of the atomic species listed in the legend.  Along with the density profile (solid line) on the right, we plot the logarithmic slope of the density, $\omega = - d \ln \rho / d \ln R$ (dashed line).  The spikes in $\omega$ correspond to layer transitions in the progenitor:  C/O--He ($2\times10^9$ cm), He--H ($6\times10^{10}$ cm), and H--wind ($3\times10^{13}$ cm).}
\label{fig:densprof}
\end{figure}

\begin{figure}
\centering
\plottwo{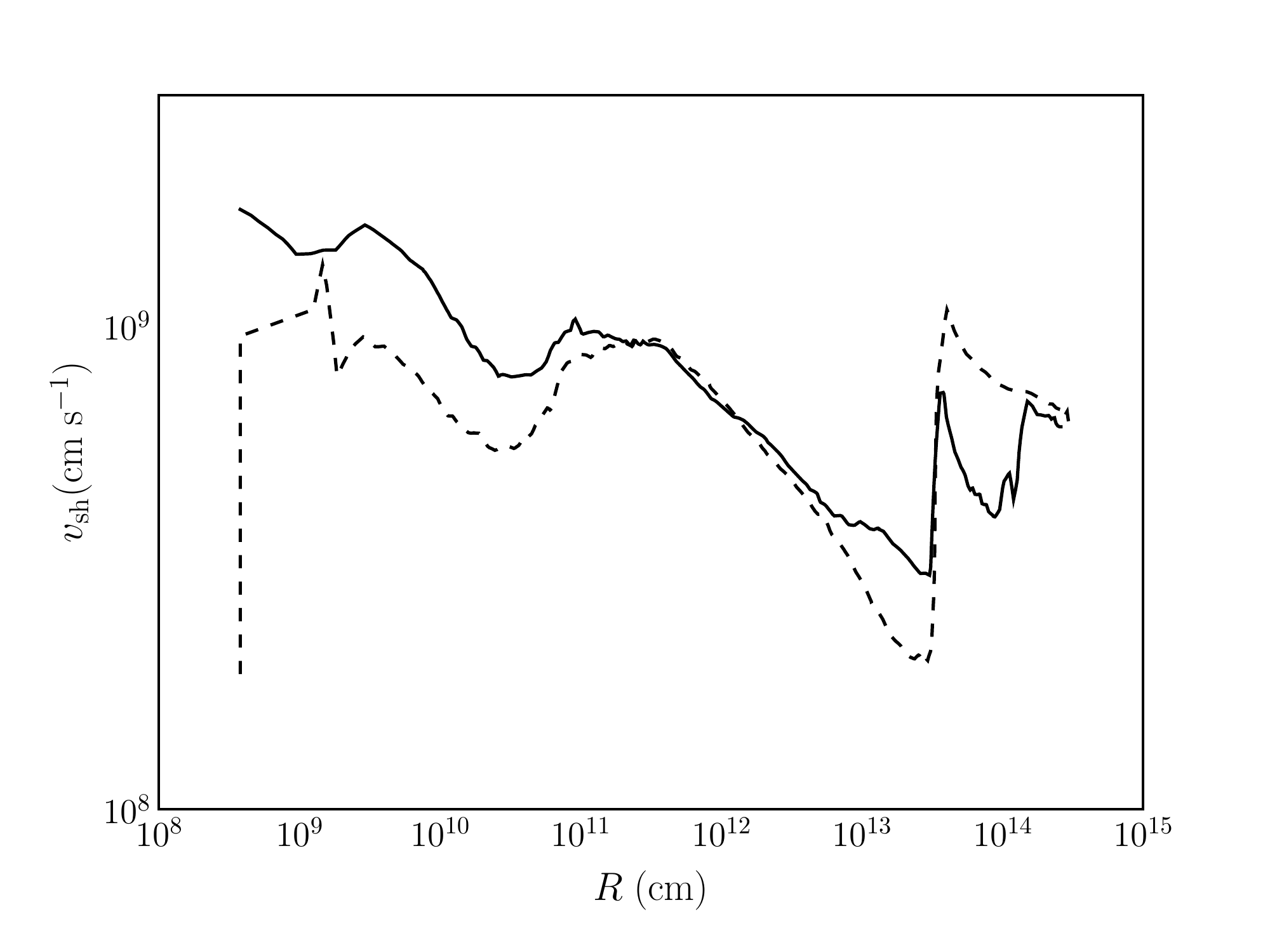}{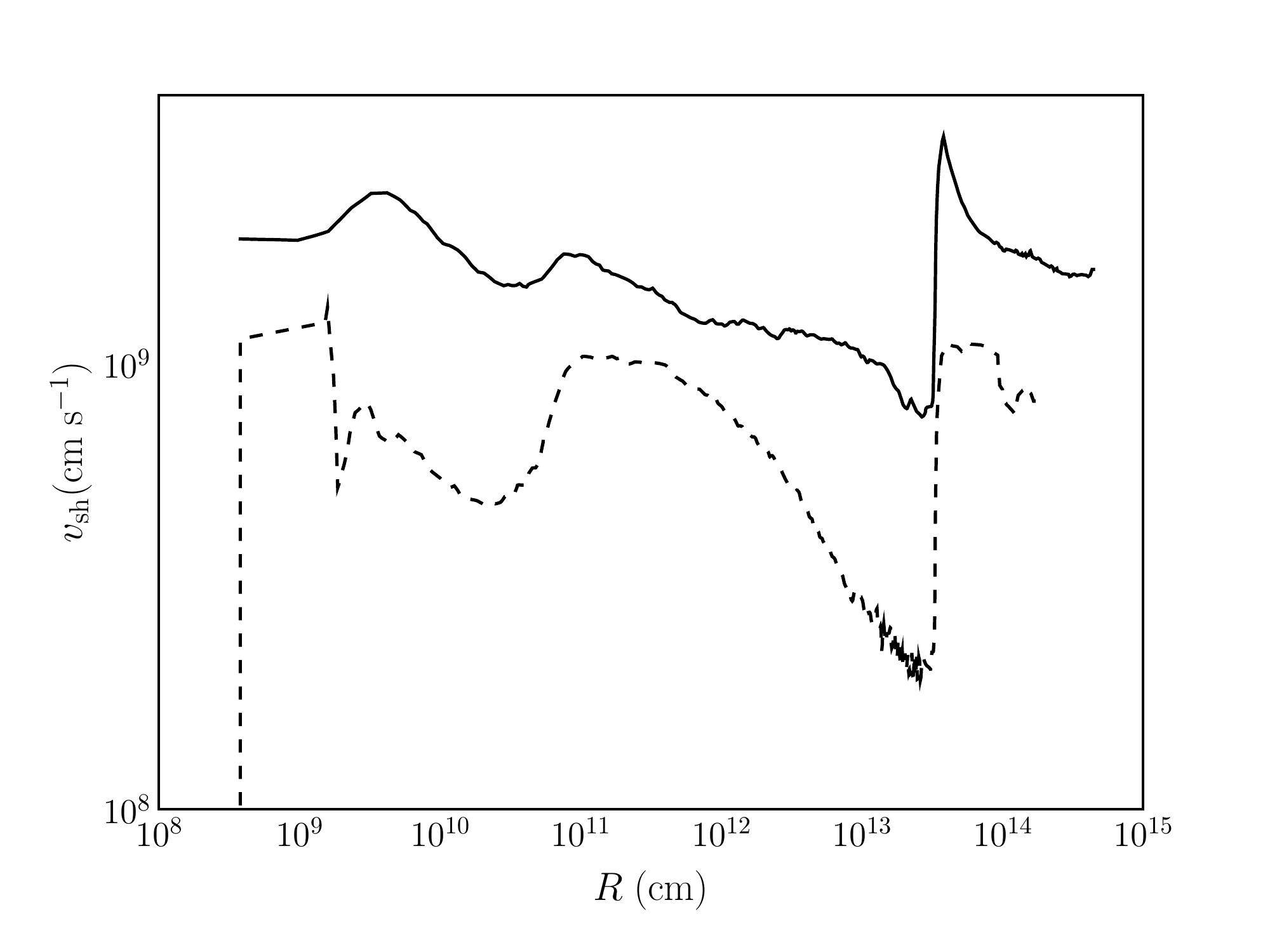}
\caption{Shock velocity as a function of radius for models v1m12 (thermal energy-dominated, left) and v3m12 (kinetic energy-dominated, right).  The solid line is the shock velocity along the pole ($\theta = 0$\degr) and the dashed lined is the shock velocity along the equator ($\theta =$ 90\degr).  For model v1m12, the shock structure becomes roughly spherical in the H envelope; the solid and dashed lines overlap until an instability finger grows up the axis and overtakes the polar shock around $5\times10^{12}$ cm.  The shock structure never becomes spherical in the case of model v3m12 and instability fingers overtake the shock much earlier, around $7\times10^{11}$ cm.  The equatorial shocks are well-defined only at radii greater than about $2\times10^9$ cm.}
\label{fig:vsh}
\end{figure}

In this section we provide a general analysis and discussion of the dynamics of our explosion models.  The dynamics of the various jet models are primarily determined by the nature of the jets themselves, however, the structure of the progenitor star also affects the behavior of the blast waves.  We show the radial chemical and density structures of the progenitor in Figure \ref{fig:densprof}.  While the models we present here are multidimensional in nature and not well described by one-dimensional models, we can elucidate the dynamics by considering the evolution in the polar and equatorial regions separately.  In Figure \ref{fig:vsh} we show the shock velocities in the polar and equatorial directions as functions of radius in the progenitor star for models v1m12 and v3m12.  

The evolution of the shock velocities in the polar and equatorial directions is described qualitatively by the similarity solution for spherical adiabatic blast waves \citep{Taylor:46, Sedov:59, chev76} in power-law atmospheres described by $\rho(r) \propto r^{-\omega}$.  Such solutions give the shock radius in time as $ R_{\rm sh}(t) = K^{1/(5-\omega)} t^{2/(5-\omega)}$, where $K$ is a constant, accleration-like term dependent on the central density of the atmosphere and the energy of the explosion.  We can then write the shock velocity as a function of radius as
\begin{equation}
\label{eq:vshock}
v_{\rm sh}(R_{\rm sh}) =  v_{\rm sh,0} \left (\frac{R_{\rm sh}}{R_{\rm sh,0}} \right )^{\omega/2 - 3/2},
\end{equation}
where $v_{\rm sh,0}$ is the shock velocity at some initial radius $R_{\rm sh,0}$.  Equation (\ref{eq:vshock}) shows that for $\omega < 3\ (\omega > 3)$ the shock decelerates (accelerates), and for $\omega =3$ the shock velocity is constant.  In Figure \ref{fig:densprof} we also plot the power-law slope of the progenitor density profile, $\omega = - d \ln \rho / d \ln R$.  Comparison with Figure \ref{fig:vsh} shows that, as expected, when $\omega>3$ the shocks accelerate and when $\omega < 3$, they decelerate.  The polar shocks in all models depart from this behavior in the hydrogen envelope due to the growth of instabilities in the post-shock flow.  We discuss how instabilities affect the shock evolution in more detail in \S\ref{sec:clumps}.

The remainder of this section is a basic description of the dynamics of our explosion simulations.  It is included for completeness, however, the evolution is non-trivial and we attempt to describe only what is most pertinent to the ensuing analysis.\footnote{The interested reader may view movies available at \url{http://www.as.utexas.edu/$\sim$smc/SNmovies.html}. }  In \S\ref{subsec:kejets} we describe the evolution of the kinetic energy-dominated jets in the progenitor core (\S\ref{subsec:kecore}), progenitor envelope (\S\ref{subsec:keenv}), and circumstellar wind (\S\ref{subsec:kewind}).  In \S\ref{subsec:thjets} we give an analogous discussion for the thermal energy-dominated jets.  In \S\ref{subsec:RMinst} we discuss the conditions that lead to instability growth that reaches the forward shock.  Readers not interested in the detailed kinematics may skip to \S\ref{sec:clumps} where we describe the development of material clumps in the jet models.  For reference, the transition from the C/O core to the He core is at a radius of about $2\times10^9$ cm and the transition from the He core to the H envelope is located at about $6\times10^{10}$ cm (see Fig. \ref{fig:densprof}).  

\subsection{Kinetic Energy Jets}
\label{subsec:kejets}
The jet models experience three major phases of evolution:  expansion through the core, expansion through the hydrogen envelope, and shock breakout and expansion into the circumstellar medium.  We illustrate these phases of evolution in Figures \ref{fig:v3m12core} -- \ref{fig:v3m12wind} for model v3m12.  The two higher-speed kinetic energy models evolve in a manner very similar to model v3m12, and we will, therefore, restrict our detailed discussion in the following sections to v3m12 alone.  We give a comparison of the final density structures of these three models in Figure \ref{fig:kineticjets}.  The only significant  difference in evolution as compared to model v3m12 comes early while the jets are still in the progenitor core.  In models v5m06 and v6m04, Kelvin-Helmholtz (KH) roll-ups grow much faster along the jet edges than in model v3m12.  This is attributed to greater shear with the progenitor structure in these models and, hence, a faster KH growth rate \citep{Chandra:61}.  Jet model v3m12 has parameters taken directly from those of \citet{kho99}, except that our progenitor model has an intact hydrogen envelope.  We can compare our results with those of \citet{kho99} up to the point when the jets exit the helium core at around $6\times10^{10}$ cm, which occurs at around 40 seconds, or the end of the first stage defined above.

\subsubsection{Evolution in the Core}
\label{subsec:kecore}

The injection of the jets into the core of the progenitor drives bow shocks that travel ahead of and laterally to the jets (see Fig. \ref{fig:v3m12core}).  Reverse shocks move away from the forward shocks.  Contact discontinuities exist between the jet material and the core material swept up by the forward shocks.  Vortices develop on the outer edges of the jet fluid (Figure \ref{fig:v3m12core}, left panel).  The edges of the jets themselves are also subject to small wavelength KH instabilities, but for the case of model v3m12, these instabilities grow slowly and do not become disruptive to the overall structure of the jets; this is in agreement with the findings of \citet{kho99}.  For the other kinetic energy models, v5m06 and v6m04, the KH instabilities grow more quickly.  After the jet injection stops at 2 s, the low-density region cleared out by the fast moving jet fluid closes, or heals, from the inner boundary out to the reverse shocks.  This healing can be understood as the result of the low-pressure ``vacuum'' created by the cessation of the jet inflow; the high pressure in the core of the star then acts to close this vacuum.  Around this same time, about 10 -- 15 seconds, the contact discontinuities show growth of Rayleigh-Taylor (RT) fingers (Figure \ref{fig:v3m12core}, middle and right panels).  Growth of these RT fingers is curtailed by shear with the surrounding post-shock flow.

Along the equatorial plane, the two bow shocks cross one another, creating a dense, high-pressure ``pancake'' of material with net positive radial velocity that develops into an outward-moving torus of twice-shocked material (see Figure \ref{fig:v3m12core}).  This torus moves supersonically with respect to the surrounding progenitor and drives a shock wave out in the equatorial plane (see Fig. \ref{fig:v3m12core}, right panel).  Crossing the entire helium core takes approximately 40 seconds for the polar shocks and about 90 s for the equatorial shocks.  

\subsubsection{Evolution in the Envelope}
\label{subsec:keenv}

The jets-torus structure that develops in the helium core is substantially altered by passage through the hydrogen envelope.  Initially, the shocks accelerate down the steep density gradient between the helium and hydrogen (see Figs. \ref{fig:densprof} \& \ref{fig:vsh}), but once in the $\omega < 3$ hydrogen envelope, a reverse shock develops as the forward shock is decelerated.  The reverse shock sweeps up the jet material and helium into a thin, dense shell.  At the jet-star contact discontinuities, Richtmyer-Meshkov (RM) fingers develop, seeded by the pre-existing RT fingers of jet material (see middle pane of Figure \ref{fig:v3m12env}).  The RM fingers grow at roughly the free-fall speed and are able to reach the forward shock and perturb it \citep[see \S\ref{sec:clumps} and][]{miles08}.  Before being reverse shocked, the edges of the equatorial torus develop KH rollers (Figure \ref{fig:v3m12env}, middle and right panels).  The strong reverse shock, however, wipes out much of this detail as it sweeps up a thin shell.  This shell is itself unstable and develops RT and RM fingers that become prominent as the explosion proceeds through the envelope. 

\subsubsection{Evolution Following Shock Breakout}
\label{subsec:kewind}

Figure \ref{fig:v3m12wind} shows the evolution of model v3m12 as the shocks exit the star and enter the circumstellar environment.  Around 3$\times 10^4$ seconds after the start of the simulation, the polar shocks break out of the surface of the progenitor model (left panel of Figure \ref{fig:v3m12wind}).  Due to slight differences between the evolution of the north and south jets, the northern shock breaks out slightly before the southern.  The source of this north-south asymmetry is artificial.  The build up of small numerical error after many time steps leads to non-uniform refinement in the the north and south.  This, in turn, causes the north-south solutions to diverge in the subsequent evolution.  The shocks accelerate down the steep density gradient in the outer envelope and enter the lower-pressure circumstellar region (see Fig. \ref{fig:densprof}).  The shocks travel laterally and nearly encompass the entire perimeter of the star before the equatorial shock breaks out, roughly $10^5$ seconds after the polar shocks break out (middle and right panels of Figure \ref{fig:v3m12wind}).  The reverse shocks sweep up another thin shell of material that is RT unstable.  The simulation ends at $5\times 10^5$ seconds (5.79 days), at which time the explosion is in approximate free expansion.  The final kinetic energy of the ejecta is $\sim5\times10^{50}$ erg for model v3m12, whereas for models v5m06 and v6m04 the final ejecta kinetic energies are both $\sim7\times10^{50}$ erg.

\subsection{Thermal Energy Jets}
\label{subsec:thjets}

Model v1m12 is the only model in which the initial energy budget of the jets is dominated by internal energy, as opposed to kinetic energy.  Since we have fixed the energy and power to be the same in all four models, the slower speed of v1m12 requires higher temperature to attain equivalent total power.  Because not all of the internal energy in the jets is converted to kinetic energy, v1m12 produces a slightly under-energetic explosion; the ejecta posses total kinetic energy of $\sim3\times10^{50}$ erg.

\subsubsection{Evolution in the Core}

Following the relatively low-speed jet inflow, the thermal energy-dominated jet fluid quickly expands in all directions as the initial internal energy is converted to kinetic energy.  The jet fluid attains a maximum radial velocity of about $3\times10^9$ cm s$^{-1}$ (see Figure \ref{fig:v1m12velx} for the early velocity field in this model).  The jets widen much more quickly compared to the other jet models due to a higher degree of lateral adiabatic expansion (see left panel of Figure \ref{fig:v1m12core}).  After the jet inflow shuts off, around two seconds, the reverse shock moves quickly down the axis to the center of the star.  The reverse shock closes the evacuated region opened by the jet at approximately four seconds.  This is in contrast with the kinetic energy models wherein the jet cavity closes mostly from the bottom up.  The less-directed nature of the energy deposition in this model leads to the spread of a significant fraction of the kinetic energy to large angles from the jets (see \S\ref{sec:geometry}).  As a result, the equatorial outflow is larger and more energetic at early times than in the kinetic energy models.  Around 50 s, the polar shocks exit the helium core, followed by the lateral shocks at $\sim$80 s (middle and right panels of Figure \ref{fig:v1m12core}).  

\subsubsection{Evolution in the Envelope and Shock Breakout}

In the hydrogen envelope, the shock structure of model v1m12 becomes effectively spherical (see Figure \ref{fig:v1m12env}).  A thin, dense shell develops behind the reverse shock that is, as in the kinetic energy models, RT unstable (Figure \ref{fig:v1m12env}, middle panel).  As the reverse shock hits the jet-star contact discontinuities, RM fingers develop but do not have enough time to catch the shock.   As shown in Figures \ref{fig:v1m12env} (right panel) and \ref{fig:v1m12wind}, RT fingers form in the thin shell and grow to large sizes before the shock breaks out of the progenitor at $\sim10^5$ s.  The shock departs the star roughly spherically and an unstable shell forms at the wind-induced reverse shock, analogous to the the kinetic energy models (Figure \ref{fig:v1m12env}, right panel).   At the simulation's end, the model is approximately freely expanding.

\subsection{High-velocity Instabilities}
\label{subsec:RMinst}

In the kinetic energy models, the shock is overtaken by downstream instabilities (see middle and right panels of Fig. \ref{fig:v3m12env}).  This can occur for RT fingers in the fast-growth phase or RM fingers with Atwood number near unity \citep[see][]{miles08}.  In model v3m12 and the other kinetic energy models, RM fingers from the jet fluid contact discontinuity are indeed able to reach the forward shock in the hydrogen envelope.  As the forward polar shock enters the H envelope, a reverse shock forms and travels downstream slowing the fast-moving post-shock material to a new post-forward-shock velocity.  The contact discontinuity between the jet fluid and the helium has initial perturbations arising from earlier RT instability.  Once reverse-shocked, the contact discontinuity becomes RM unstable and the initial RT perturbations grow into RM fingers.  In the frame of the contact discontinuity, the impulsive instability velocity is \citep{youngs86}
\begin{equation}
\label{eq:vinst}
v_{\rm RM} = \frac{2\pi a_0}{\lambda} \Delta U \frac{\rho_1 - \rho_2}{\rho_1 + \rho_2},
\end{equation}
where $\Delta U$ is the change in velocity caused by the reverse shock, $\lambda$ is the wavelength of the initial perturbation, and $a_0$ is the amplitude of the initial perturbation.  Prior to encountering the reverse shock, the contact discontinuity has perturbations characterized approximately by $a_0 / \lambda \sim0.5$.  In the frame of the contact discontinuity, for model v3m12 we estimate that the reverse shock causes an impulse $\vert \Delta U \vert \sim  4\times10^7$ cm s$^{-1}$.  The contact discontinuity has a post-reverse-shock Atwood number $(\rho_1 - \rho_2)/(\rho_1 + \rho_2) \sim 3/4$, and so $v_{\rm RM} \sim 1\times10^8$ cm s$^{-1}$.  Transforming back into the lab frame, this implies that being reverse-shocked {\it speeds up} the initial perturbations in the contact discontinuity relative to its pre-reverse-shocked velocity by about $1\times10^8$ cm s$^{-1}$, meaning that the dense RM fingers are now moving at $v_0 +\ v_{\rm RM} \sim 1.2\times10^9\ {\rm cm}\ + 1\times10^8\ {\rm cm} \sim 1.3\times10^9$ cm s$^{-1}$ in the lab frame.  This is approximately the speed of the forward shock at the time the reverse shock impacts the contact discontinuity.

If drag on the RM fingers is negligible, they will move ballistically with the speed given by equation (\ref{eq:vinst}).  As the shock encounters the shallow density profile of the hydrogen envelope (see Fig. \ref{fig:densprof}), it will slow below the speed of the RM fingers and they will overtake the shock.  This is consistent with what is seen in our simulations.  In Figure \ref{fig:vsh} the shock velocity departs from the similarity solution given by equation (\ref{eq:vshock}) at a radius of about $6\times10^{11}$ cm, where the RM fingers catch up.  The new shock velocity is then set by the speed of the fingers which initially agrees with the estimate given above.  The RM fingers and, hence, the forward shock do not continue at the constant free-fall velocity, but undergo deceleration due to drag and shock dynamics.

Fast-moving RM fingers also develop at the jet-star contact discontinuity in model v1m12, however the separation between the contact discontinuity and the forward shock is much greater in this model (Fig. \ref{fig:v1m12env}, left panel) and the fingers catch the shock at a much larger radius ($\sim7\times10^{12}$ cm).  Also, only one finger catches the shock in this model and it grows immediately along the axis.  Growth of this finger may be artificially enhanced by the numerical coordinate singularity along the pole. 

\section{Clumps in Jet-Driven Supernovae}
\label{sec:clumps}

\subsection{Nickel Clumps}
Jet-driven supernovae have a robust mechanism that transports heavier elements beyond lighter:  the jets themselves.  It is plausible that the jets contain a large fraction of nickel, or even, as we have assumed, are entirely nickel.  This, then, permits the effective transport of nickel to velocities and radii where it can contribute to hydrogen and helium excitation observed at early times in SNe.  In Figure \ref{fig:mofr} we show the final radial distribution of the masses of H, He, and Ni for the four models.  As can be seen for model v1m12, a large mass of nickel has been carried out into the hydrogen envelope where it can excite the H$\alpha$ and He I 10830\AA\ lines through radioactive decay.  The presence of a bipolar arrangement of nickel clumps has been inferred from spectral modeling of SNeII-P \citep{Elmhamdi:03, Chugai:05, Chugai:06}, and our jet-driven models produce such a configuration (by {\it Ansatz}).  We plot the Ni, H, and He mass distributions as functions of $z$-direction velocity ($v_z = v_r \cos\theta$) in Figure \ref{fig:dmdvz}.  As can be seen, the nickel distribution of v1m12 consists of two oppositely directed clumps moving at a typical velocity of 1500 km $s^{-1}$.  Each clump is composed of about 0.05 $M_\odot$, though this is certainly an upper limit because of our assumption that the jets are composed entirely of nickel.  Figure \ref{fig:ni56_v1} shows a plot of the $^{56}$Ni mass fraction for model v1m12 at the end of the simulation, illustrating the nickel clumps.

The mass distributions for the kinetic energy models (v3m12, v5m06, and v6m04) are quite different than that of model v1m12.  The mass distributions for these models are given in Figures \ref{fig:mofr} -- \ref{fig:dmdvz}.  The kinetic energy models drive a large fraction of the nickel mass beyond the hydrogen and helium.  This may reduce the amount of nickel available to excite hydrogen and helium spectral lines.  Only about half of the Ni mass of each model overlaps in radius with H and He, or about 0.05 $M_\odot$ for v1m12, 0.03 $M_\odot$ for v5m06, and 0.02 $M_\odot$ for v6m04.  The existence of bare nickel beyond any hydrogen may also have implications for the gamma-ray signal from such explosions, as a higher fraction of gamma-rays from the decay of nickel would escape.  Figure \ref{fig:dmdvz} shows that the bipolar nickel clumps reach much higher relative velocities ($\sim 5000$ km s$^{-1}$) than for model v1m12.  The final nickel distribution is shown for v3m12 in Figure \ref{fig:ni56_v3}.

\subsection{Helium Clumps}

\citet{Fassia:98} show that He I 10830\AA\ emission from SN 1995V requires the transport of nickel to high velocities {\it and} the presence of pristine helium in the hydrogen envelope.  In our models, RT fingers that develop at the He/H interface produce pure-He clumps in the H envelope (see Figures \ref{fig:v3m12env} and \ref{fig:v1m12env}, middle and right panels).  In Figure \ref{fig:dclumpdv} we show the He clump mass fraction $\chi_{\rm He} = M_{\rm He, clumped} / M_{\rm He}$ for our four models along with the H mass distribution.  We consider a cell to be a pristine helium clump if it is more than 90\% helium by mass.  While the clumped fraction at certain velocities can be quite high ($\chi_{\rm He,max}\sim0.1-0.2$), the total fraction of He mass in clumps within the H envelope is only 1-2\% across all four models. This is roughly a factor of ten less than what \citet{Fassia:98} find gives a good fit to the He I emission.  We note, however, that our clumped fractions are lower limits due to numerical resolution.  Any He clumps that fall below the minimum cell size become numerically mixed with other material.  Despite the low clumped fractions, the strong He I emission may still be recovered from our models.  We find a nickel mass at large radii that is a factor of two larger than that assumed by \citet{Fassia:98}, and if this amount of nickel is included in the spectral modeling, the exposure of the He to a larger amount of Ni might enhance the He I emission to levels in accord with the observations.





\section{Aspherical Explosions and Implications for Supernova Polarization}
\label{sec:geometry}

The current observational data show that SNeII-P have time-dependent polarization signals, indicating that the shapes of SN photospheres change as the explosions expand.  In this section we discuss the shapes of our simulated explosions and their implications for SNeII-P observations.  The final shapes of our models are controlled by the effectiveness of the jets to transport kinetic energy from the jet direction to more equatorial angles.  In Figure \ref{fig:ekv1}, we show the angular distribution of kinetic energy at various times for the jet models.  The kinetic energy distribution of model v1m12 at the end of the simulation is nearly flat, indicating a roughly spherically-symmetric energy distribution.  The other three models leave significant amounts of kinetic energy in angles associated with the jets.  This implies that the kinetic energy-dominated jets are ``ballistic'' and plow their way through the progenitor without losing much kinetic energy to more equatorial angles through fragmentation and expansion.  

Carrying our simulations to quasi-homologous expansion allows us to explore the overall geometry of the explosions at times when supernovae are actually observed.  As dramatically demonstrated by \citet{Leonard:06}, there is compounding evidence that SNeII-P exhibit very low polarization through the plateau phase but, as the photosphere recedes below the hydrogen envelope, the polarization increases greatly.  Since the continuum polarization is primarily the result of electron scattering, we may compare the shapes of surfaces of constant electron scattering optical depth, $\tau_{\rm es}$, in our models with SN geometries inferred from polarization measurements.  To compute $\tau_{\rm es}$, we first expand our simulation results to the desired time using the relation for homologous expansion, $r=vt$.  We calculate the new temperature of our models by assuming that the expansion is adiabatic.  Considering only radiation and gas pressure, the internal energy is then
\begin{equation}
\label{eq:eint}
\epsilon = \frac{a_{\rm rad} T^4}{\rho} + \frac{3}{2} \frac{k_{B} T}{\mu m_{\rm H}} = {\rm const.}
\end{equation}
Assuming adiabatic expansion neglects radiative losses, however, this will be adequate for qualitatively estimating the photosphere shapes.  The new gas temperature is found by solving equation (\ref{eq:eint}), which is then used to calculate the electron fraction arising from thermal ionization of hydrogen, $x_T$, using Saha's equation.  We include contributions to the electron fraction due to gamma-ray ionizations from the radioactive decay of $^{56}$Ni in a manner similar to that of \citet{Utrobin:95}.  We first calculate the local gamma-ray luminosity on a cell-by-cell basis as
\begin{equation}
\label{eq:Lni}
L_{\rm Ni} = (6.45 e^{-t/8.8\ {\rm day}} + 1.45 e^{-t/111.3\ {\rm day}}) \times 10^{43} (m_{\rm Ni} / M_\odot),
\end{equation}
where $m_{\rm Ni}$ is the mass of nickel in a given cell.  Assuming that the gamma-ray luminosity contributes only to hydrogen ionization, and entirely locally within a cell, we find the electron fraction due to nickel decay by iteratively solving
\begin{equation}
\label{eq:efrac}
\alpha_{32} x_\gamma^2 n_{\rm H}^2 dV = (1-x_\gamma^{0.28}) \frac{\gamma_i L_{\rm Ni}}{E_{13}}
\end{equation}
for $x_\gamma$, the electron fraction arising from the gamma-rays, where $\alpha_{32} =3.37\times10^{-13}$ cm$^3$ s$^{-1}$ is the de-excitation rate at 5000 K, $n_{\rm H}$ is the hydrogen concentration, $dV$ is the cell volume, $\gamma_i=0.39$ is a hydrogen ionization branching ratio, and $E_{13}=12.1$ eV is the transition energy from the second excited state to the ground.  Assuming that all gamma-ray energy is locally deposited within a cell is valid because, in our case, the typical gamma-ray mean free path, $l = 1/\kappa_\gamma \rho$, is small compared to the simulation cell size for a gray gamma-ray opacity of $\kappa_\gamma \sim 0.03\ {\rm cm^2\ g^{-1}}$ due to electron scattering and absorption \citep{Sutherland:84}.  The total electron fraction is taken to be $x = {\rm max}[1,\ x_T + x_\gamma]$.  The electron scattering optical depth is then \citep[see][]{Jeffery:90}
\begin{equation}
\label{eq:taues}
\tau_{\rm es} = 0.4 \int_\infty^R x \rho_{\rm H} dr,
\end{equation}
 where $\rho_{\rm H}$ is the hydrogen mass density and $r$ is along the observer's line of sight.  

In Figure \ref{fig:tau_v1}, we show the evolution of the electron scattering surfaces as a function of time for model v1m12, calculated using the method described above.  Early-on, about 5 days after shock breakout, the photosphere ($\tau_{\rm es} \approx 1$) is slightly oblate.  This is a result of the {\it slightly} aspherical breakout.  The polar shocks erupt from the star about a third of a day before the equatorial shocks and have that much time to expand laterally, and thus the kinetic energy of the polar shocks is diluted and the shock slows.  On day 50, the photosphere is roughly spherical, in agreement with the early, plateau-phase polarization measurements of SN 2004dj \citep{Leonard:06}.  As the photosphere recedes, its shape elongates along the jet axis.  By day 100, the photosphere is a prolate spheroid of axis ratio $\sim 1.25$.  This is about the epoch associated with the end of the plateau phase in SNeII-P and about the time \citet{Leonard:06} saw a dramatic rise in the polarization signal from SN 2004dj.  \citet{Leonard:06} estimate an axis ratio of about 1.4 for SN 2004dj based upon the electron scattering spheroid polarization calculations of \citet{hoflich91}.  Given the many uncertainties in the interpretations of both the observations and our calculations, there is striking agreement between the photospheric evolution of model v1m12 and that of SN 2004dj. 

The kinetic energy models are far more asymmetric at all times, as is demonstrated for model v3m12 in Figure \ref{fig:tau_v3}.   Axis ratios for this model range from 2 to 3 from the earliest epochs until the explosion begins the transition to the nebular phase.  The shape of the photosphere on day 300 is actually more spherical compared with earlier times.  The dramatic asphericity of model v3m12 is inconsistent with spectropolarimetry of SNeII-P \citep{Leonard:01, Leonard:06, WW08}.  The shapes of the photospheres are similar for models v5m06 and v6m04.  This seems to indicate that if SNeII-P are driven by bipolar jets, such jets must be ``stalled'' by passage through a thick hydrogen envelope, meaning that they must efficiently lose kinetic energy to the envelope driving a more nearly spherical explosion.  A similar conclusion was reached by \citet{hof01}.  Our thermal energy-dominated jet model satisfies this criterion.  

\section{Shock Breakout}
\label{sec:breakout}

Recently, the {\it GALEX} satellite has observed two SNeII-P at, or near, the moment of shock breakout \citep{Schawinski:08, Gezari:08}.  \citet{Gezari:08} model the light curve as a spherical breakout, but as we show here, the breakout light curve will vary according to the asphericity of the shock breakout.  To illustrate this, we have computed breakout light curves of our models by first finding the photosphere ($\tau_{\rm es} =1$; via the method outlined in \S\ref{sec:geometry}) and then calculating the bolometric blackbody emission at the photospheric temperature in a piecewise fashion across the photosphere and summing these blackbody contributions over the photosphere.  The resultant light curves, along with the angled-averaged photospheric temperature, are shown in Figure \ref{fig:breakout}.  The significantly aspherical breakout of model v3m12 is apparent in the ``double-peaked'' light curve.  The breakout of the jets around 0.5 days, measured in time from the beginning of the simulation, is followed by a period of near-constant luminosity associated with the expansion of the jet cocoons.  The equatorial shock breakout leads to the second peak at 1.5 days.  The light curve of model v1m12 is not significantly different from that of a spherical shock breakout model \citep[see Figs. 1 and 2 of][]{Gezari:08}.  The slightly aspherical breakout of model v1m12 is evident in the light curve, but the first peak in this case is much smaller and is much closer in time to the large peak than in model v3m12.  Considering the {\it GALEX} observations, the early peak and plateau in the light curve of model v3m12 would have been detected for SNLS 04D2dc if the breakout were strongly aspherical.  The absence of this early peak and plateau in the breakout light curve indicates the breakout was indeed nearly spherical.  We do note, however, that the temporal resolution of the {\it GALEX} data is roughly one hour, such that the early peak in the light curve of v3m12 would have been only marginally detected.  The light curve of model v1m12 is in much better agreement with the {\it GALEX} data.  While our method of computing the light curves is crude and neglects many important aspects of shock breakout (such as the radiative precursor), our theoretical light curves do illustrate the significant differences that can arise from a strongly aspherical breakout and indicate that the characteristics of such a breakout could have been detected by {\it GALEX}, if they were present.




\section{Discussion and Conclusions}
\label{sec:discussion}

We have performed 2D axisymmetric, high-resolution simulations of stellar explosions driven by bipolar jets.  We have followed the evolution of four different jet models to near free expansion.  The energy budgets of  three of our jet models (v3m12, v5m06, and v6m04) are dominated by kinetic energy whereas that of one jet model (v1m12) contains a large fraction of thermal energy, and all jet models inject a total of $10^{51}$ erg.  This distinction in the four models drastically affects the resulting structures of the models.  The kinetic energy models remain highly elongated as they expand into the wind surrounding our progenitor model, whereas model v1m12 produces an effectively spherical outer envelope with large departures from symmetry in the core.   These shapes correspond to how efficiently the jet kinetic energies are isotropized.  Only model v1m12 results in an effectively spherically-symmetric kinetic energy distribution (see Fig. \ref{fig:ekv1}).  An anisotropic distribution of kinetic energy may have implications for the measurement of supernova energies from observations.   While radiative transport calculations of our results would be required to reproduce typical spectral lines used in estimating supernova energies, it is likely that the kinetic energy models, with their asymmetric energy distributions, would yield viewing angle-dependent kinetic energy estimates.  If a spherical explosion model were assumed when analyzing the observations of an aspherical supernova, errors in the estimate of the kinetic energy could arise.

We resolve the growth of fluid instabilities in the explosions.  We show that the jet fluid is unstable and develops several RT fingers that, when reverse-shocked in the H envelope, serve as initial perturbations for rapid RM growth.  The RM fingers travel at nearly the free-fall velocity and are able to overtake the shock in the kinetic energy models.  The edges of the jets are also KH unstable, but the growth rate depends sensitively on the speed and density of the jet fluid.  Model v3m12 does not show significant KH growth, but all three other models do.  The equatorial outflow in all four models is also subject to KH growth along the top and bottom surfaces, as predicted by \citet{Wheeler:08}.

\begin{figure}
\centering
\begin{tabular}{ccc}
\includegraphics[width=2.in, trim= .15in 0 3.23in 0in, clip]{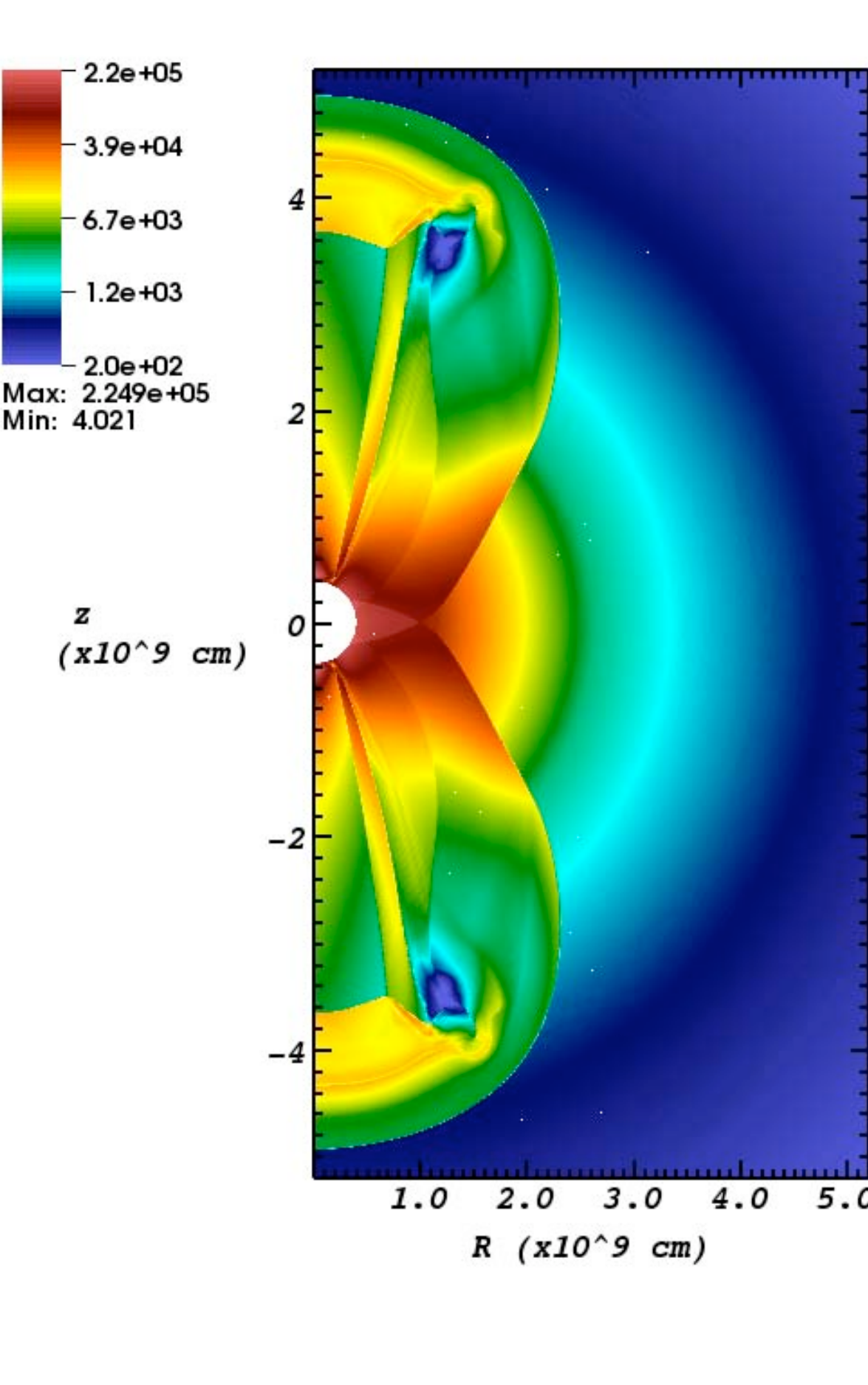} &
\includegraphics[width=2.in, trim= .15in 0 3.23in 0in, clip]{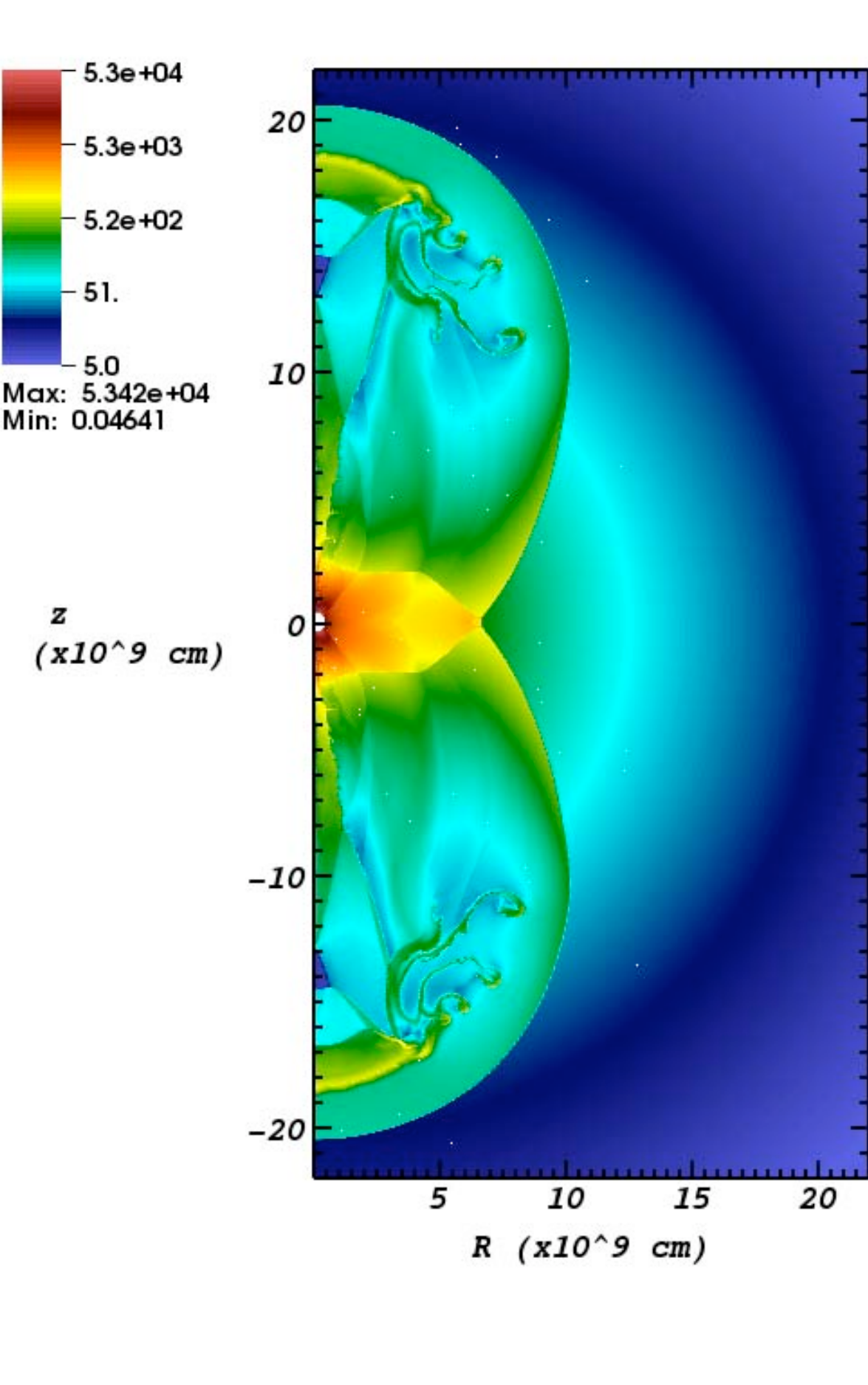} &
\includegraphics[width=2.in, trim= .15in 0 3.23in 0in, clip]{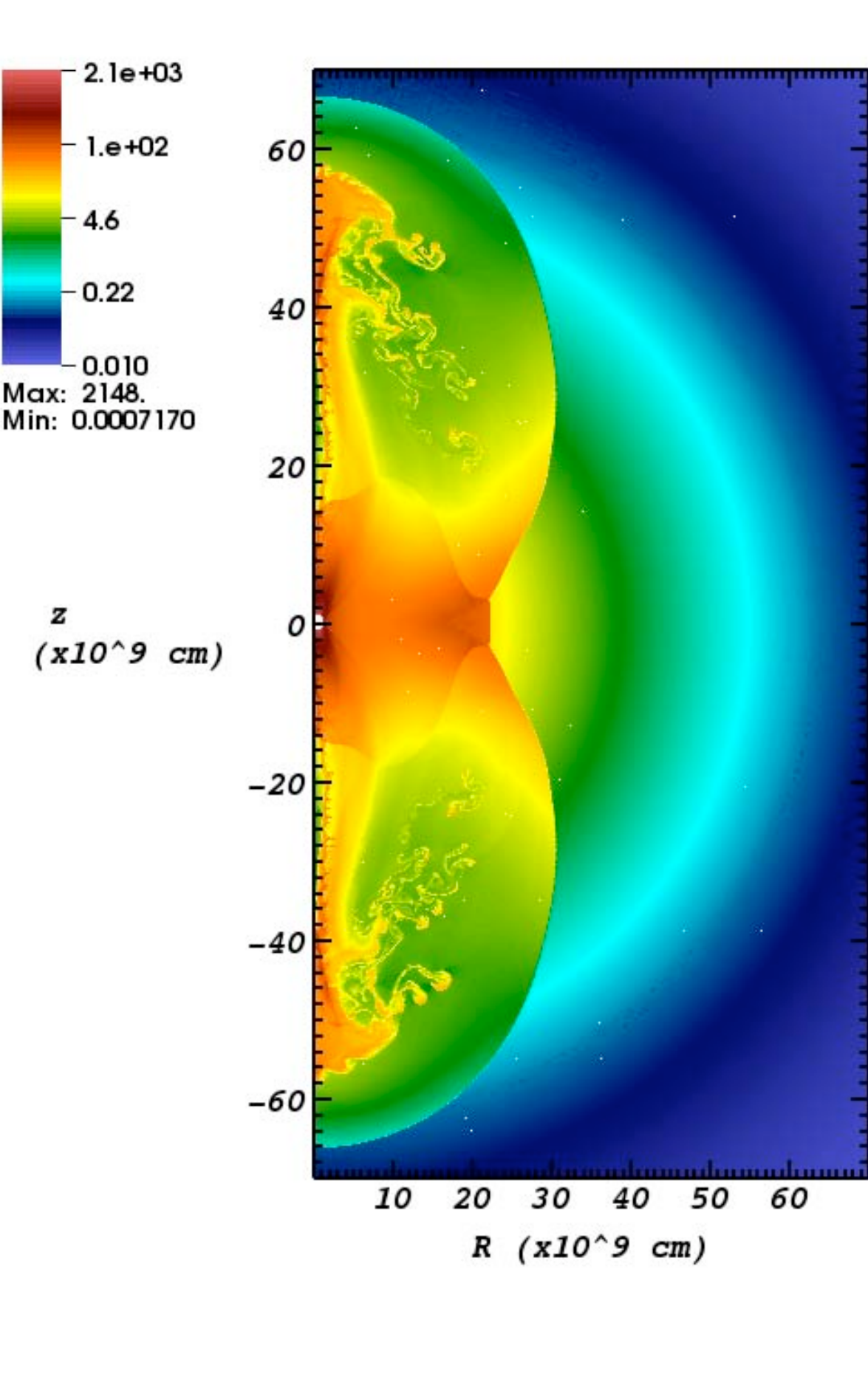}
\end{tabular} 
\caption[Model v3m12:  core]{Snapshots of the density evolution in kinetic energy-dominated model v3m12 before leaving the helium core of the star, which has a radius of $\sim 5\times10^{10}$ cm.  The axis lengths are given in centimeters and the density is plotted in g cm$^{-3}$.  The simulation times are, from left to right, 2 s, 10.6 s, and 40 s.  Note the stripping of jet material caused by shear with the stellar material and the subsequent development of small-scale, complex structure in the right panel.  The three panels also show the development and evolution of the equatorial structure resulting from the collision of the jet-induced shocks at the midplane.  The middle and right panels also show the beginning growth of RT fingers at the interface between the jet material and the shocked star material.} 
 \label{fig:v3m12core}
\end{figure}

\begin{figure}
\centering
\begin{tabular}{ccc}
\includegraphics[width=2.in, trim= .15in 0 3.23in 0in, clip]{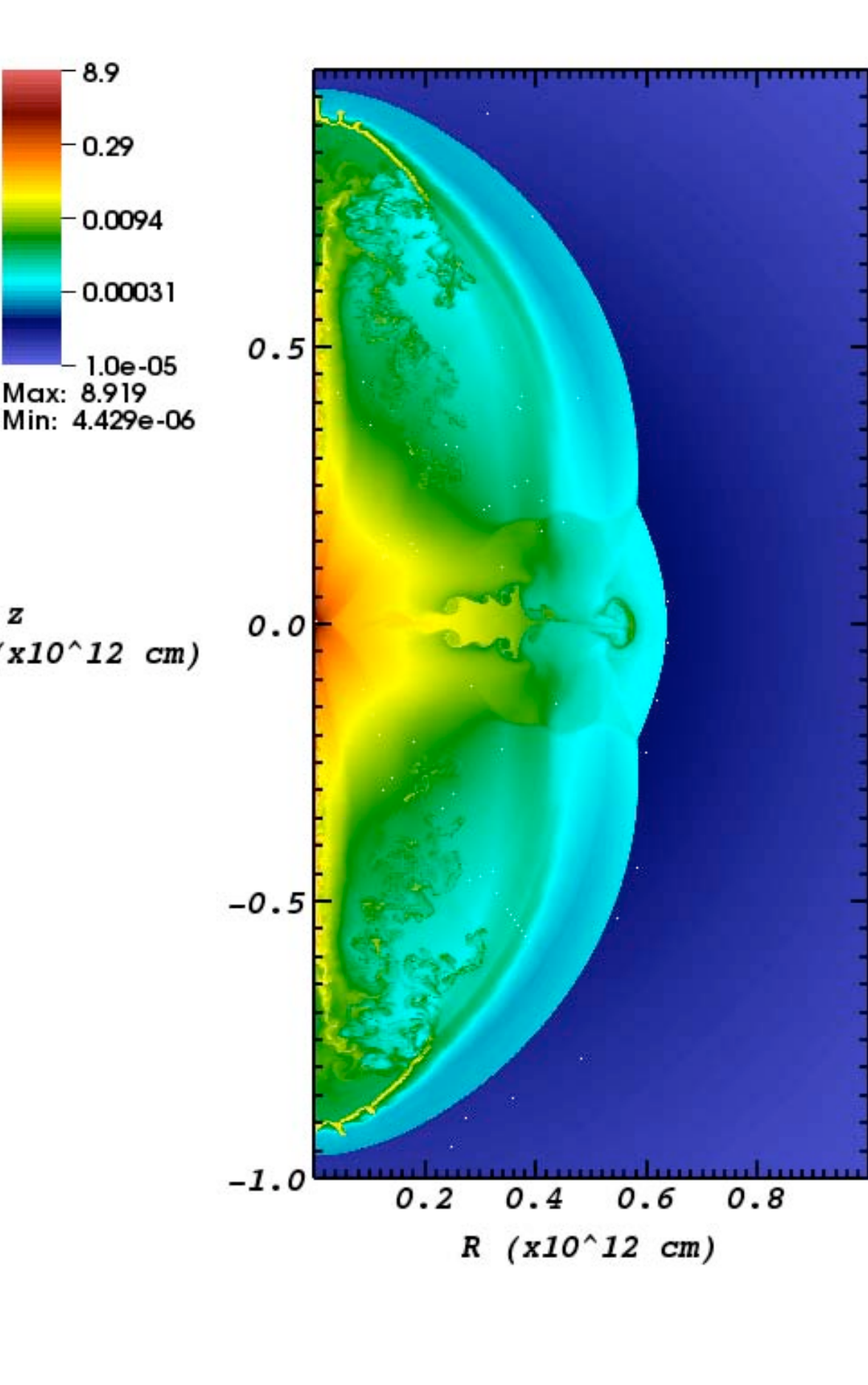} &
\includegraphics[width=2.in, trim= .15in 0 3.23in 0in, clip]{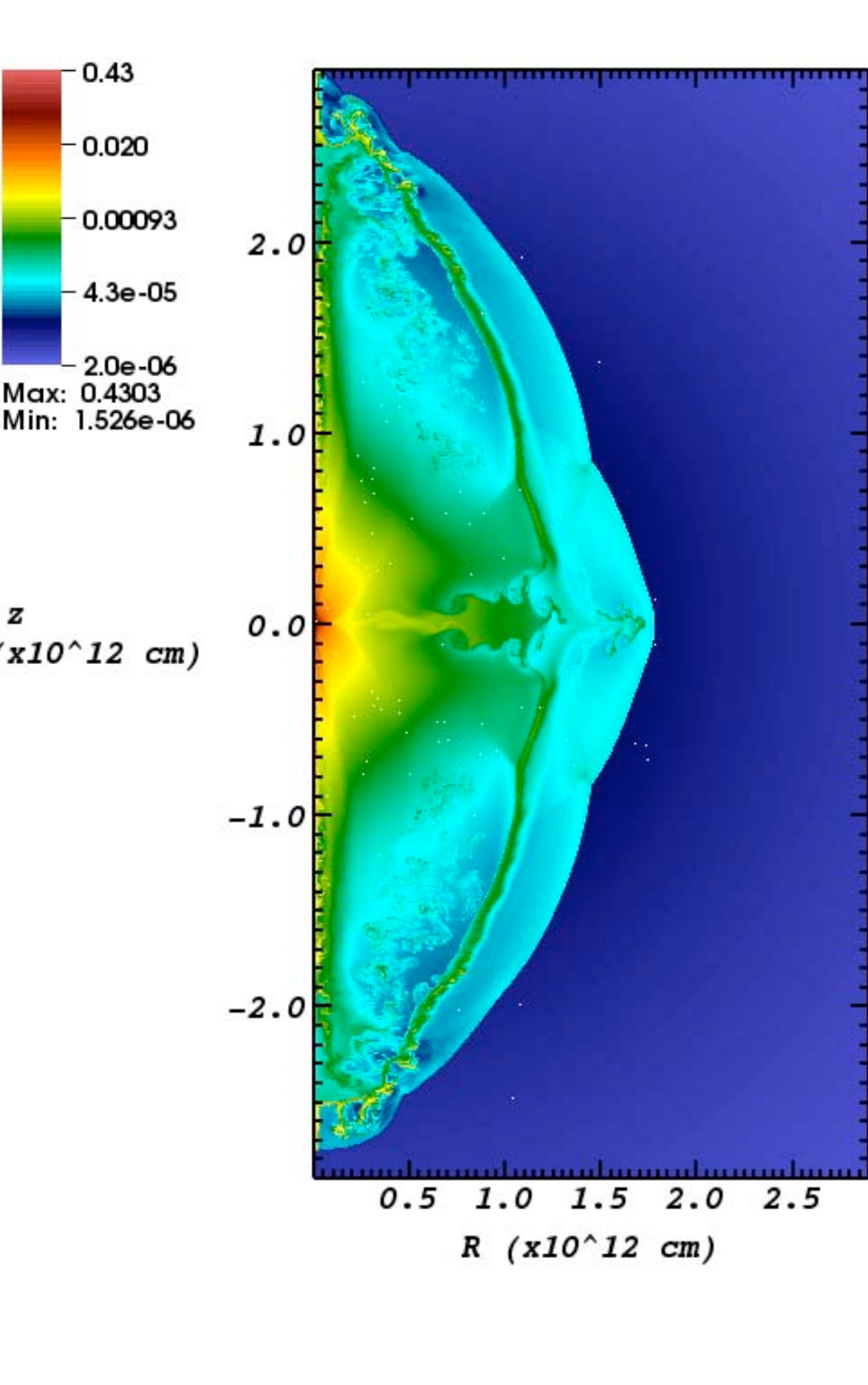} &
\includegraphics[width=2.in, trim= .15in 0 3.23in 0in, clip]{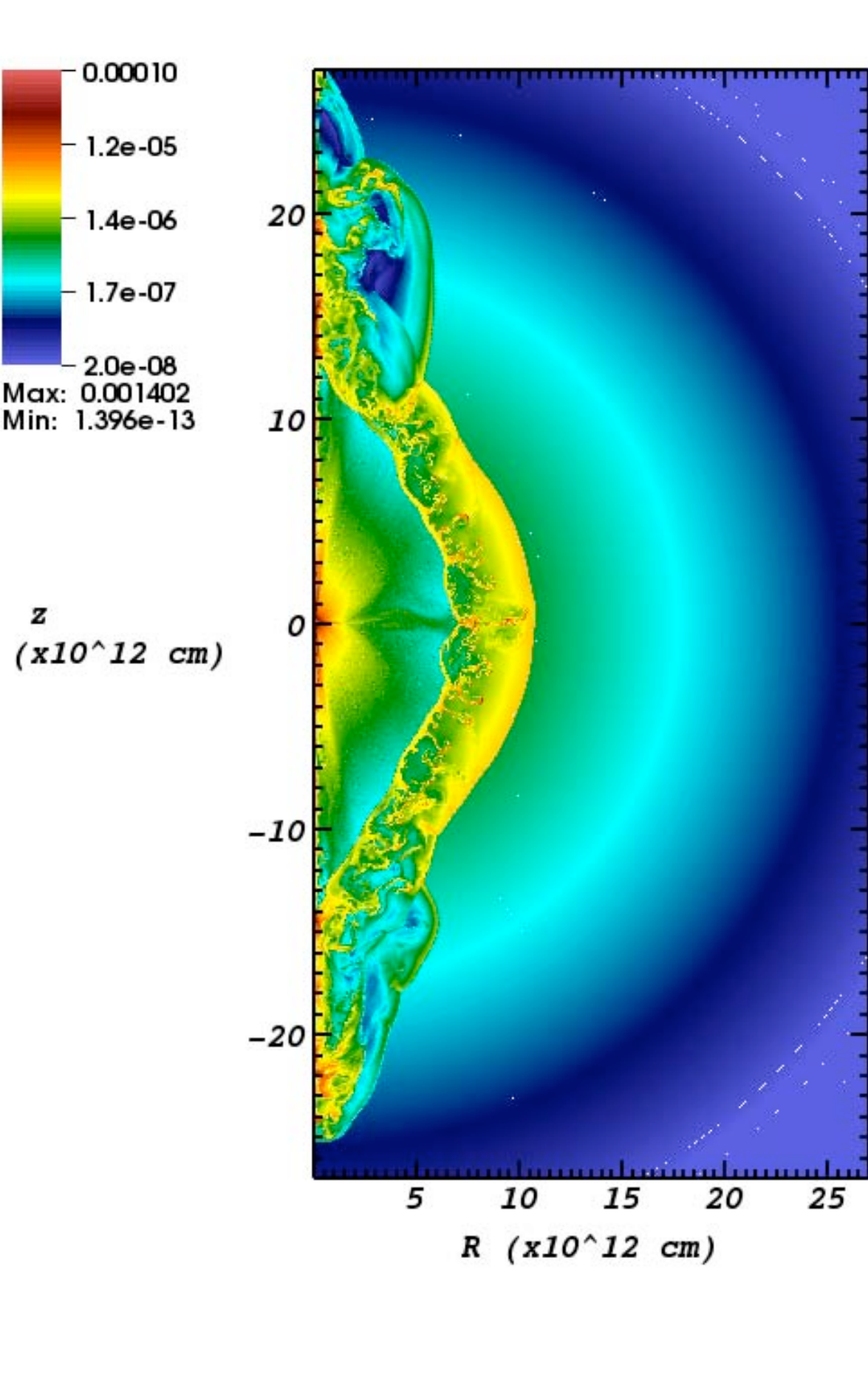}
\end{tabular} 
\caption[Model v3m12:  envelope]{Snapshots of the density evolution in kinetic energy-dominated model v3m12 while transiting the stellar envelope.  The axis lengths are given in centimeters and the density is plotted in g cm$^{-3}$.  The simulation times are, from left to right, 700 s, 2200 s, and 26,000 s.  In the equator, the ``pancake''-like outflow has developed KH roll-ups as it has sheared through the envelope of the progenitor.  The leading KH roll-up detaches from the bulk of this flow and impinges on the equatorial shock, as shown in the middle panel.  The RT fingers of jet fluid are swept up by the reverse shock into a thin, dense shell.  The fingers serve as an effective initial perturbation to instability growth in this shell that, as seen in the middle and right panels, is able to reach up to and ahead of the forward shock, thus perturbing the forward shock structure.}
\label{fig:v3m12env}
\end{figure}

\begin{figure}
\centering
\begin{tabular}{ccc}
\includegraphics[width=2.in, trim= .15in 0 3.23in 0in, clip]{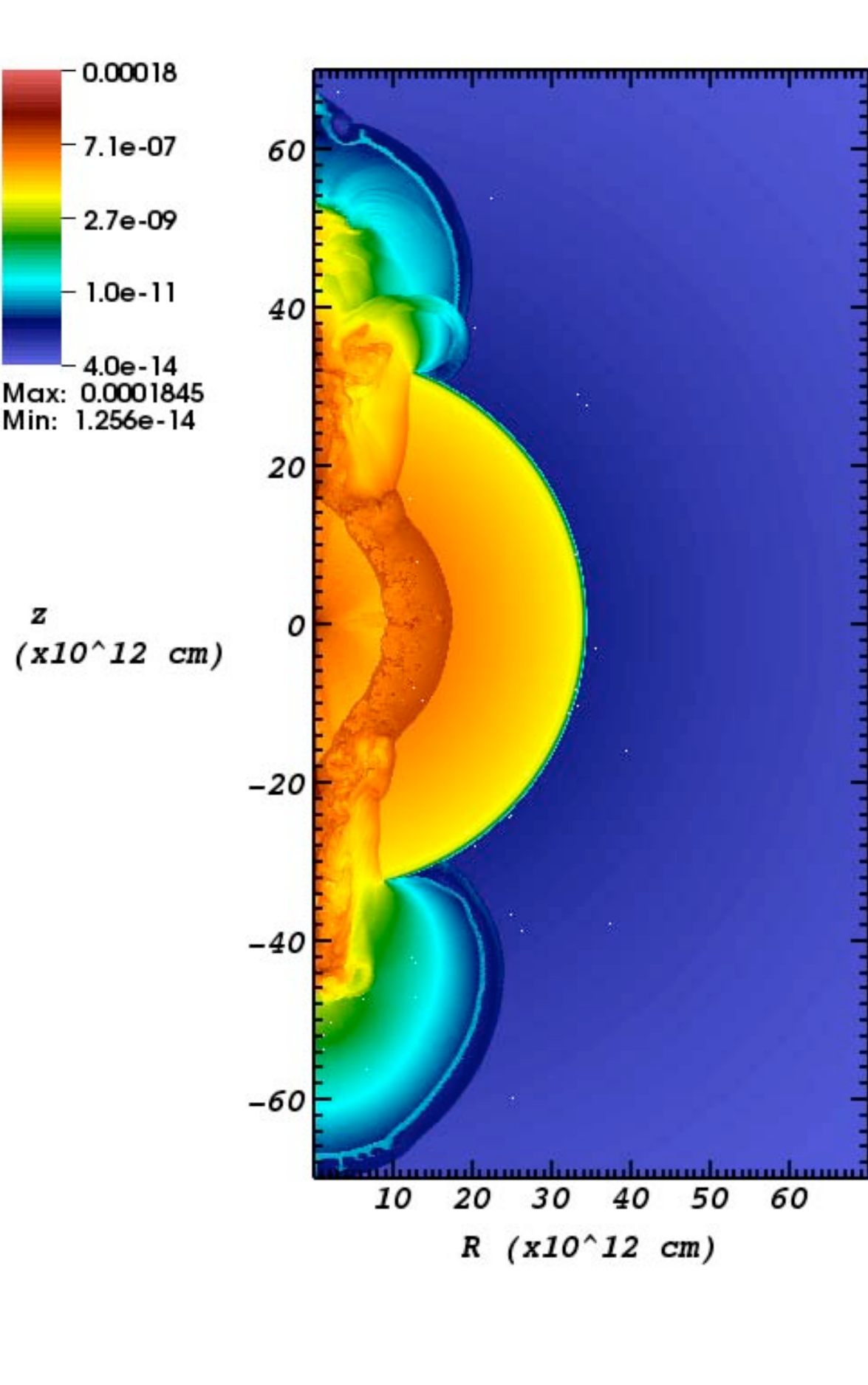} &
\includegraphics[width=2.in, trim= .15in 0 3.23in 0in, clip]{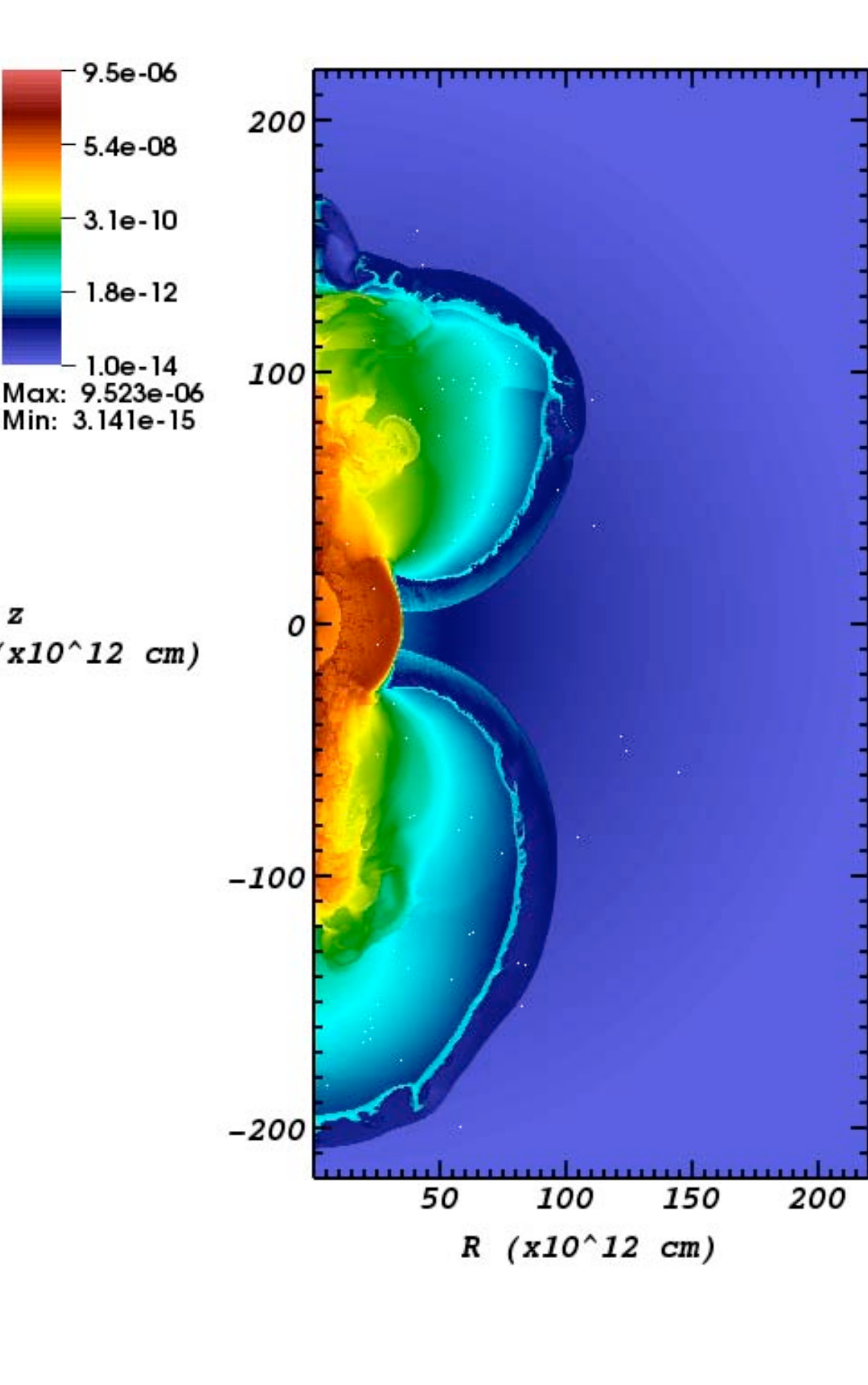} &
\includegraphics[width=2.in, trim= .15in 0 3.23in 0in, clip]{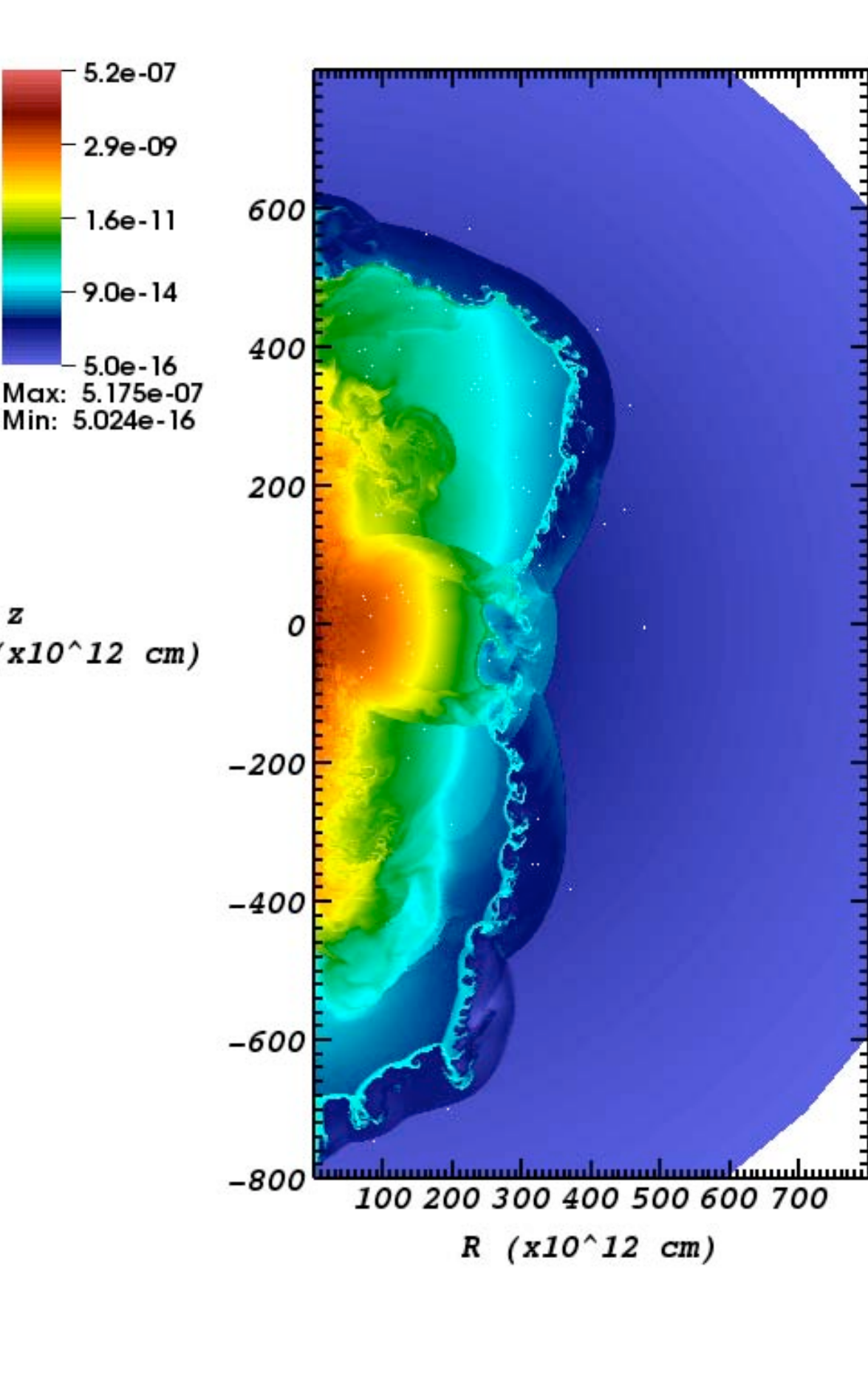}
\end{tabular} 
\caption[Model v3m12:  wind]{Snapshots of the density evolution in kinetic energy-dominated model v3m12 after shock breakout from the star.  The axis lengths are given in centimeters and the density is plotted in g cm$^{-3}$.  The simulation times are, from left to right, 52,000 s, 130,000 s, and 500,000 s.  This figure shows the significant lag between the breakout of the polar shocks and that of the equatorial shock.  A thin, unstable shell of wind material is swept up by the forward shocks, analogous to that shown in Figure \ref{fig:v3m12env}.  The final structure (right panel) is significantly prolate, inconsistent with early polarization measurements of SNeII-P.  Note that the slight north-south asymmetries are artifacts of numerical inaccuracies and are amplified by shock breakout.}
\label{fig:v3m12wind}
\end{figure}

\begin{figure}
\centering
\begin{tabular}{ccc}
\includegraphics[width=2.in, trim= .15in 0 3.23in 0in, clip]{v3m12_dens_1123.pdf} &
\includegraphics[width=2.in, trim= .15in 0 3.23in 0in, clip]{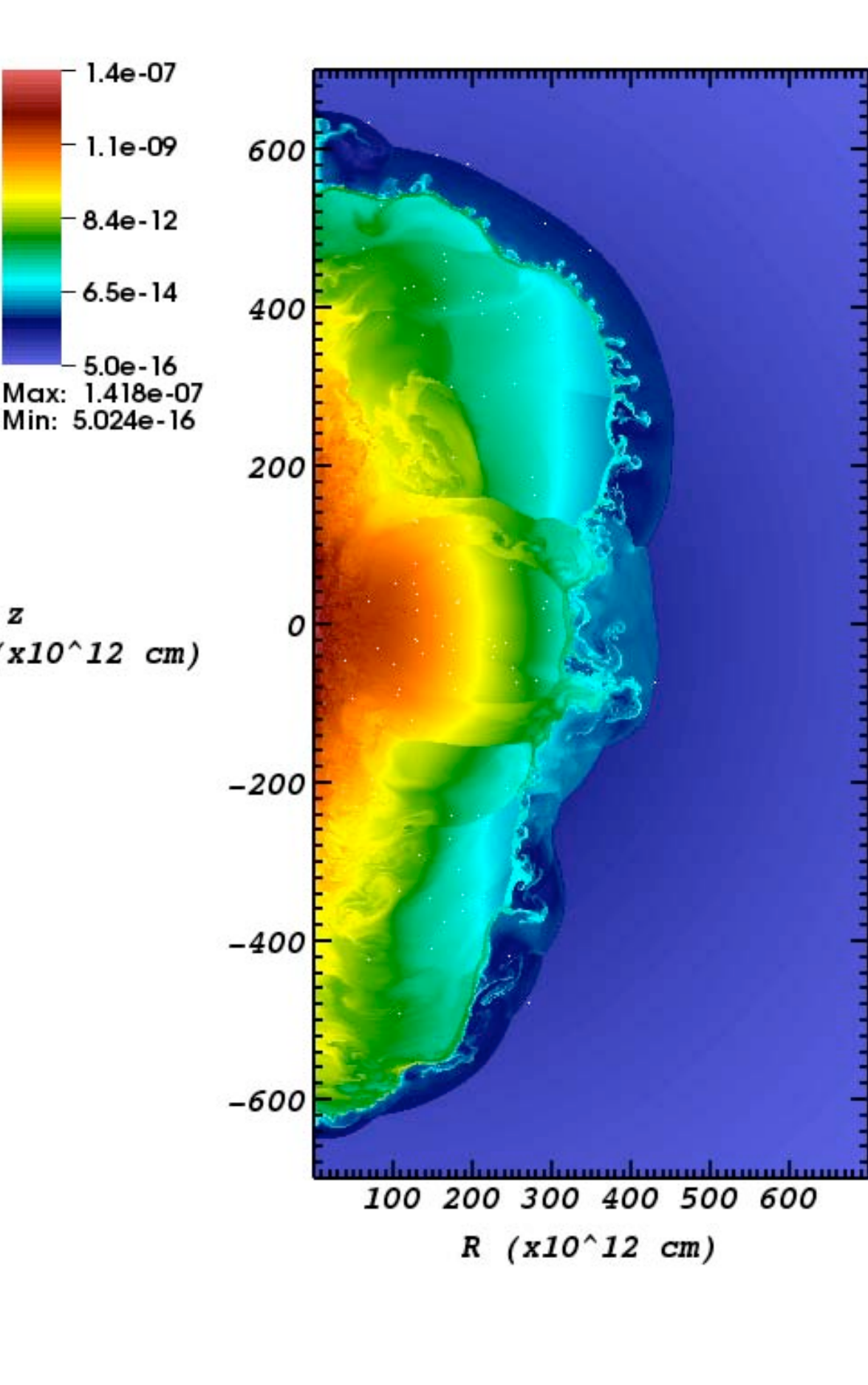} &
\includegraphics[width=2.in, trim= .15in 0 3.23in 0in, clip]{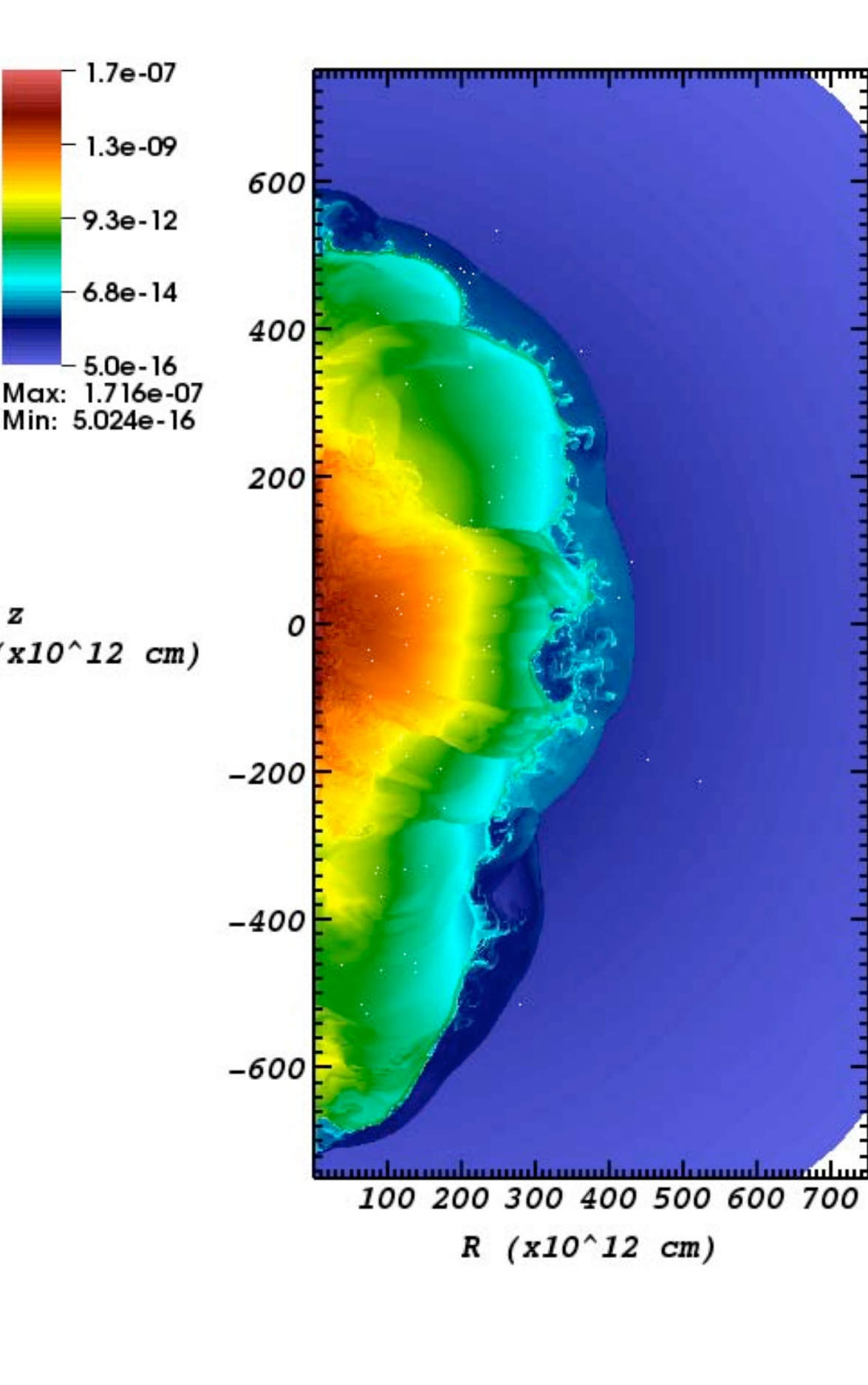}
\end{tabular} 
\caption{The density structures of the three kinetic energy-dominated models v3m12, v5m06, and v6m04 (left to right), at the end of the simulations, 5.79 days.  The final structures are remarkably similar given the differences in jet injection speeds and jet masses.}
\label{fig:kineticjets}
\end{figure}

\begin{figure}
\centering
\includegraphics[width=3in, trim= 0.25in 0.25in 1.in 0.25in, clip]{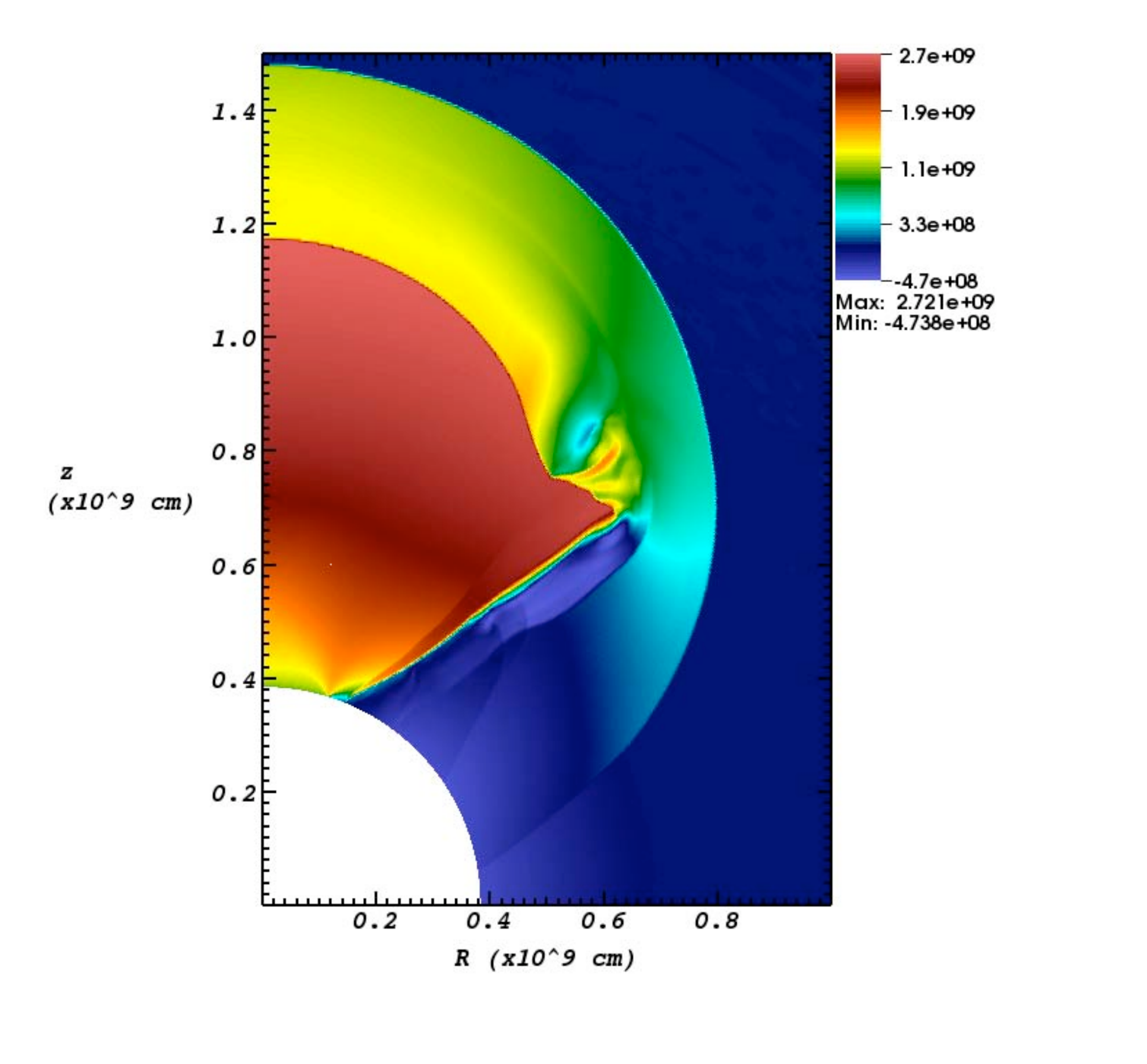}
\caption[Model v1m12:  velocity]{The radial velocity field of thermal energy-dominated model v1m12 at 1 s.  The maximum velocity attained of $\sim 3\times10^9$ cm s$^{-1}$ is much greater than the input velocity of $10^9$ cm s$^{-1}$ due to the conversion of thermal energy to kinetic energy in the jet fluid.}
\label{fig:v1m12velx}
\end{figure}

\begin{figure}
\centering
\begin{tabular}{ccc}
\includegraphics[width=2.in, trim= .15in 0 3.23in 0in, clip]{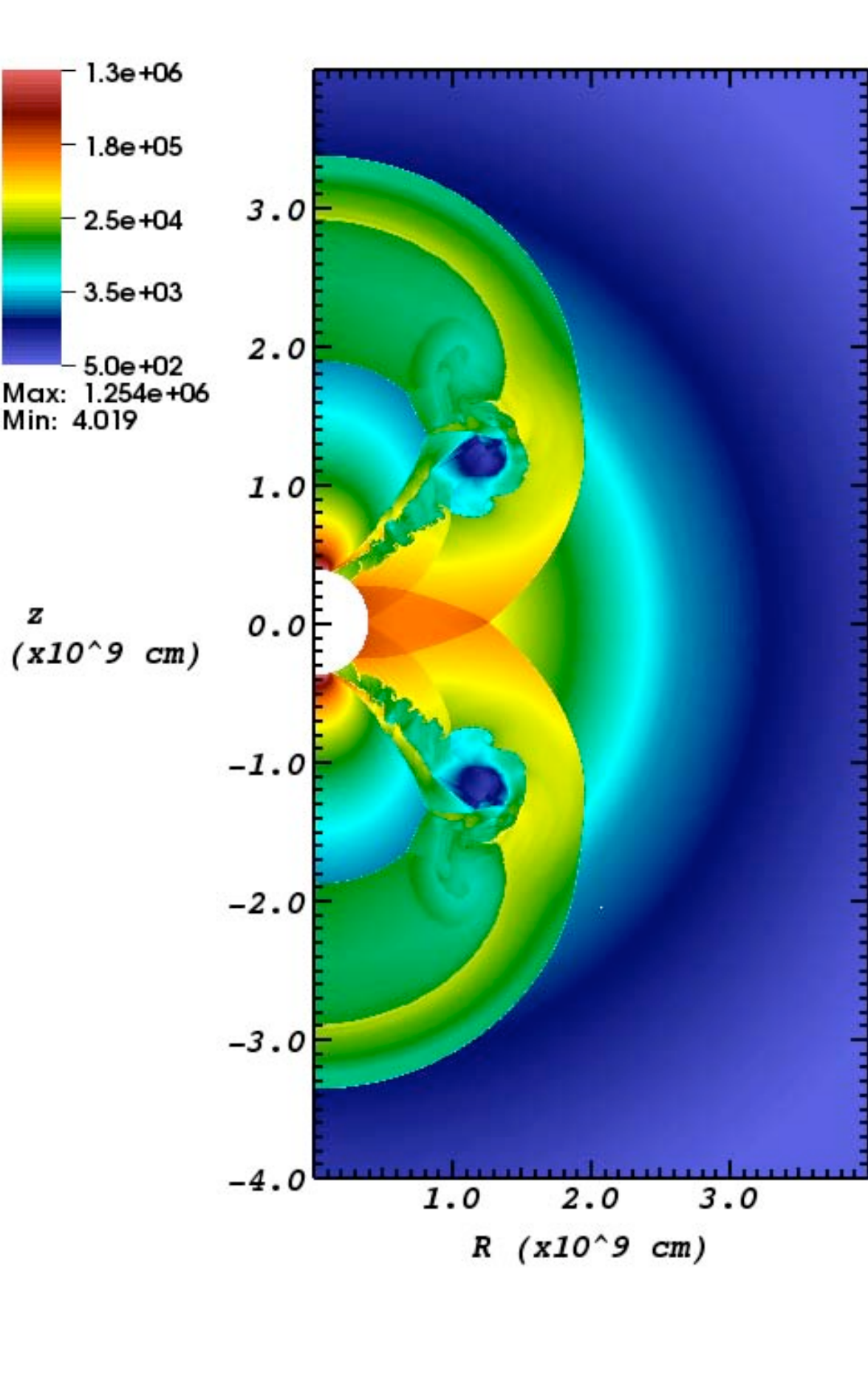} &
\includegraphics[width=2.in, trim= .15in 0 3.23in 0in, clip]{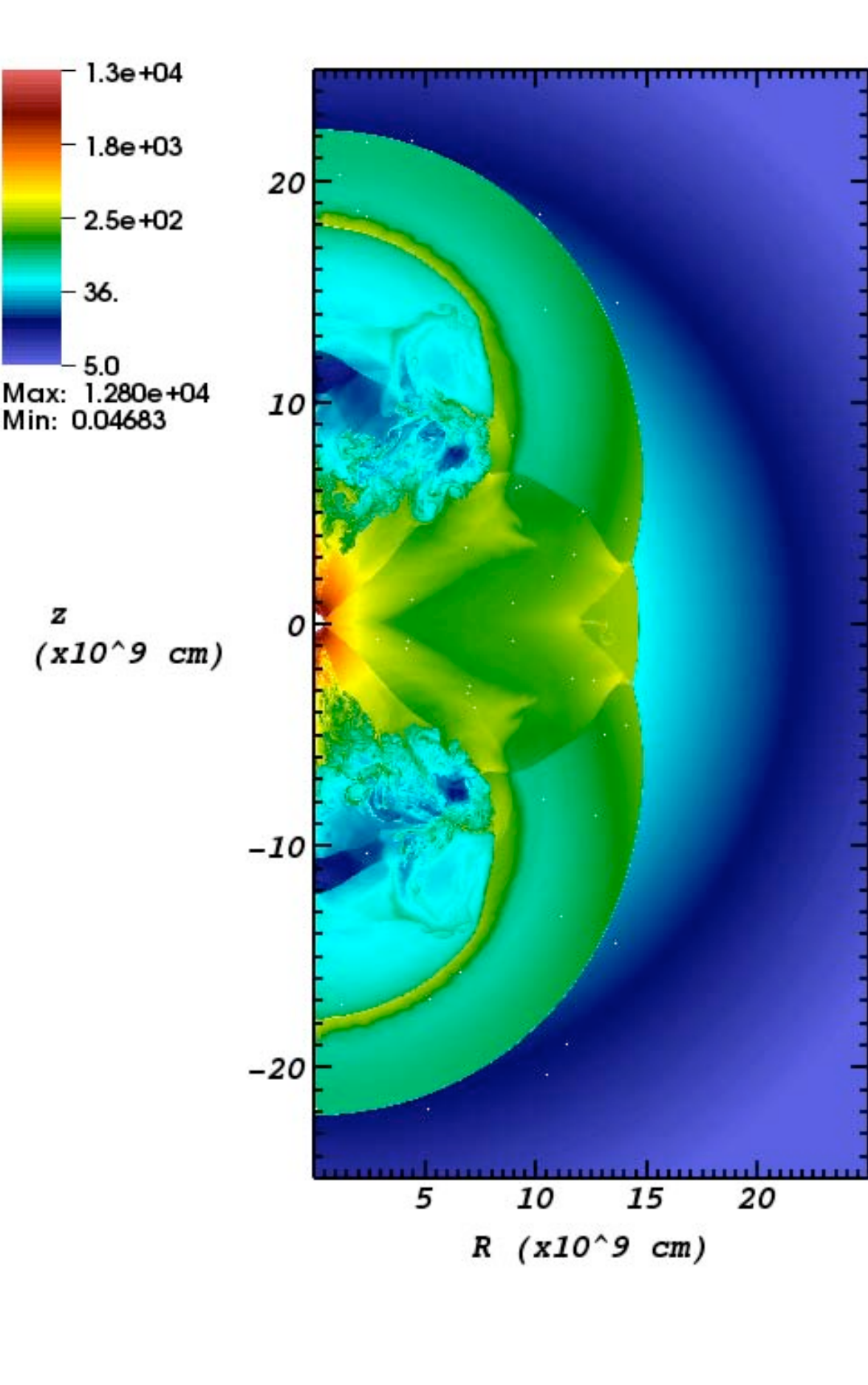} &
\includegraphics[width=2.in, trim= .15in 0 3.23in 0in, clip]{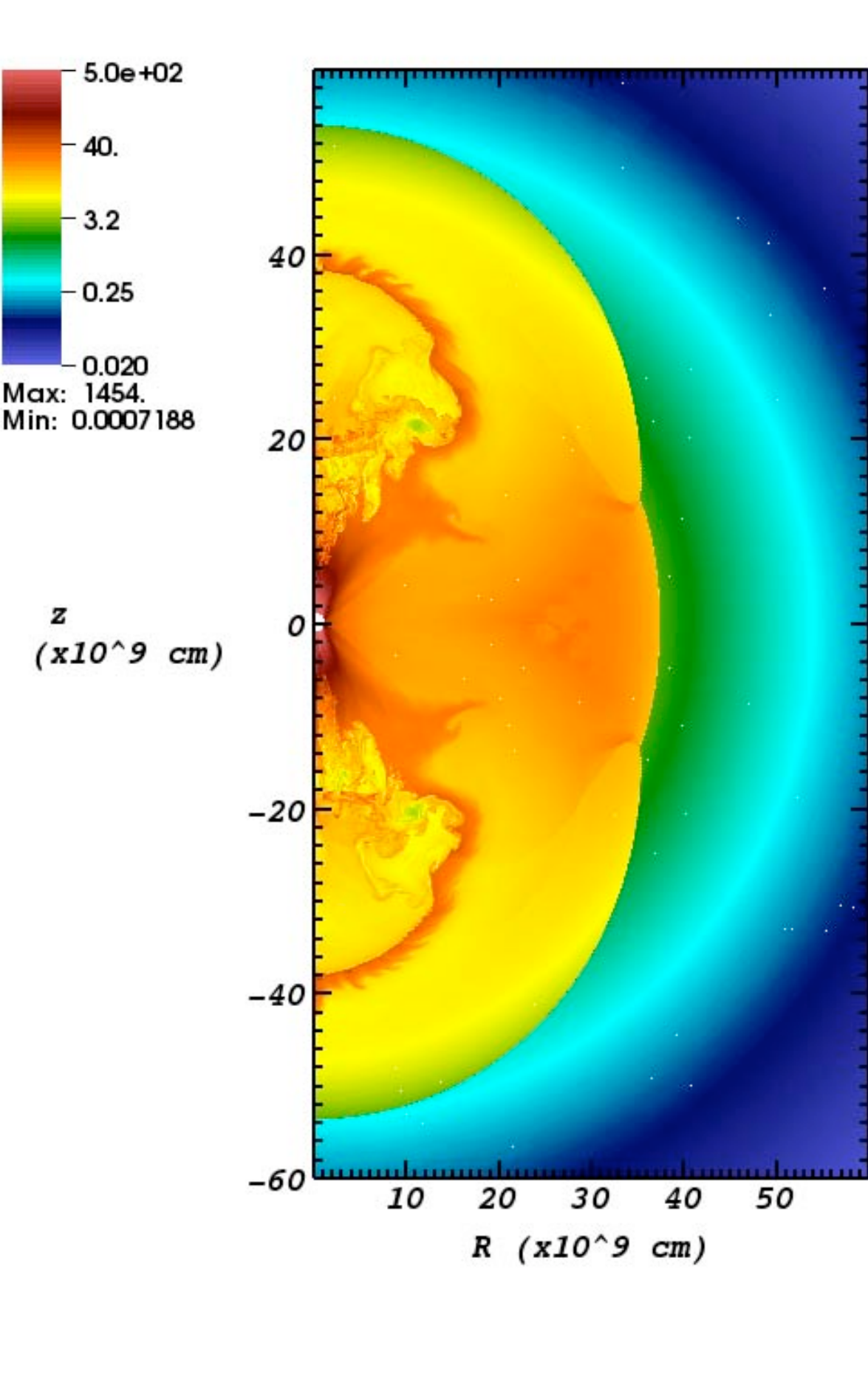}
\end{tabular} 
\caption[Model v1m12:  core]{Snapshots of the density field for thermal energy-dominated model v1m12 before leaving the helium core.  The axis lengths are given in centimeters and the density is plotted in g cm$^{-3}$.  The times of the snapshots are, from left to right, 2 s, 20 s, and 60 s.  The left panel shows the rapid lateral expansion of the jet structure in this model.  Shown also are large vortices that develop into turbulent structures, seen in the middle and right panels.  The right panel shows the beginning growth of RT fingers in the interface between the jet and the shocked star material.  The equatorial structure is apparent in the middle panel and is significantly broader than in the kinetic energy models.  The edges of the equatorial structure are less sharp than in the other models, but still subject to short-wavelength KH instability (see Fig. \ref{fig:v1m12env}).}
\label{fig:v1m12core}
\end{figure}

\begin{figure}
\centering
\begin{tabular}{ccc}
\includegraphics[width=2.in, trim= .15in 0 3.23in 0in, clip]{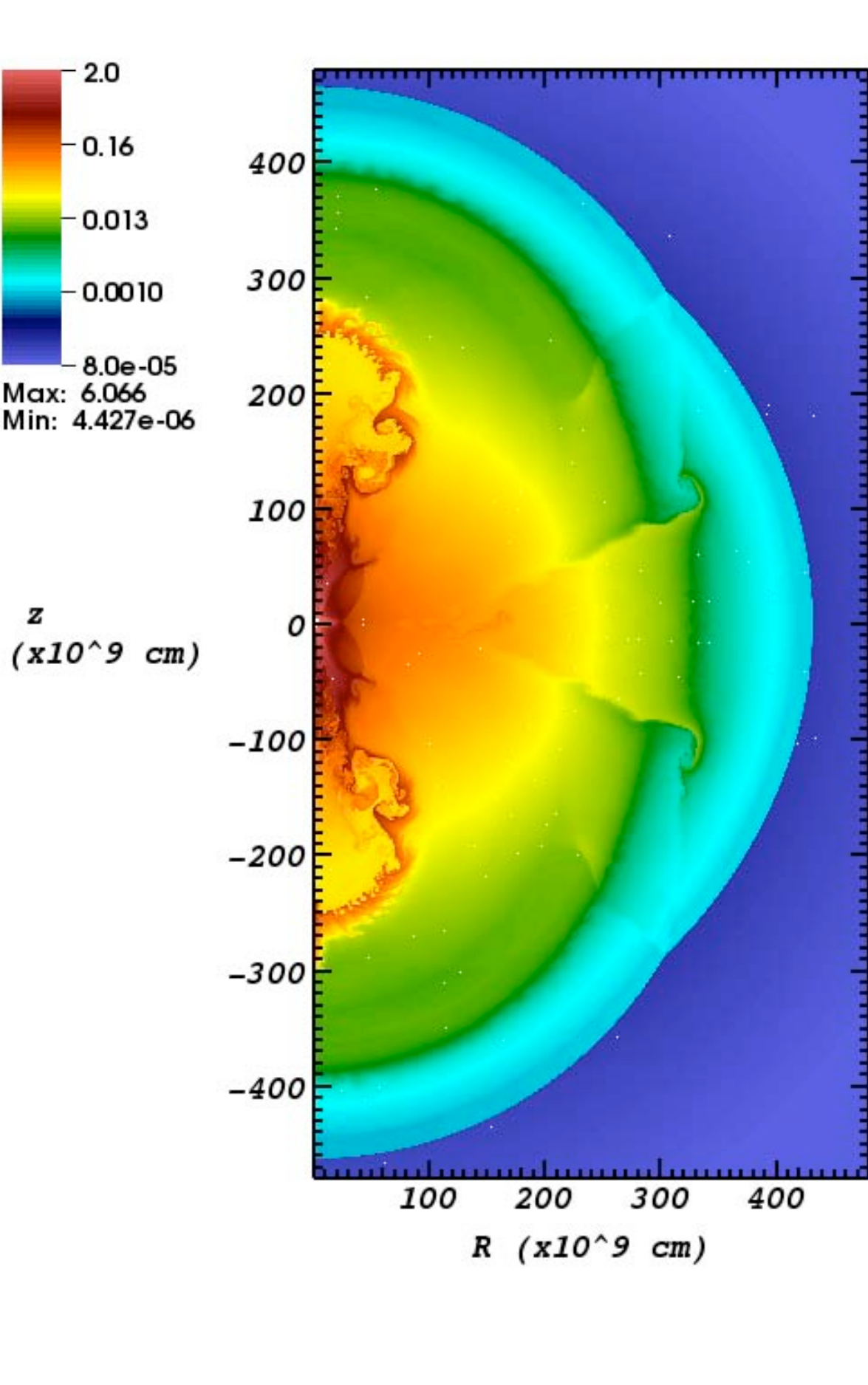} &
\includegraphics[width=2.in, trim= .15in 0 3.23in 0in, clip]{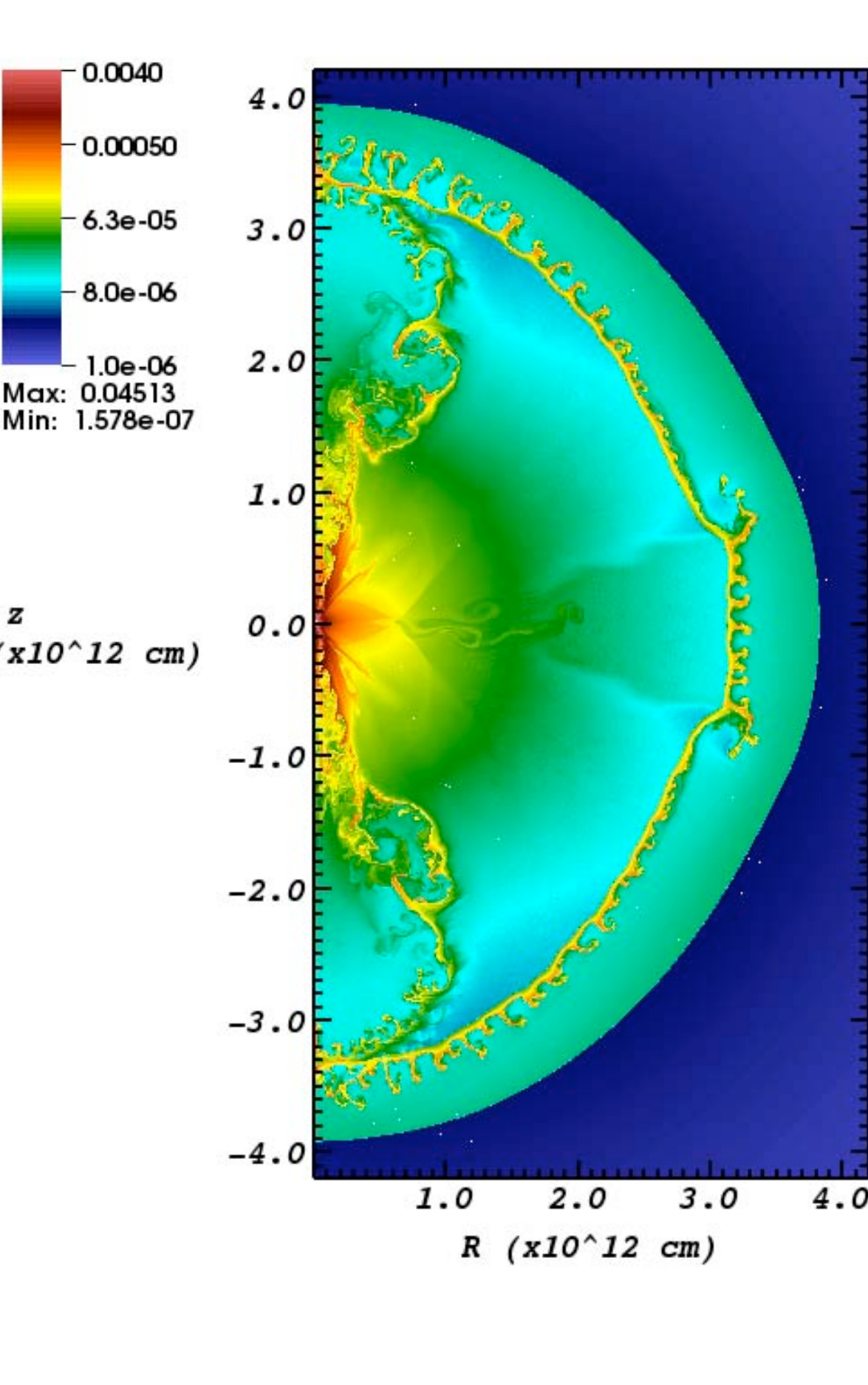} &
\includegraphics[width=2.in, trim= .15in 0 3.23in 0in, clip]{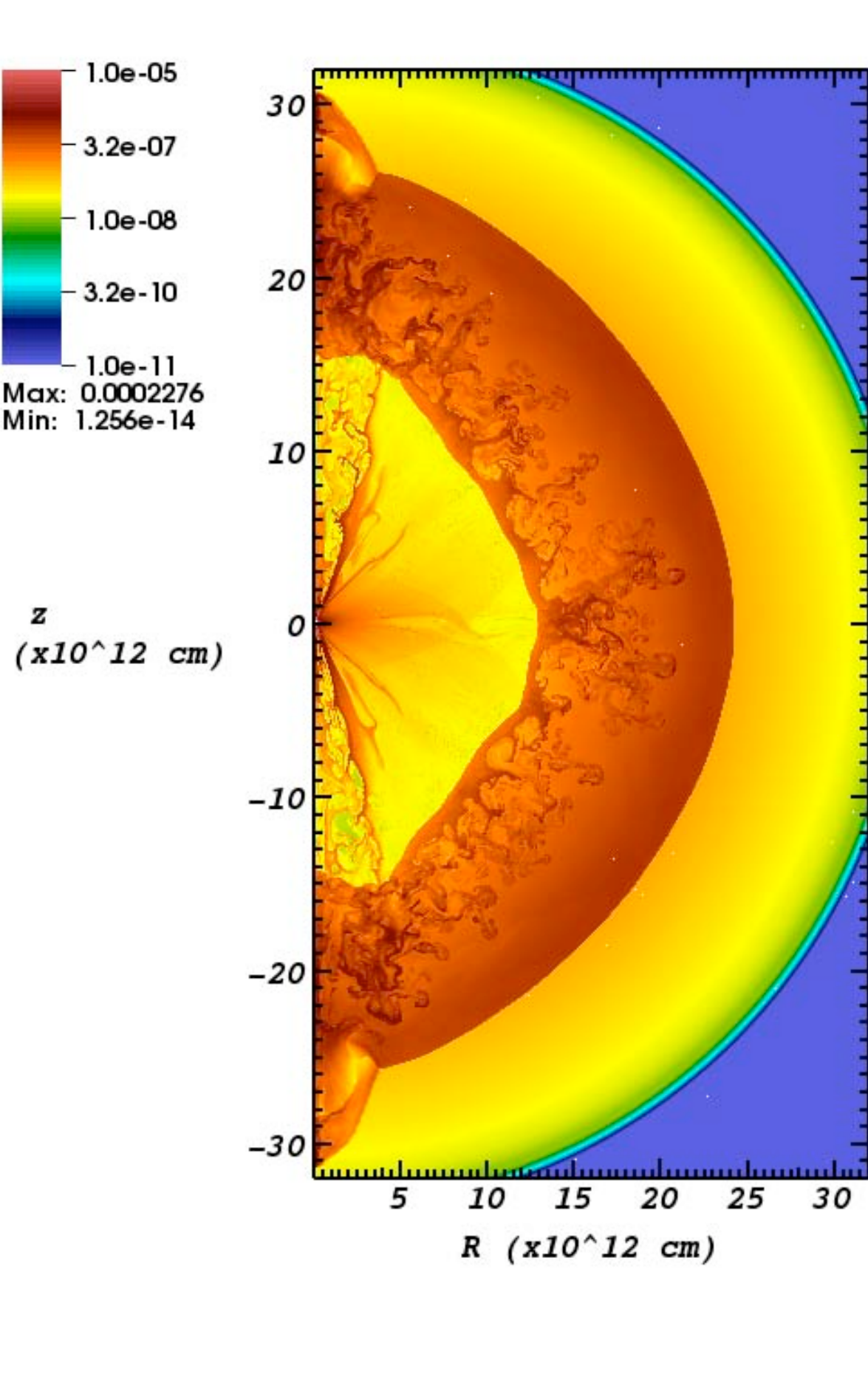}
\end{tabular} 
\caption[Model v1m12:  envelope]{Snapshots of the density field for thermal energy-dominated model v1m12 in the progenitor envelope.  The axis lengths are given in centimeters and the density is plotted in g cm$^{-3}$.  The times of the snapshots are, from left to right, 500 s, 6500 s, and 85,000 s.  The left panel shows that the equatorial outflow is much broader than in the kinetic energy models and is still subject to the KH instability along its surfaces.  As also seen in the kinetic energy models, the reverse shock sweeps up a thin, RT unstable shell that wipes away much of the complex structure that has developed up to this point.  The right panel shows the RM finger of jet fluid that grows up the axis and catches the shock.}
\label{fig:v1m12env}
\end{figure}

\begin{figure}
\centering
\begin{tabular}{ccc}
\includegraphics[width=2.in, trim= .15in 0 3.23in 0in, clip]{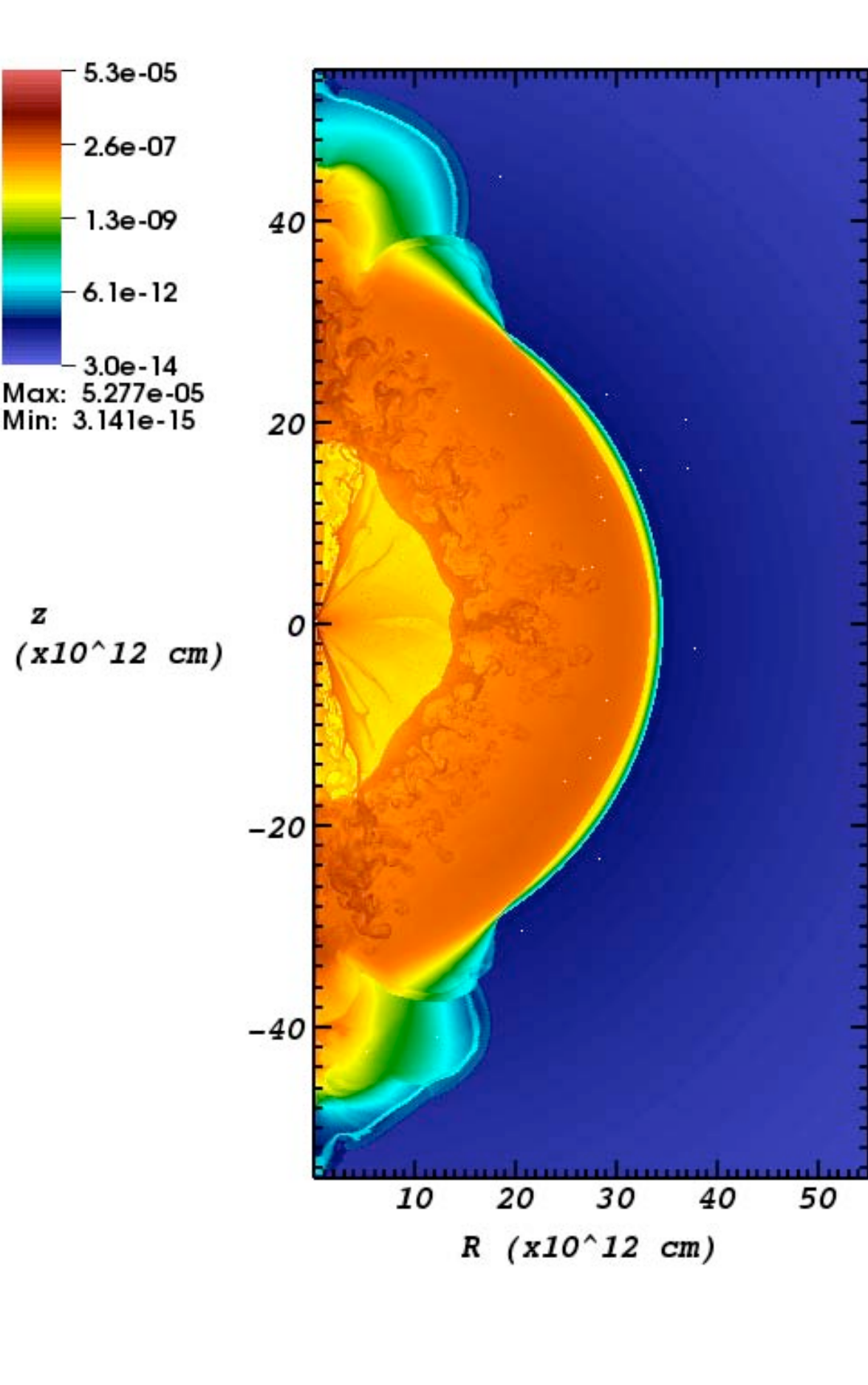} &
\includegraphics[width=2.in, trim= .15in 0 3.23in 0in, clip]{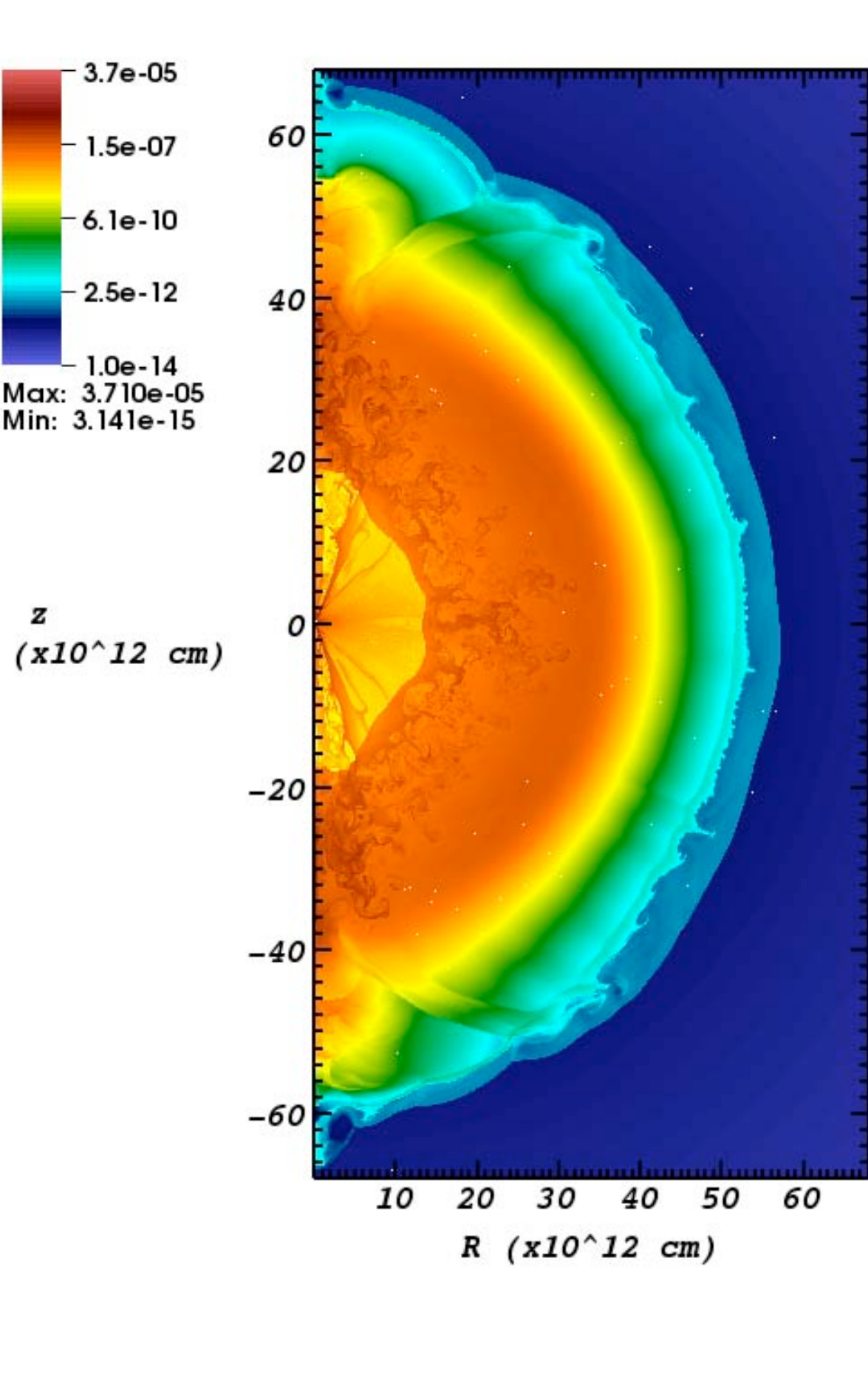} &
\includegraphics[width=2.in, trim= .15in 0 3.23in 0in, clip]{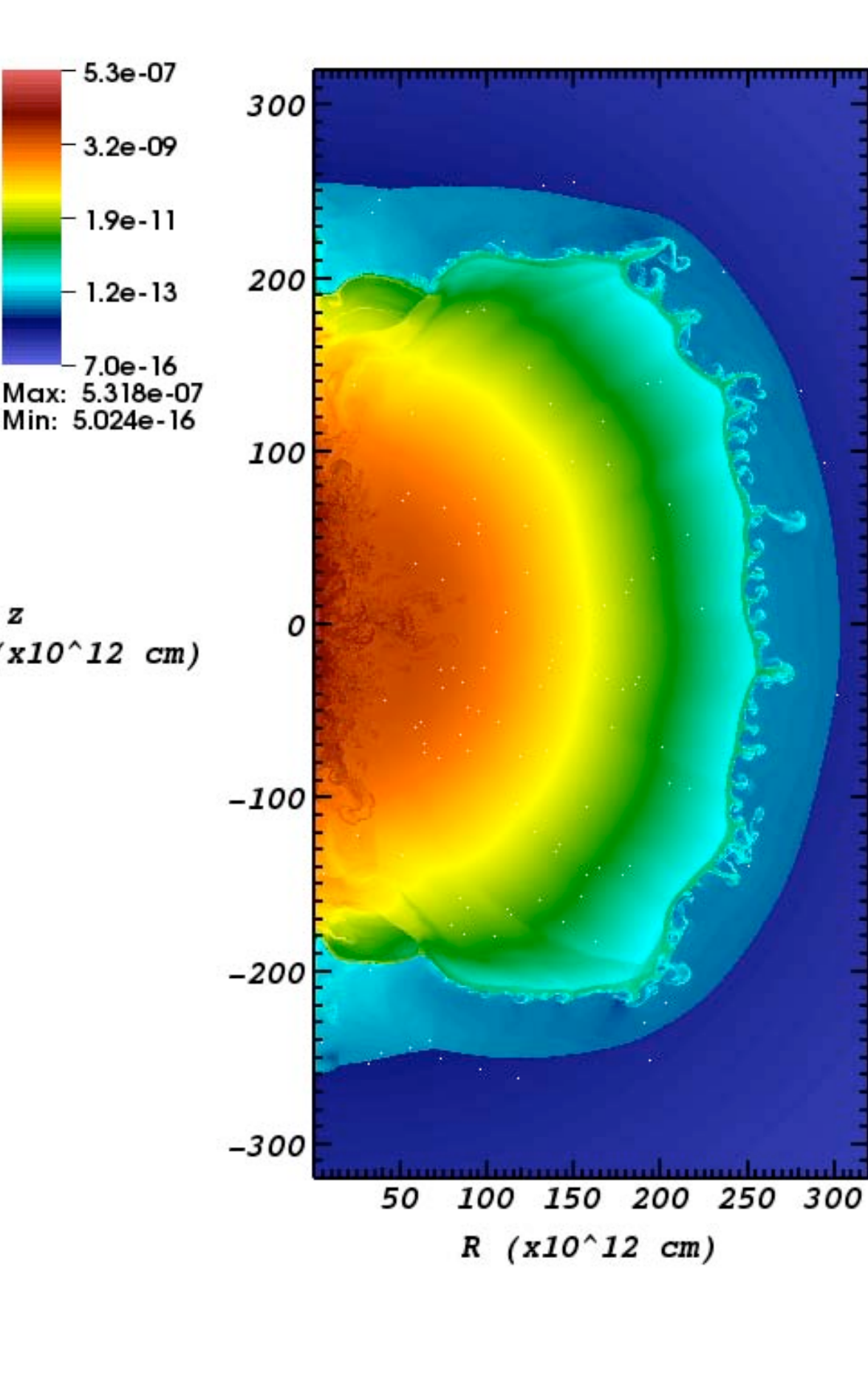}
\end{tabular} 
\caption[Model v1m12:  wind]{Snapshots of the density field for thermal energy-dominated model v1m12 as the shock exits the star.  The axis lengths are given in centimeters and the density is plotted in g cm$^{-3}$.  The times of the snapshots are, from left to right, 130,000 s, 155,000 s, and 500,000 s.  This figure shows that the lag between polar and equatorial shock breakouts is much smaller than in the kinetic energy models.  The resulting structure is nearly spherical.  As in the case of model v3m12, the reverse shock arising from shock breakout into the wind sweeps up a dense shell that is RT unstable.}
\label{fig:v1m12wind}
\end{figure}

\begin{figure}
\centering
\begin{tabular}{cc}
\includegraphics[width=3in]{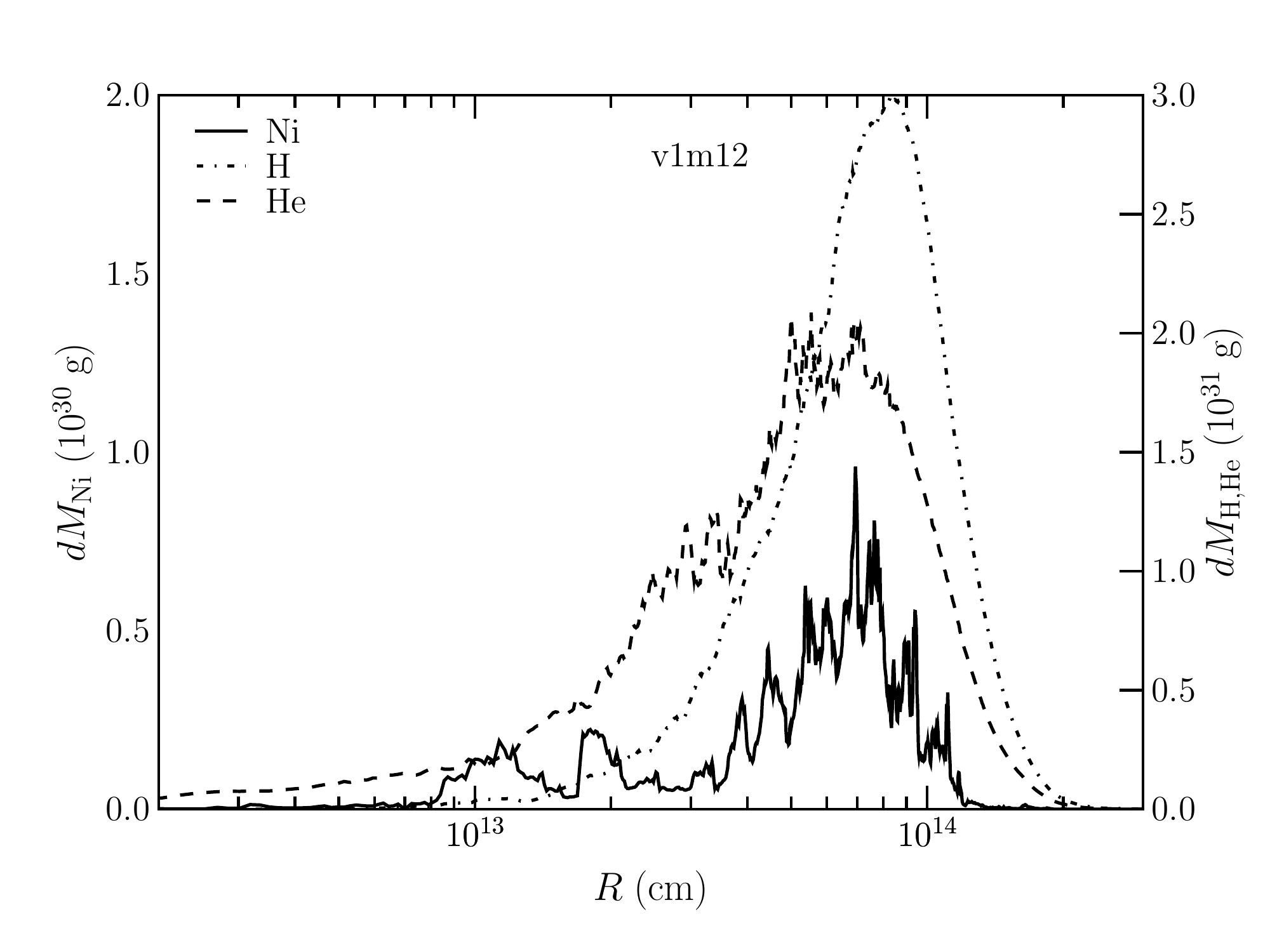} &
\includegraphics[width=3in]{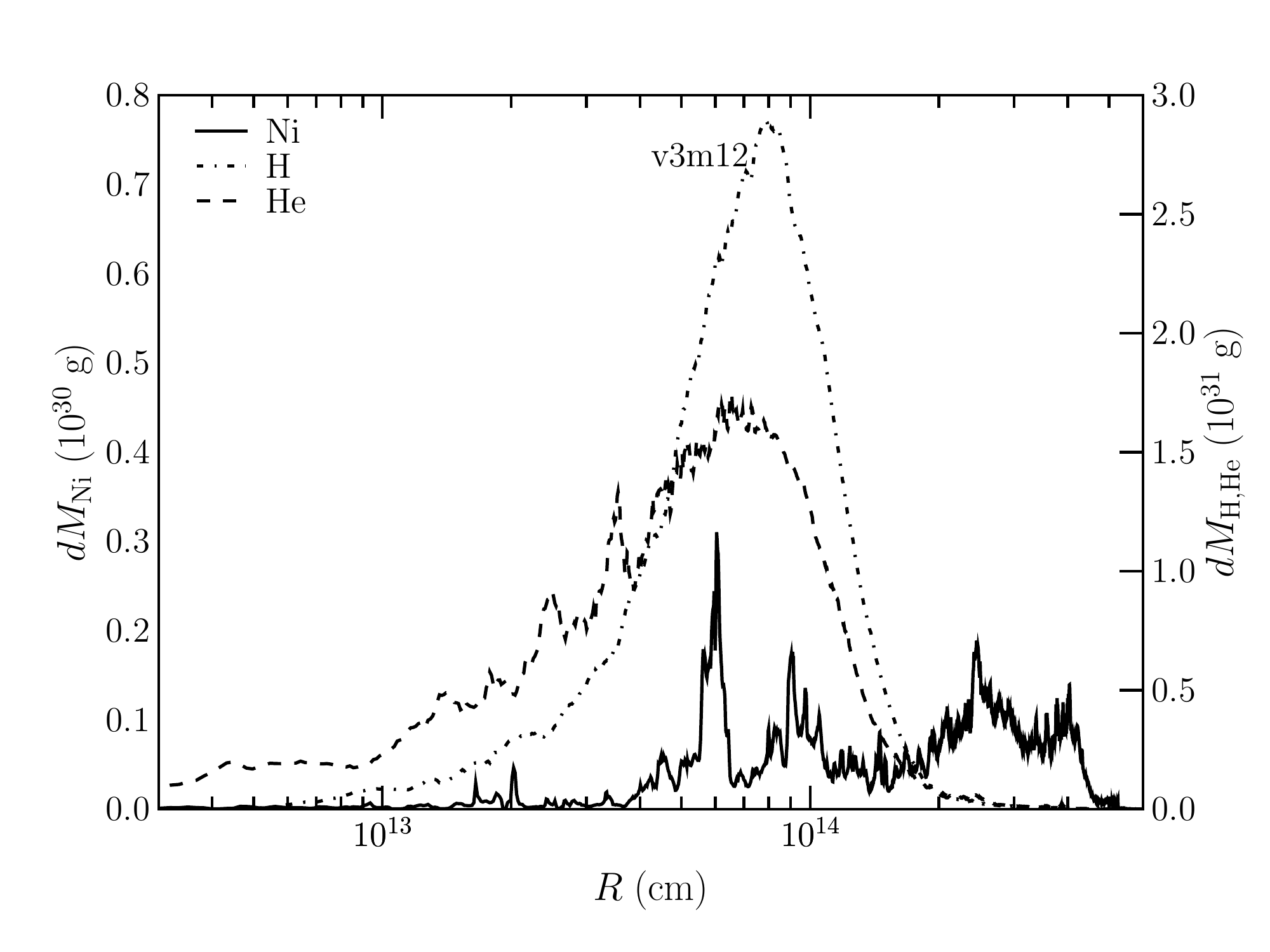} \\
\includegraphics[width=3in]{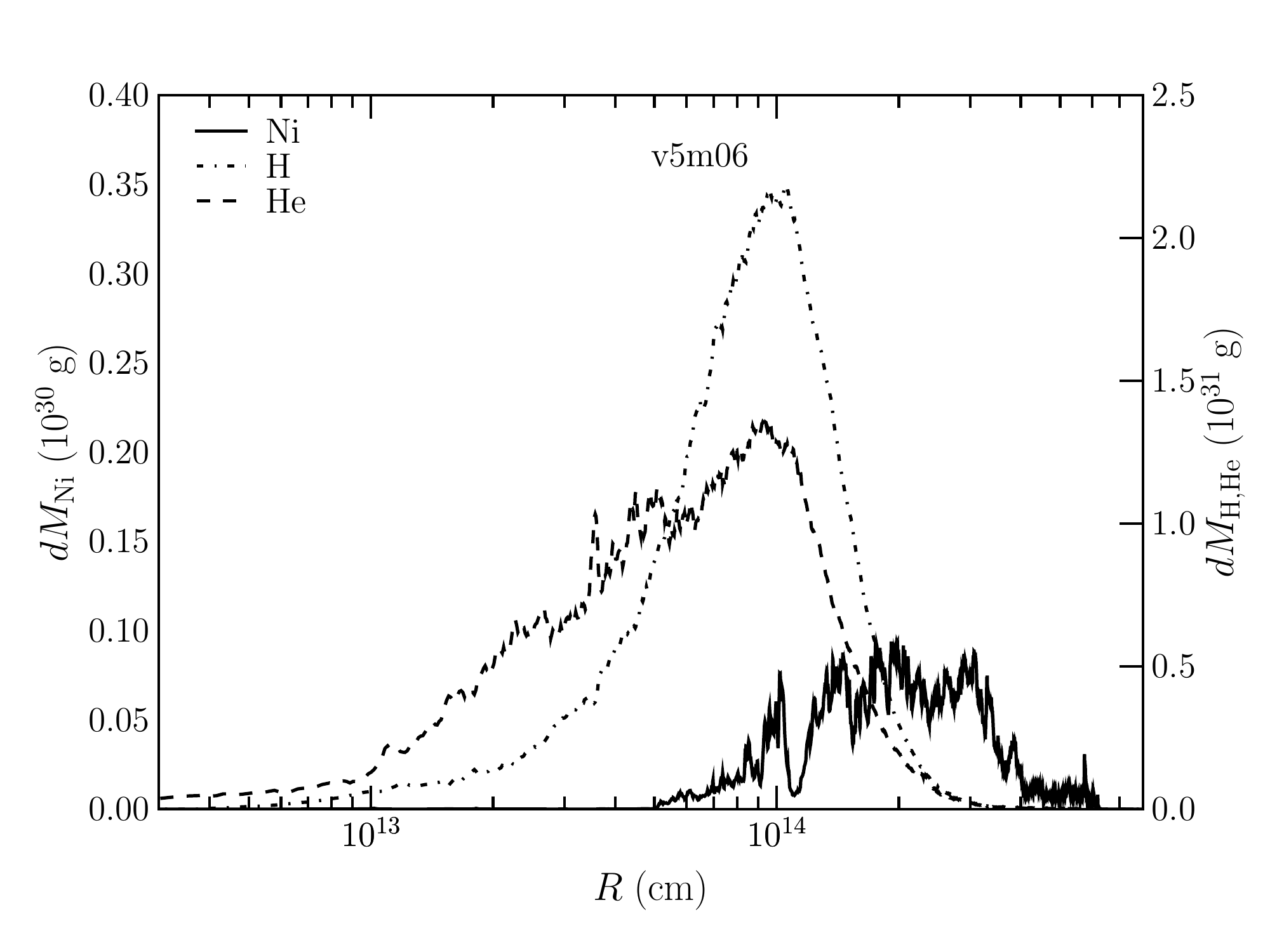}  &
\includegraphics[width=3in]{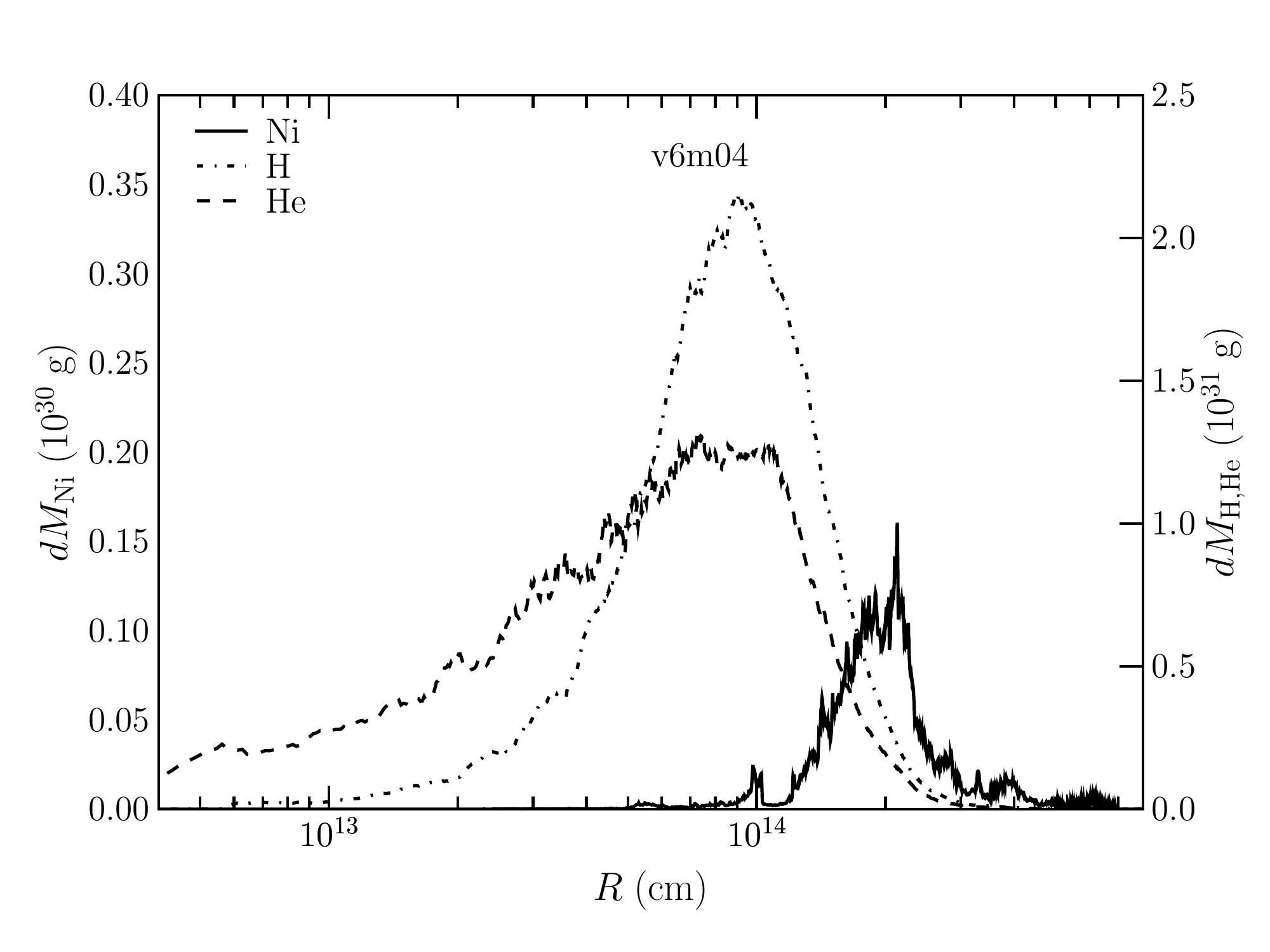}  
\end{tabular}
\caption[]{Mass distributions versus radius for the four explosion models at the end of the calculations (5.79 days).  The three kinetic energy models drive nickel ahead of the hydrogen and helium while the thermal model, v1m12 (upper left), ends with a nickel distribution very similar to that of hydrogen.  Driving nickel beyond the hydrogen and helium may reduce the nickel available to excite spectral lines via radioactive decay.}
\label{fig:mofr}
\end{figure}

\begin{figure}
\centering
\begin{tabular}{cc}
\includegraphics[width=3in]{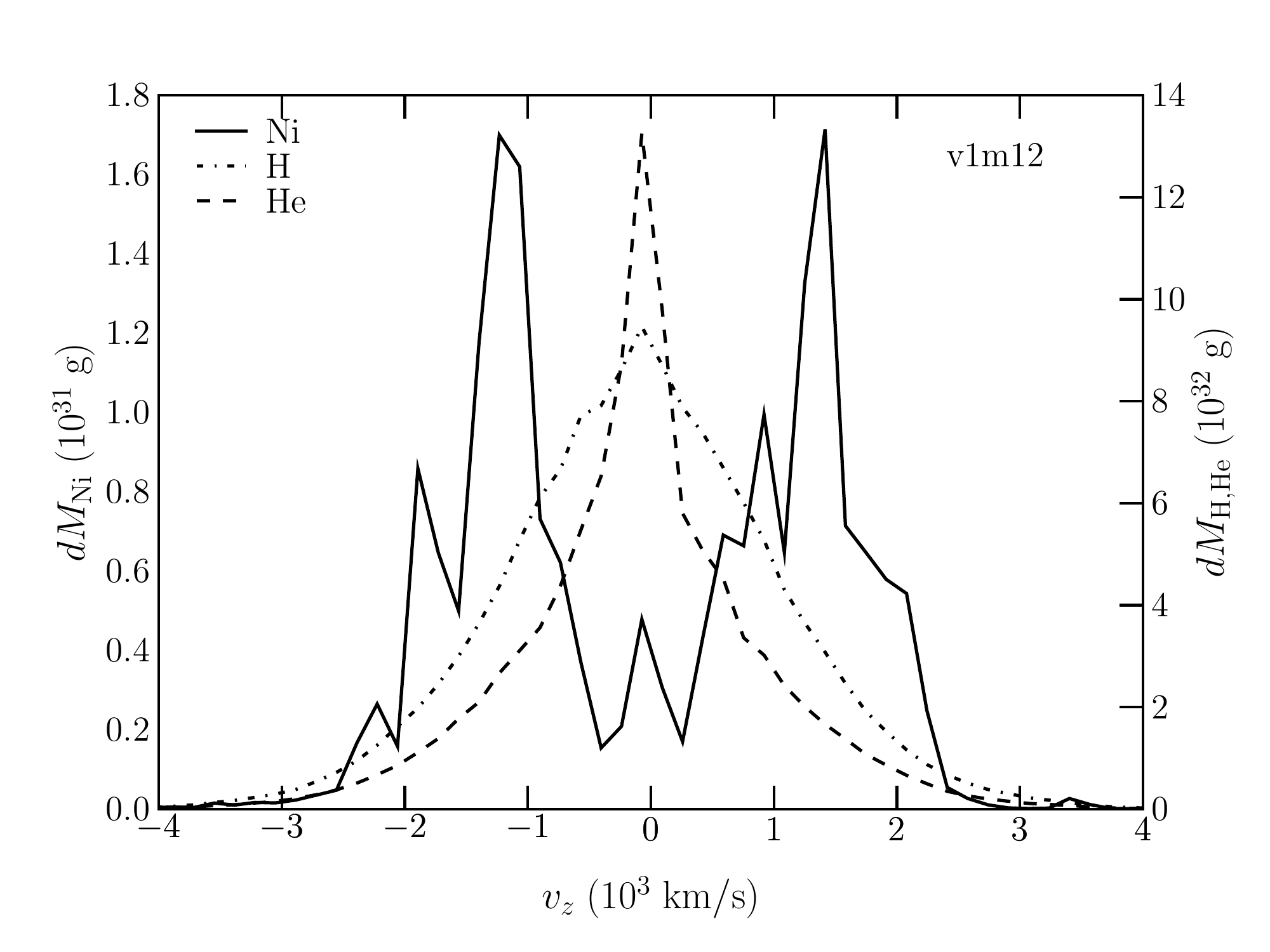} &
\includegraphics[width=3in]{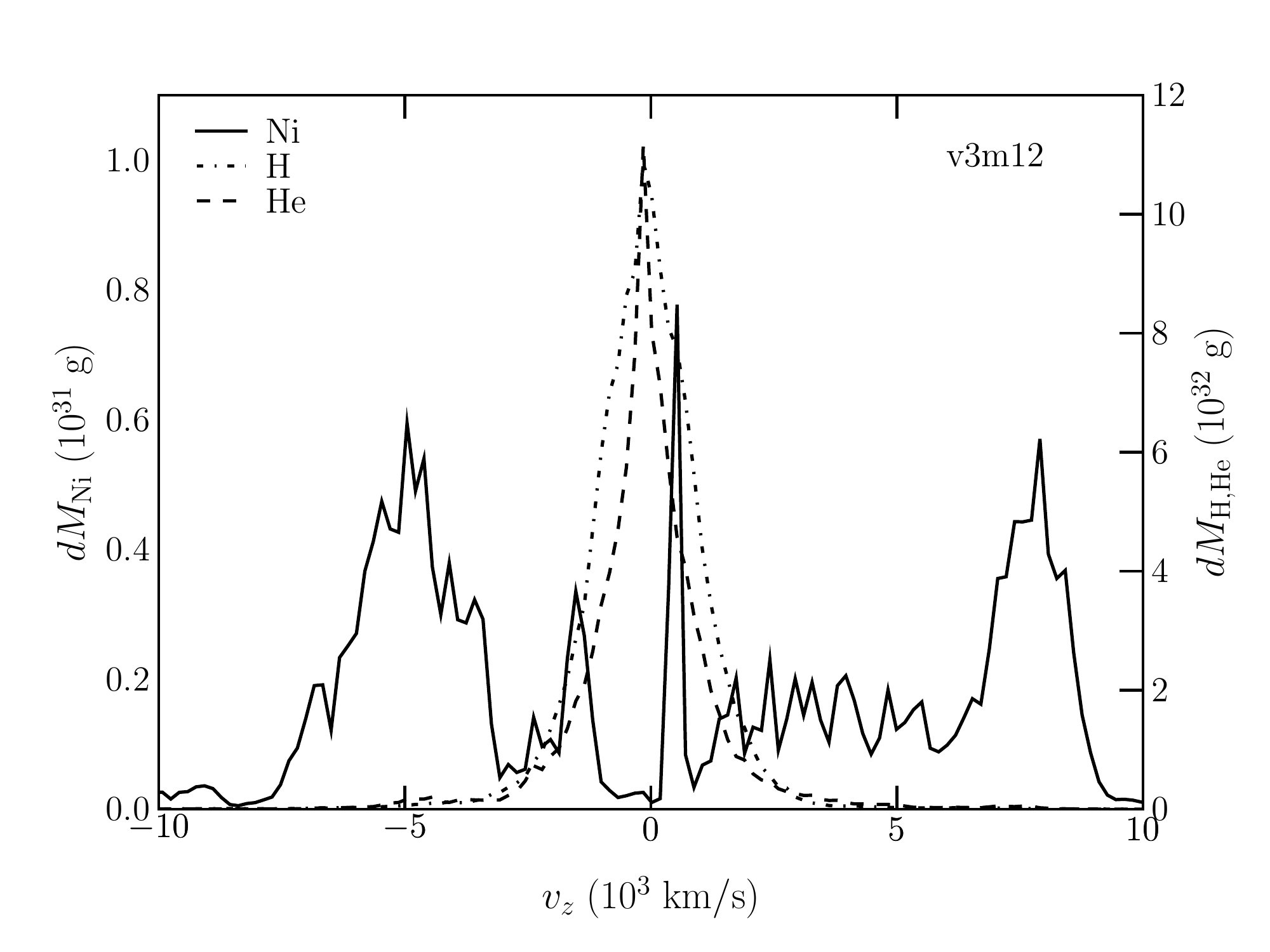} \\
\includegraphics[width=3in]{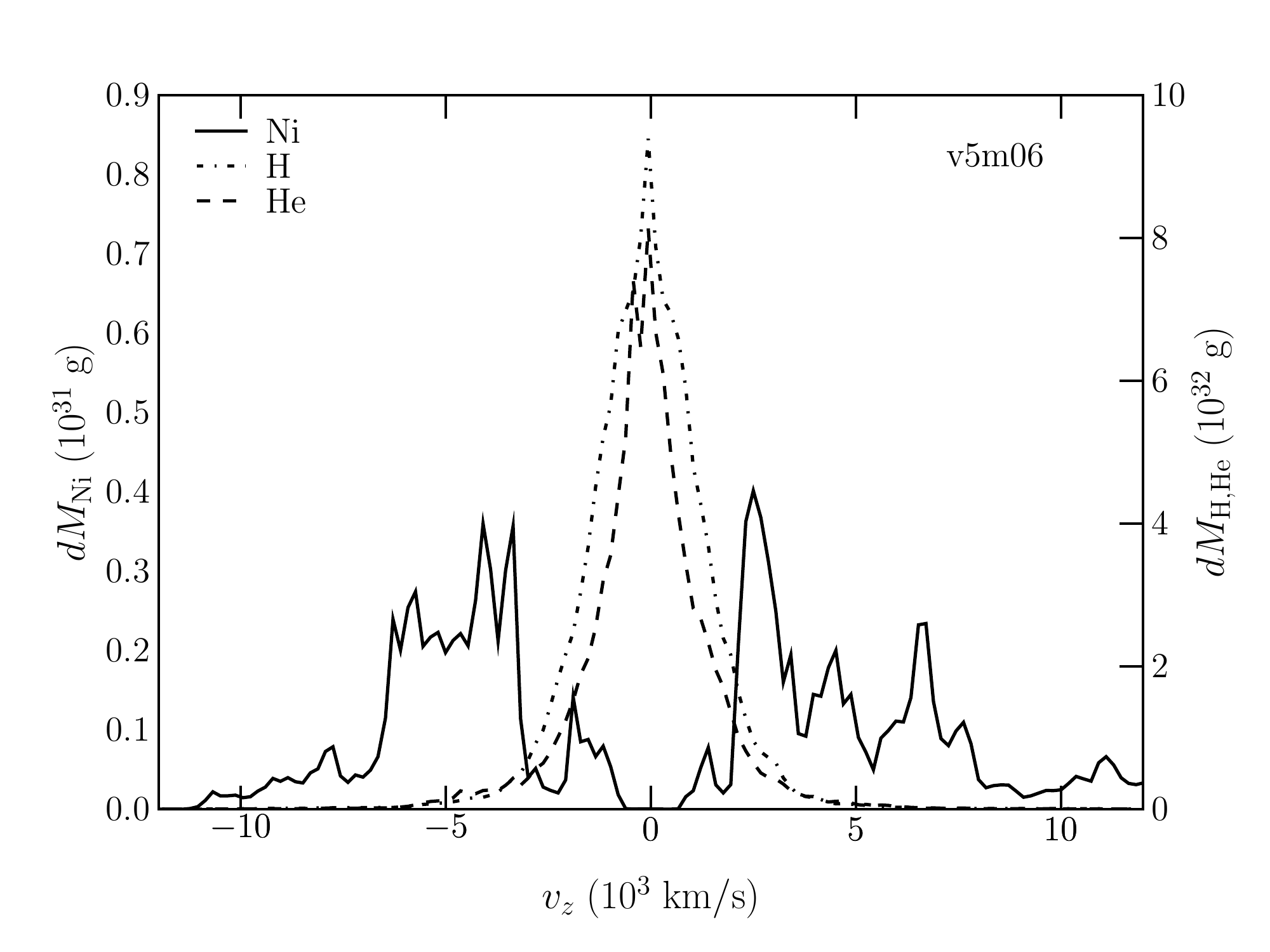} &
\includegraphics[width=3in]{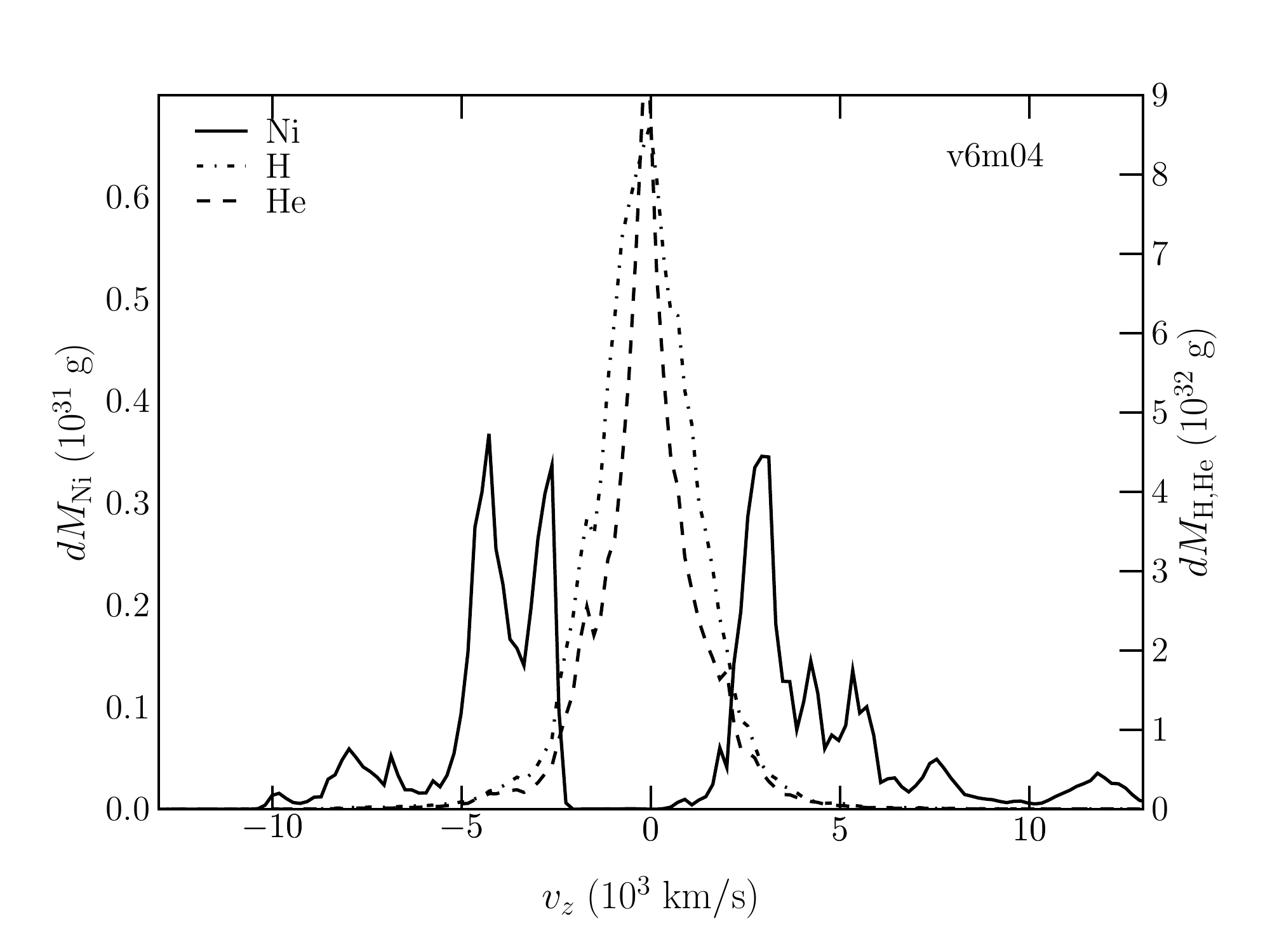} 
\end{tabular}
\caption[]{Mass distributions versus z-velocity ($v_z = v_r \cos \theta$) for the four models.  The oppositely-directed, bipolar clumps of nickel are evident.  The kinetic energy models yield nickel clumps with typical velocities of 5000 km s$^{-1}$ whereas the clump velocities in model v1m12 are $\sim1500$ km s$^{-1}$.}
\label{fig:dmdvz}
\end{figure}

\begin{figure}
\centering
\includegraphics[width=6in]{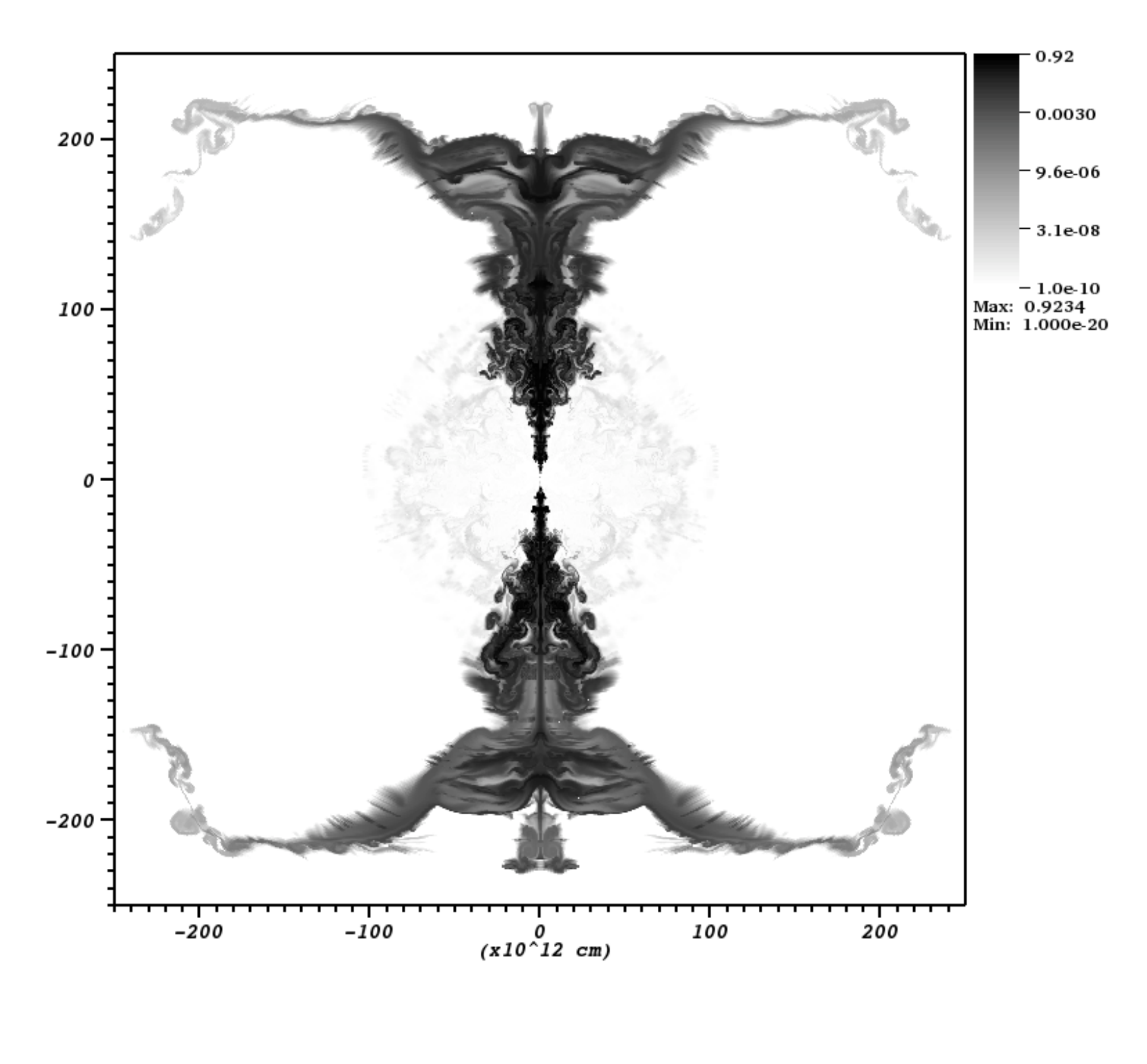}
\caption{The $^{56}$Ni mass fraction at 5.79 days for thermal energy-dominated model v1m12.  The nickel clumps, remnants of the jets, are clearly shown. The nickel at $2\times10^{14}$ cm is moving at approximately 2000 km s$^{-1}$.}
\label{fig:ni56_v1}
\end{figure}

\begin{figure}
\centering
\includegraphics[width=6in]{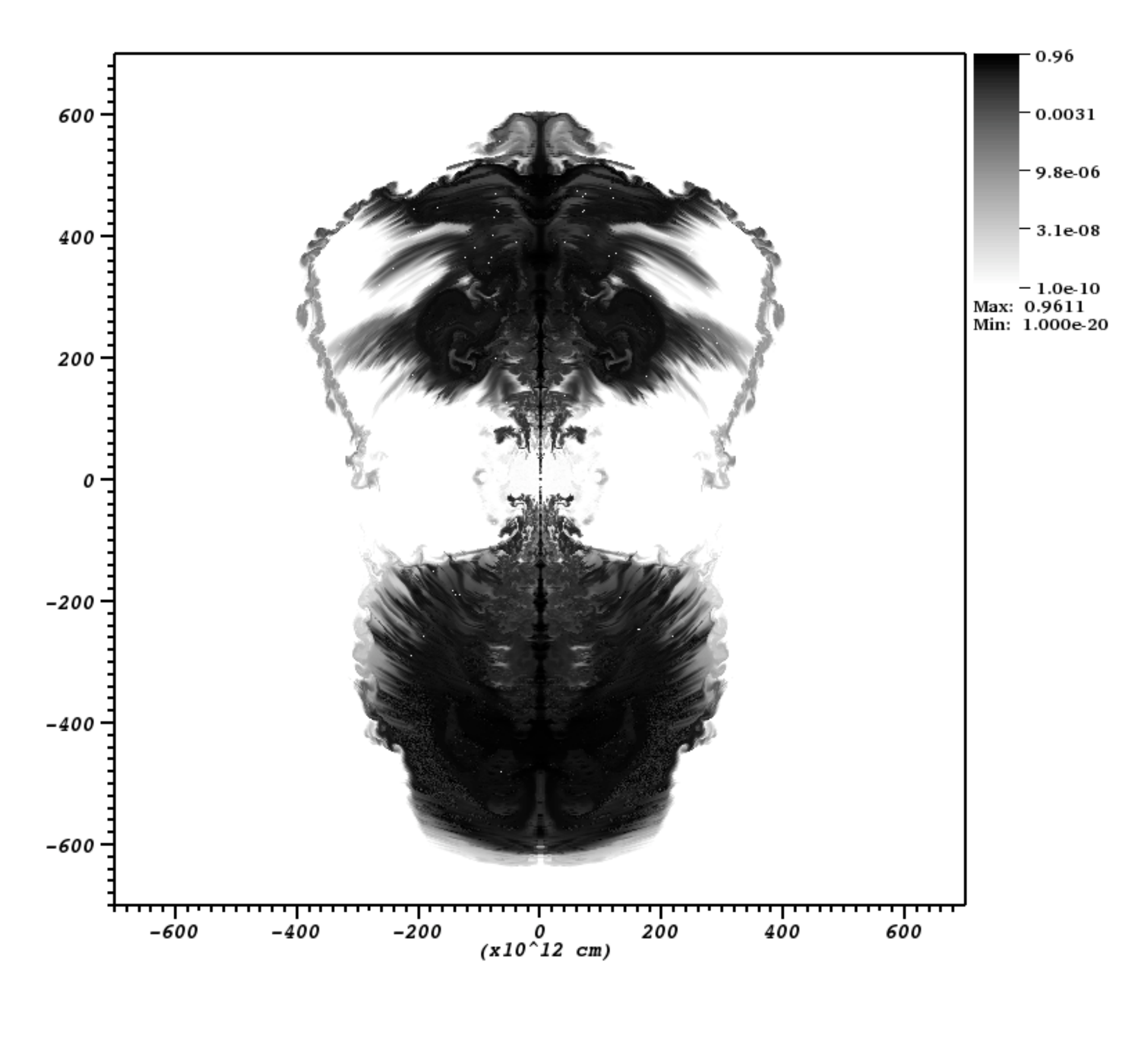}
\caption{The $^{56}$Ni mass fraction at 5.79 days for kinetic energy-dominated model v3m12.  The nickel clumps, remnants of the jets, are clearly shown. The nickel at $6\times10^{14}$ cm is moving at about 9000 km s$^{-1}$.}
\label{fig:ni56_v3}
\end{figure}

\begin{figure}
\centering
\begin{tabular}{cc}
\includegraphics[width=3in]{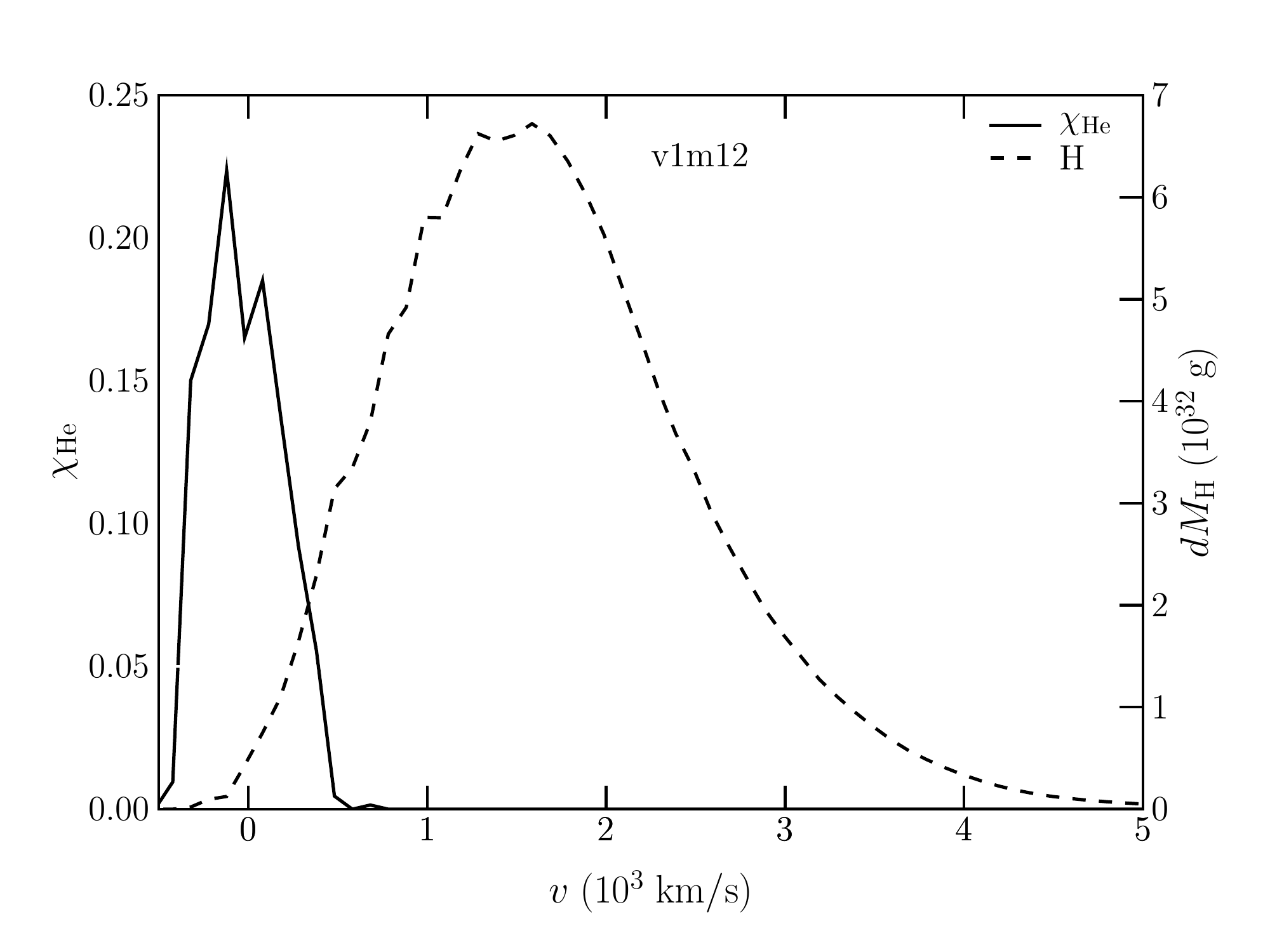} &
\includegraphics[width=3in]{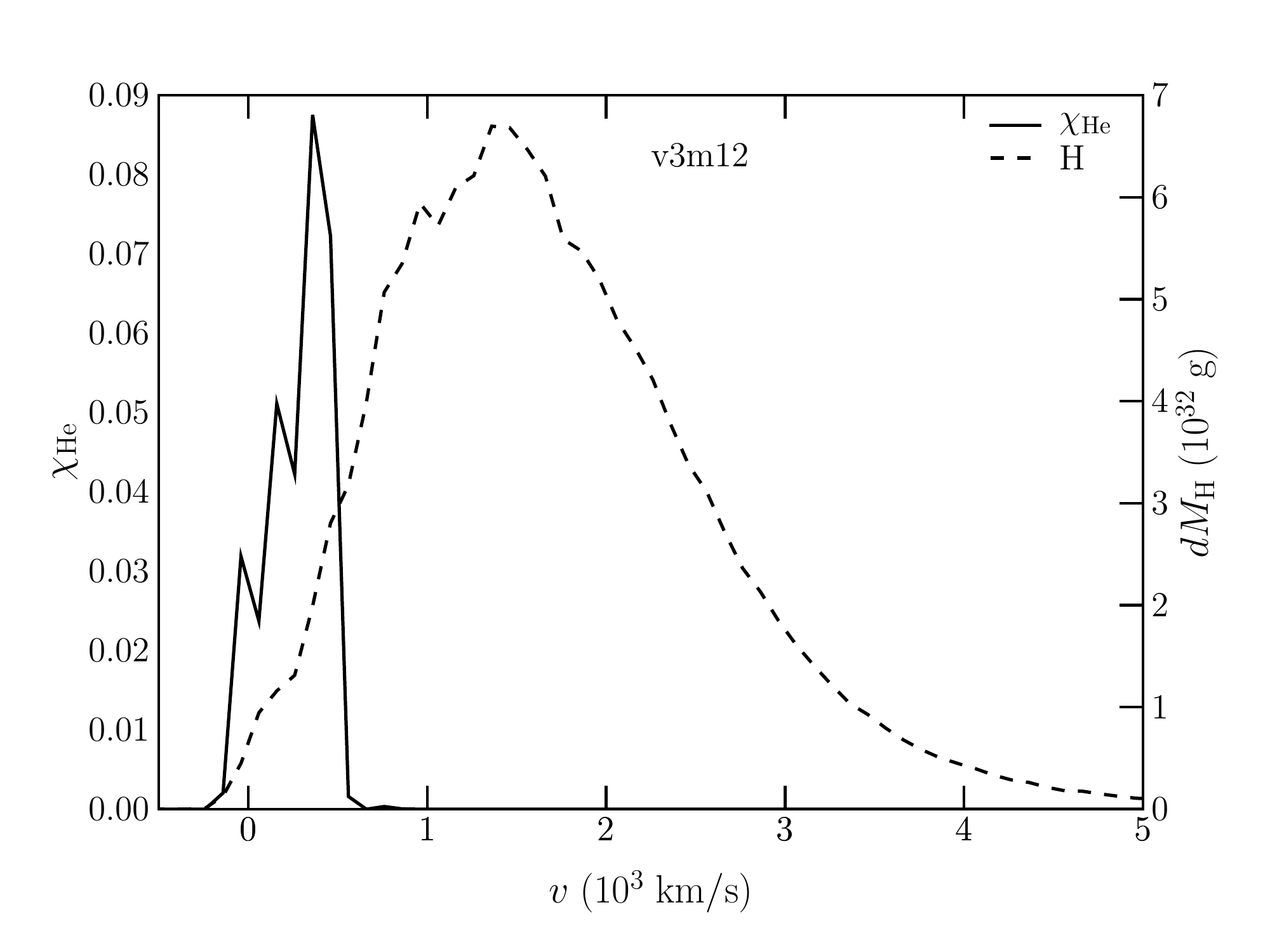} \\
\includegraphics[width=3in]{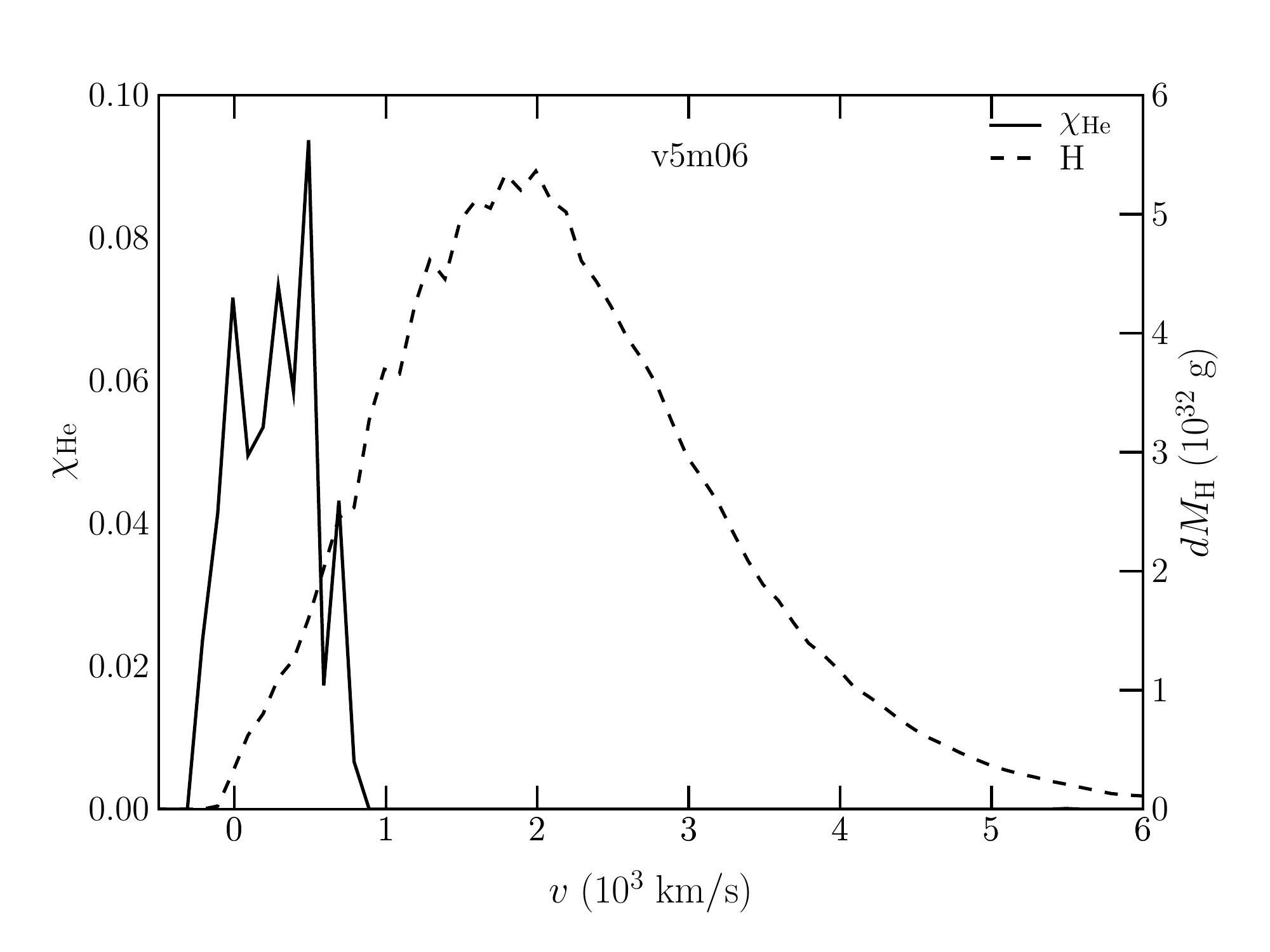} &
\includegraphics[width=3in]{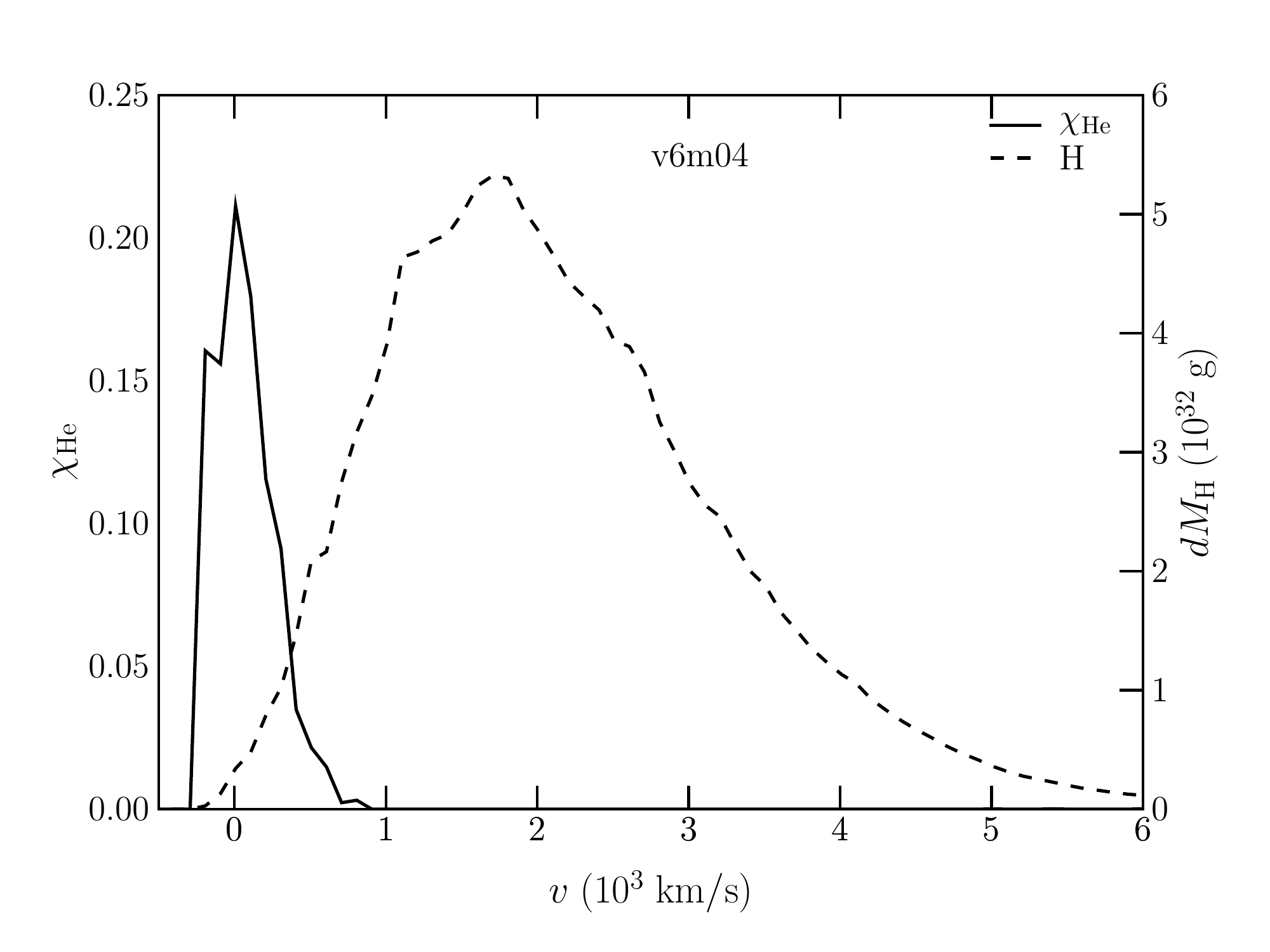} 
\end{tabular}
\caption[]{Distribution of the clumped He mass fraction ($\chi_{\rm He} = M_{\rm He, clumped} / M_{\rm He}$) and H mass distribution in the four models.  In all models, a small fraction of the He mass in the outer hydrogen envelope is clumped.  \citet{Fassia:98} find that a much higher clumped fraction is needed to reproduce the He I 10830\AA\ line in SN 1995V; however, their modeling did not account for significant nickel clumps at high velocities (see figs. \ref{fig:mofr} - \ref{fig:dmdvz}). }
\label{fig:dclumpdv}
\end{figure}

\begin{figure}
\centering
\begin{tabular}{cc}
\includegraphics[totalheight=0.2\textheight]{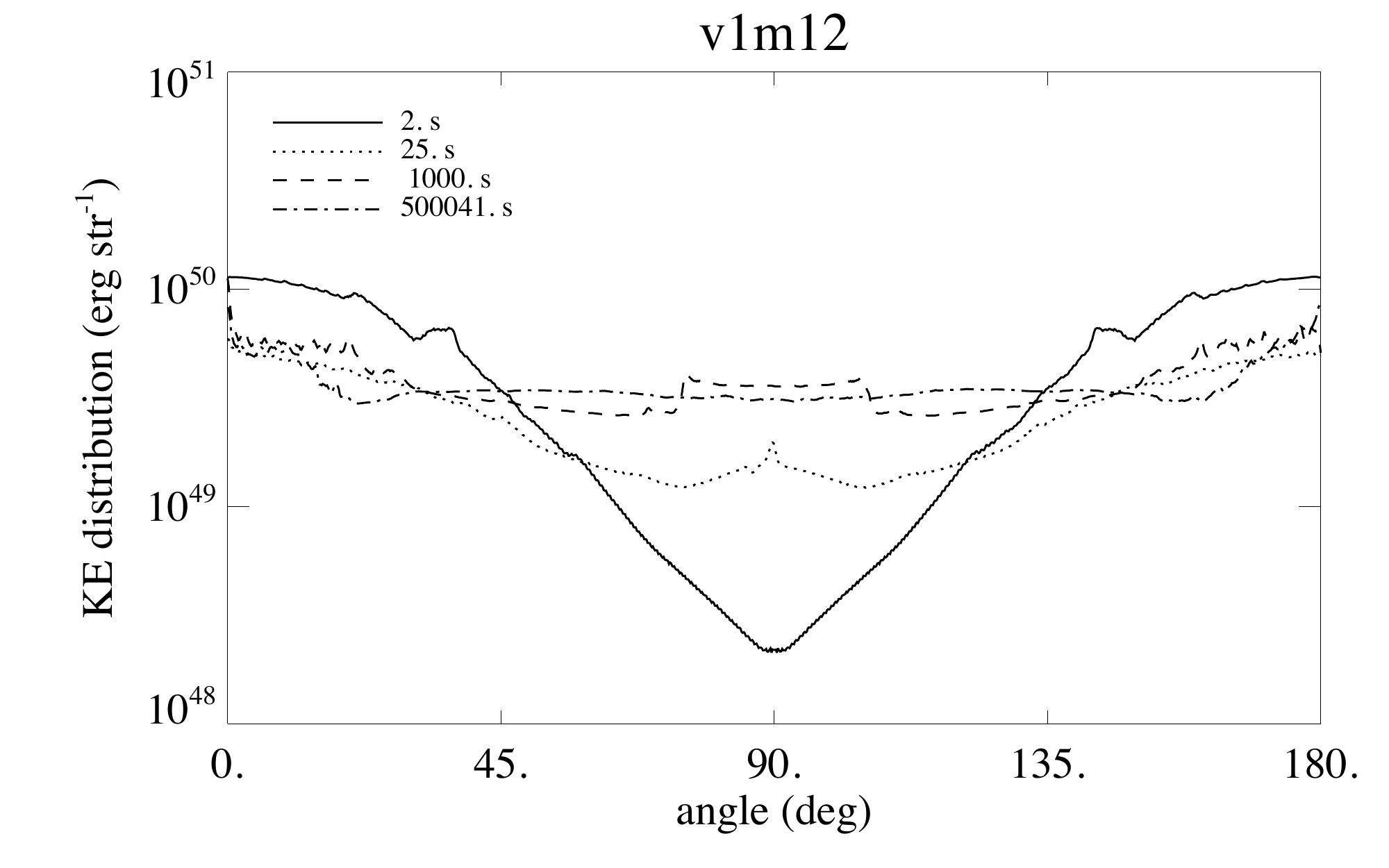} &
\includegraphics[totalheight=0.2\textheight]{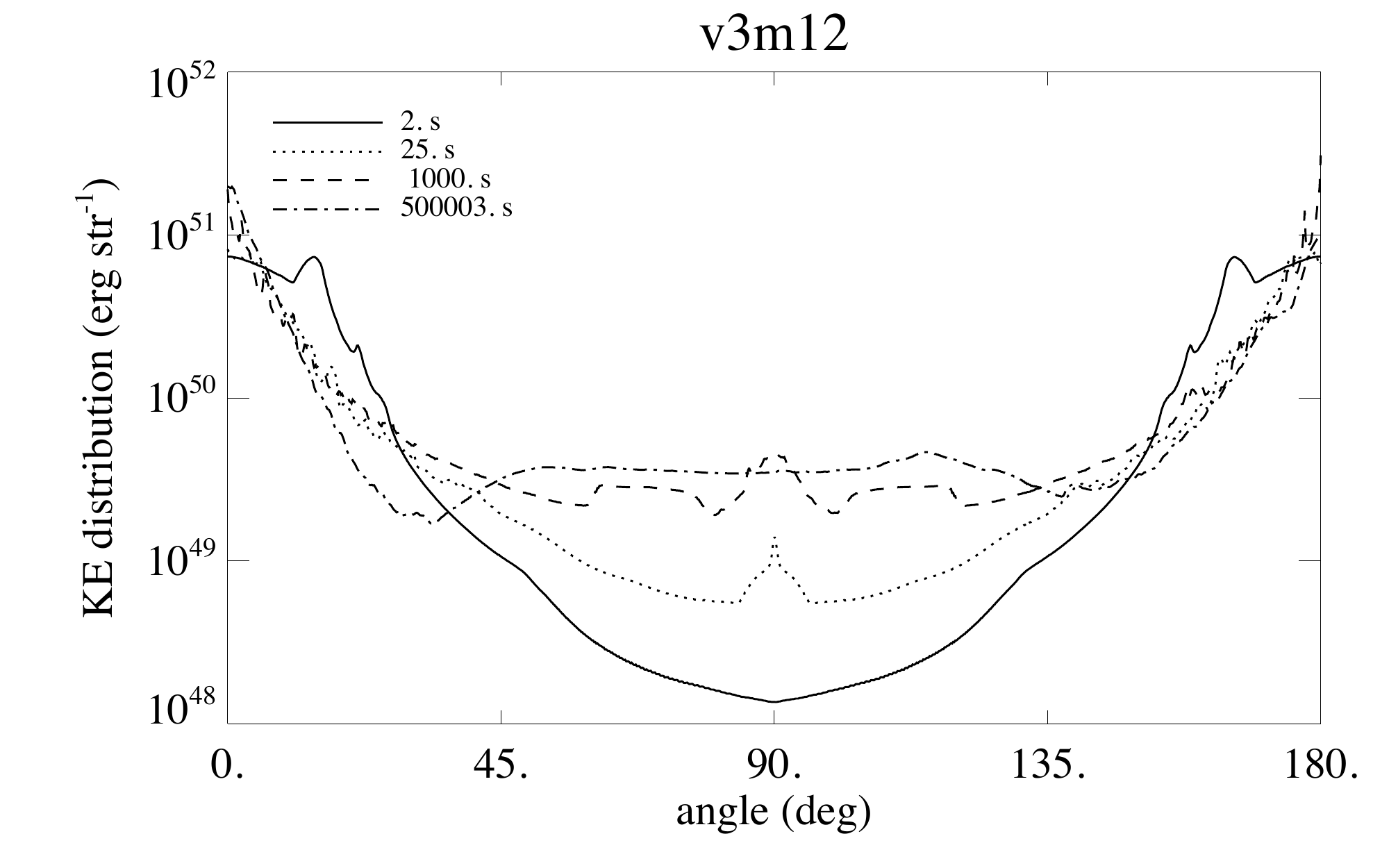} \\
\includegraphics[totalheight=0.2\textheight]{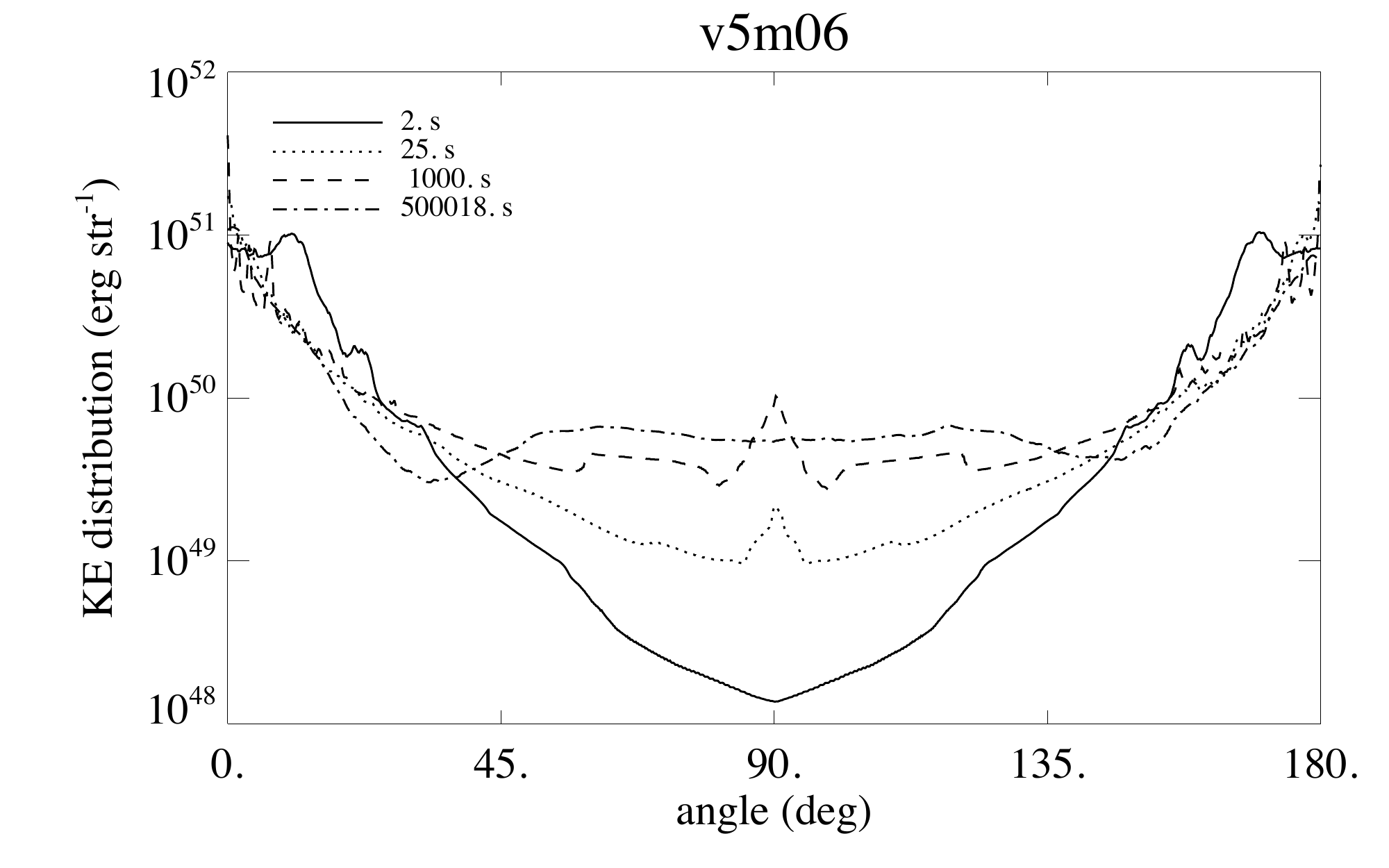} &
\includegraphics[totalheight=0.2\textheight]{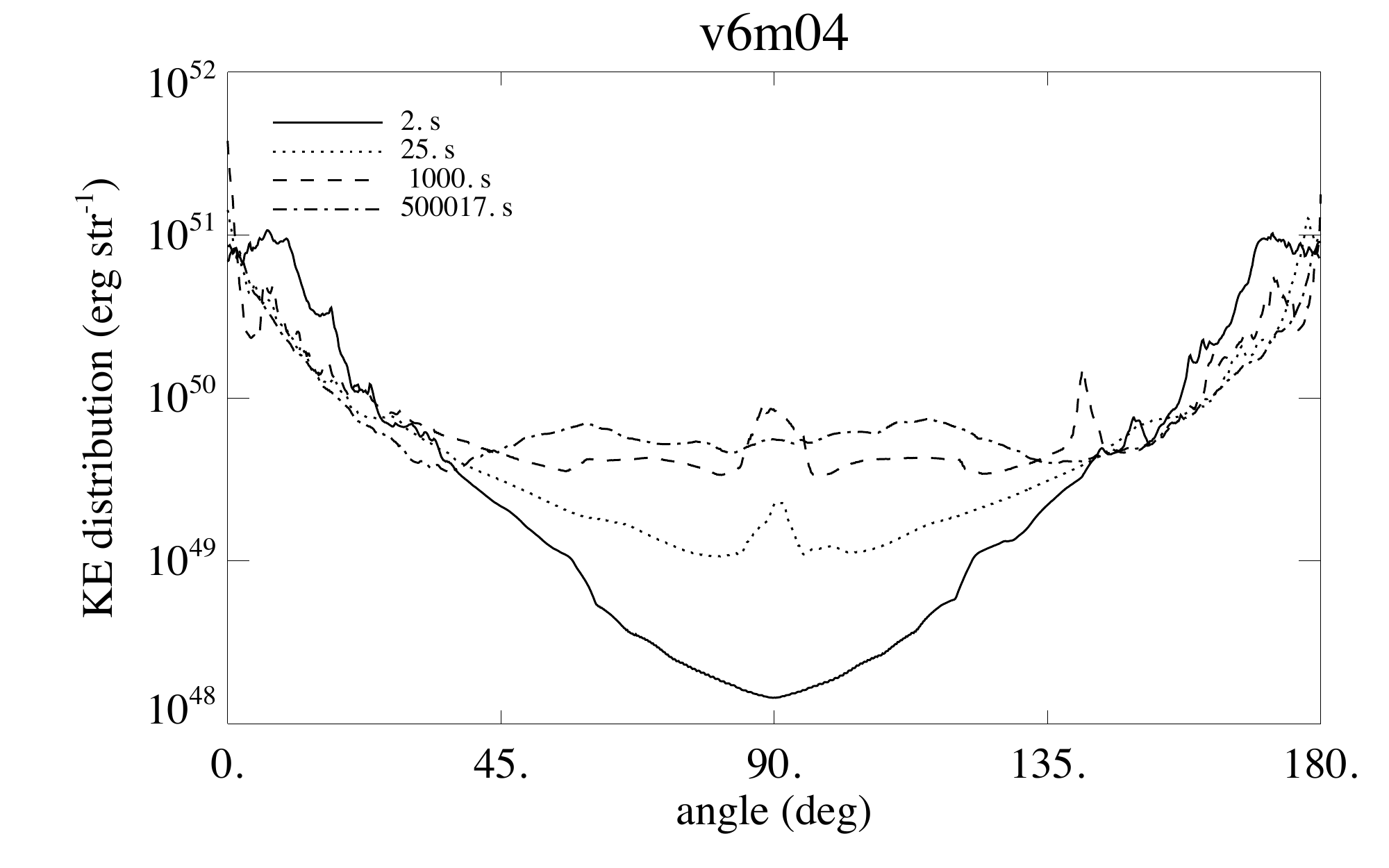}
\end{tabular}
\caption{Angular distribution of kinetic energy for the four models at various times.  Note that the thermal energy-dominated model (v1m12, upper left) has distributed its energy nearly uniformly in angle by the end of the simulation whereas the three kinetic energy-dominated models (other panels) retain significant kinetic energy near the jet axis.}
\label{fig:ekv3}
\label{fig:ekv1}
\label{fig:ekv5}
\label{fig:ekv6}
\end{figure}

In comparison with \citet{kho99}, model v3m12 shows a similar large-scale structure up to the time the shocks depart the helium core (see their Figure 2).  The extension in the jet direction is similar and the development of the equatorial outflow is comparable.  In contrast, the higher resolution employed in our model ($\Delta r_{\rm min} \sim 3\times10^6$ cm versus $\Delta r_{\rm min} \sim 4\times10^7$ cm in Khokhlov et al.) allows us to see the development and growth of fluid instabilities, particularly in the jet fluid (see the right panel of Figure \ref{fig:v3m12core}).

\begin{figure}
\centering
\begin{tabular}{cccc}
\includegraphics[width=1.25in, trim= .55in 0 .55in 0in, clip]{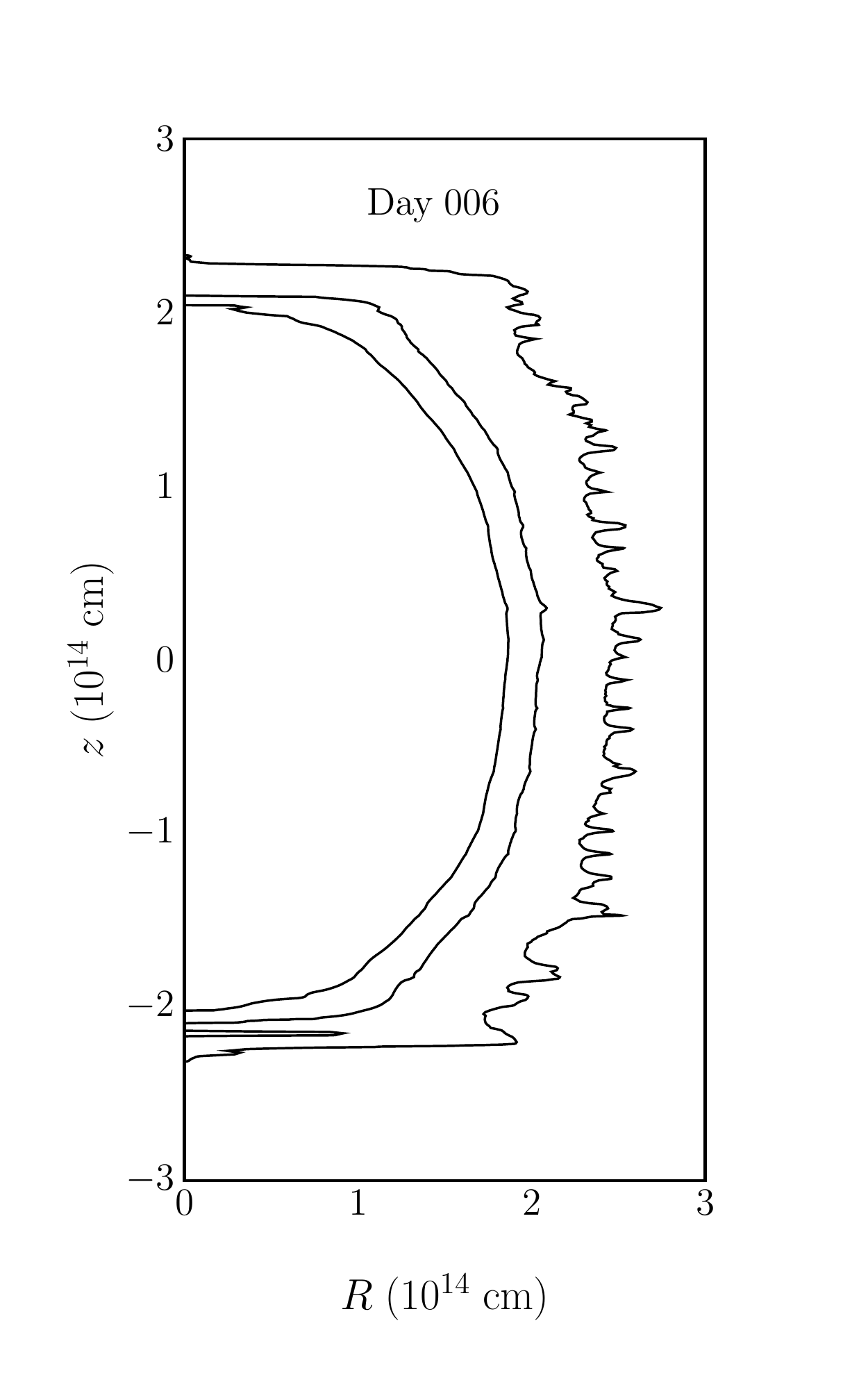} &
\includegraphics[width=1.25in, trim= .55in 0 .55in 0in, clip]{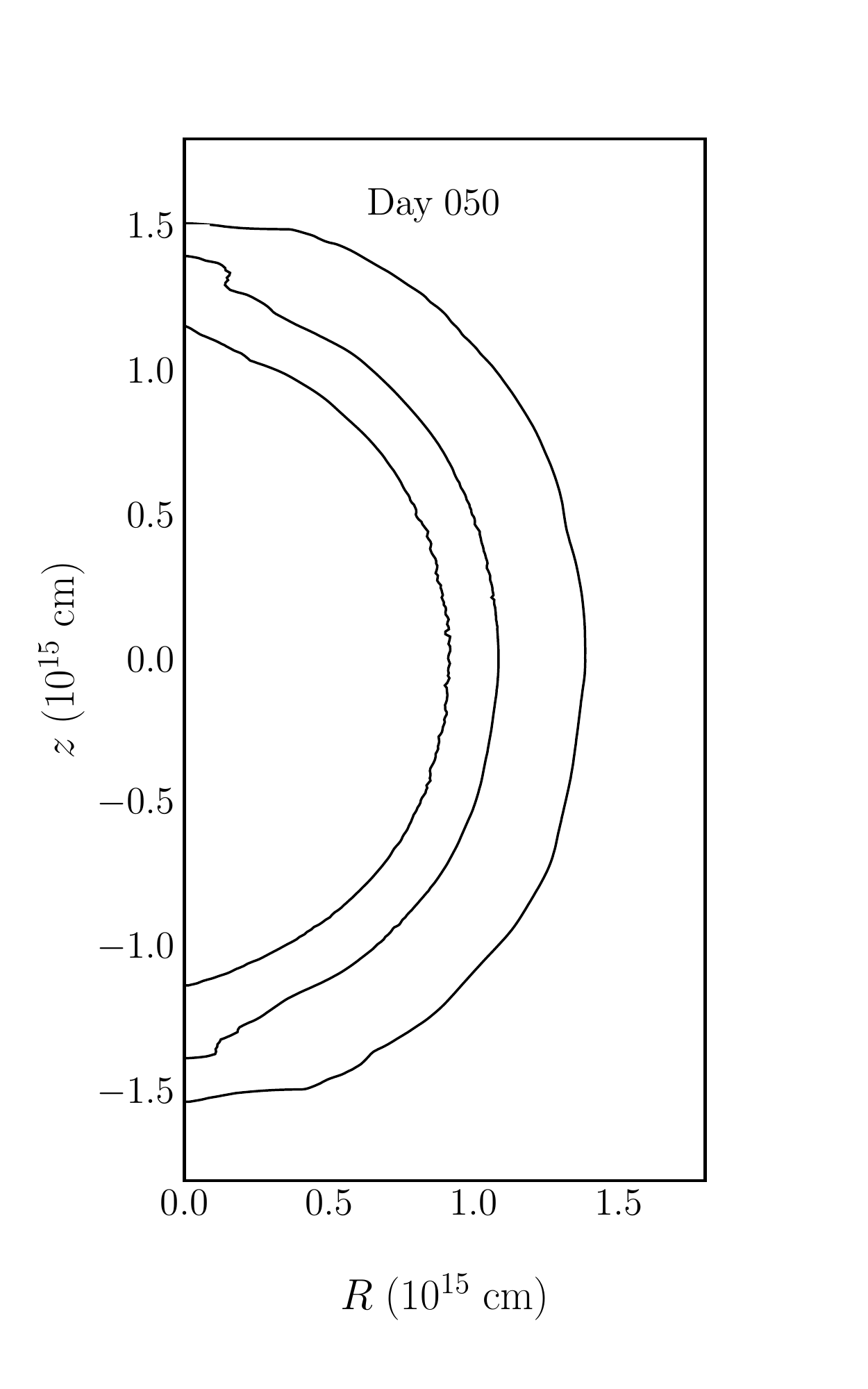} &
\includegraphics[width=1.25in, trim= .55in 0 .55in 0in, clip]{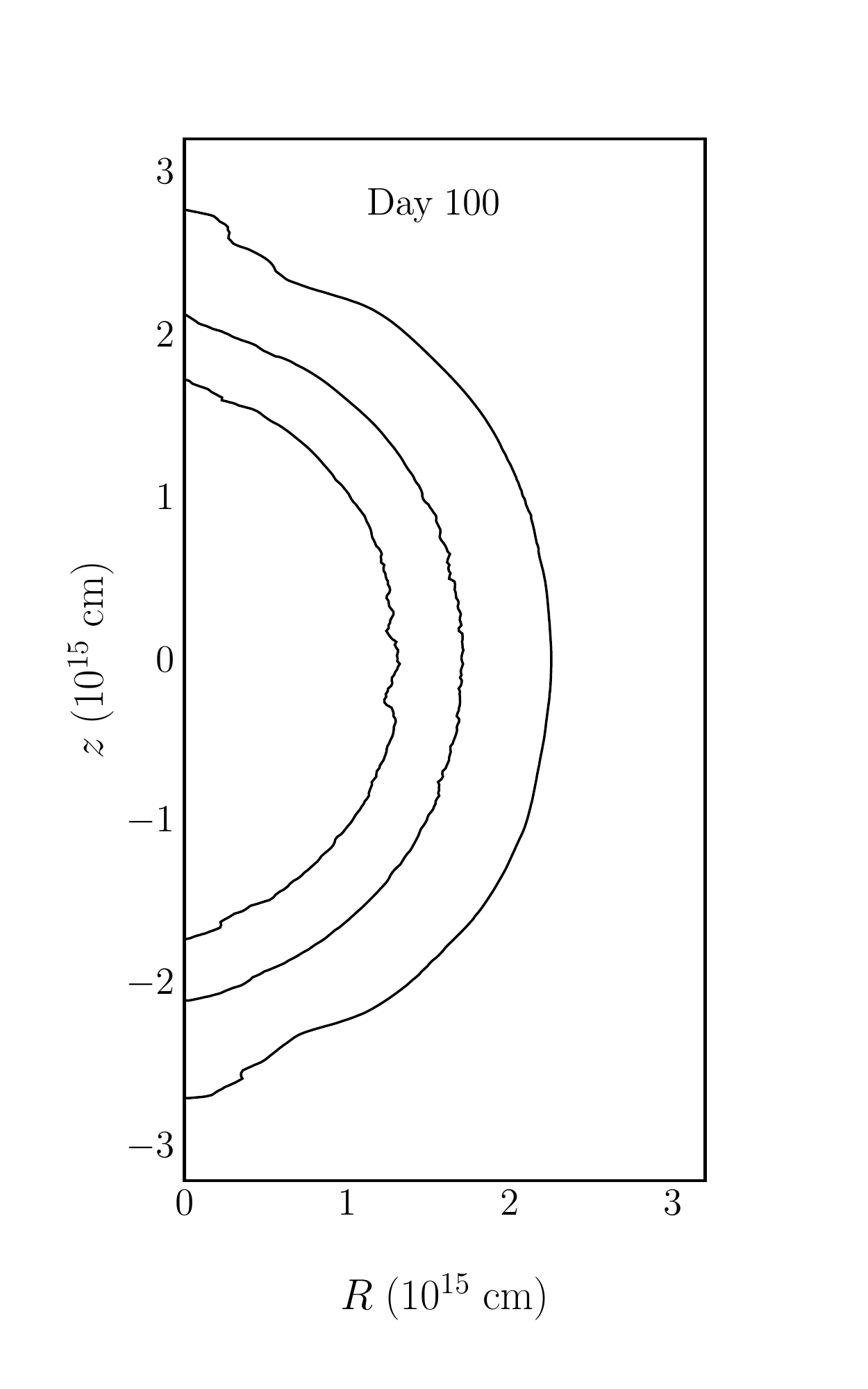} &
\includegraphics[width=1.25in, trim= .55in 0 .55in 0in, clip]{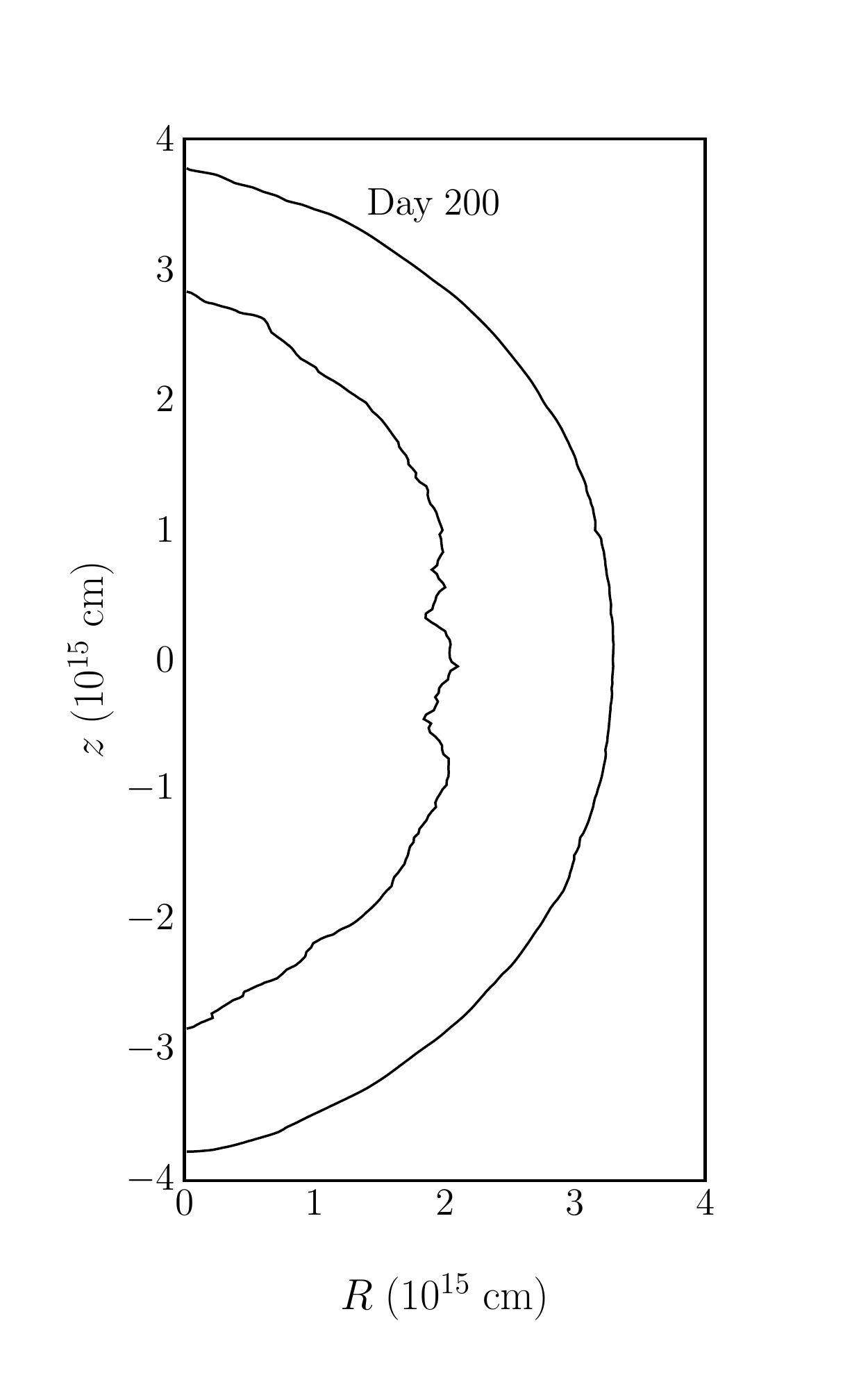}
\end{tabular}
\caption[]{Surfaces of constant electron scattering optical depth $\tau_{\rm es}$ using a constant opacity for thermal energy-dominated model v1m12 at several epochs: 6 (end of simulation), 50, 100, and 200 days.  The optical depths were calculated for a line-of-sight in the equatorial plane and the levels shown are $\tau_{\rm es} = 1, 10, 30$.  On day 6, the photospheric shape is slightly oblate with axis ratio of $\sim0.95$, but by day 50 the photosphere is essentially spherical.  On day 100, around the time of the observed jump in polarization in SN 2004dj, the axis ratio of v1m12 is $\sim1.25$ and the photosphere is prolate.  On day 200, the elongation of the inner core is apparent and the contour for $\tau_{\rm es}=30$ is absent as the model is making the transition to the nebular phase.}
\label{fig:tau_v1}
\end{figure}

\begin{figure}
\centering
\begin{tabular}{cccc}
\includegraphics[width=1.25in, trim= .55in 0 .55in 0in, clip]{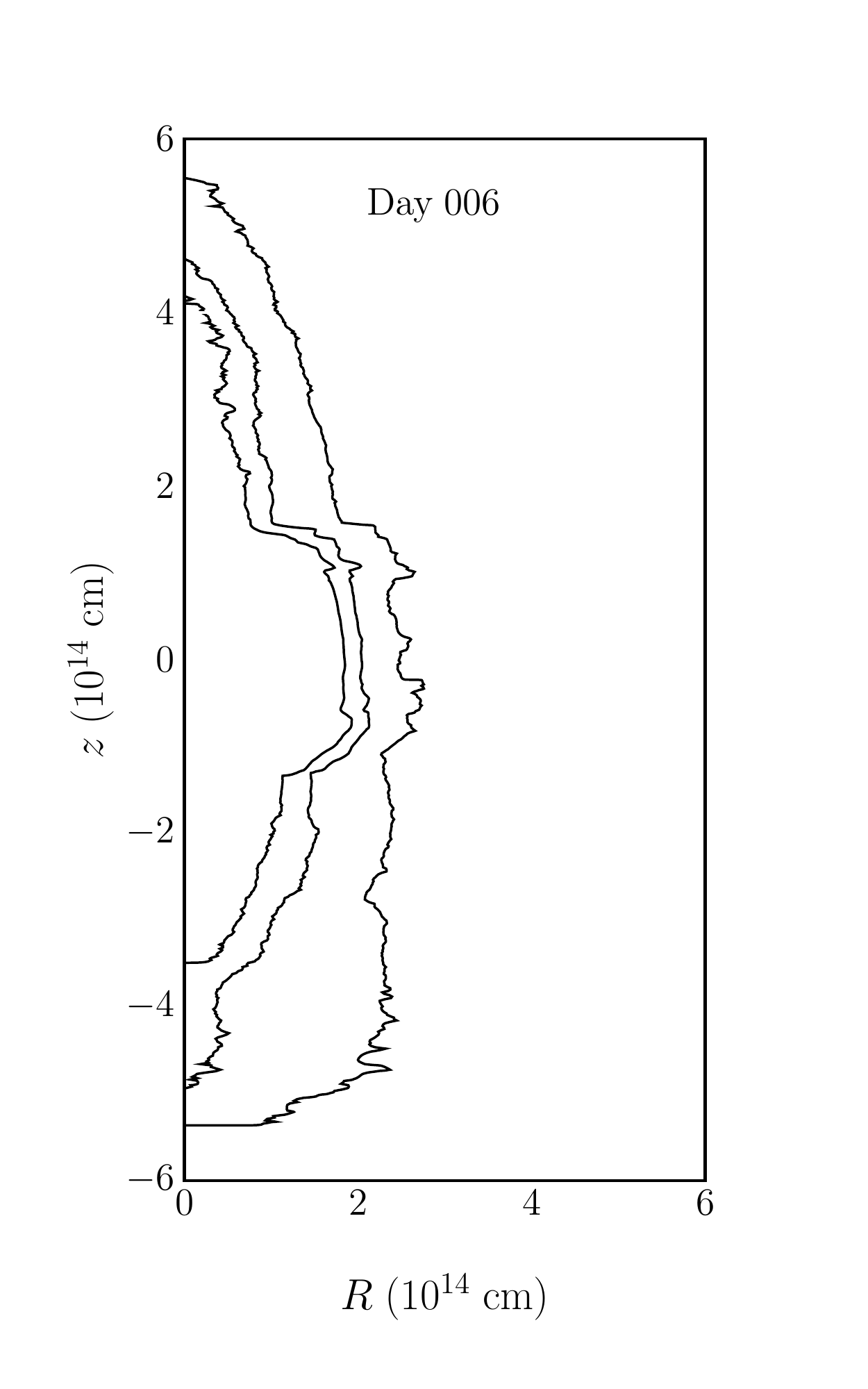} &
\includegraphics[width=1.25in, trim= .55in 0 .55in 0in, clip]{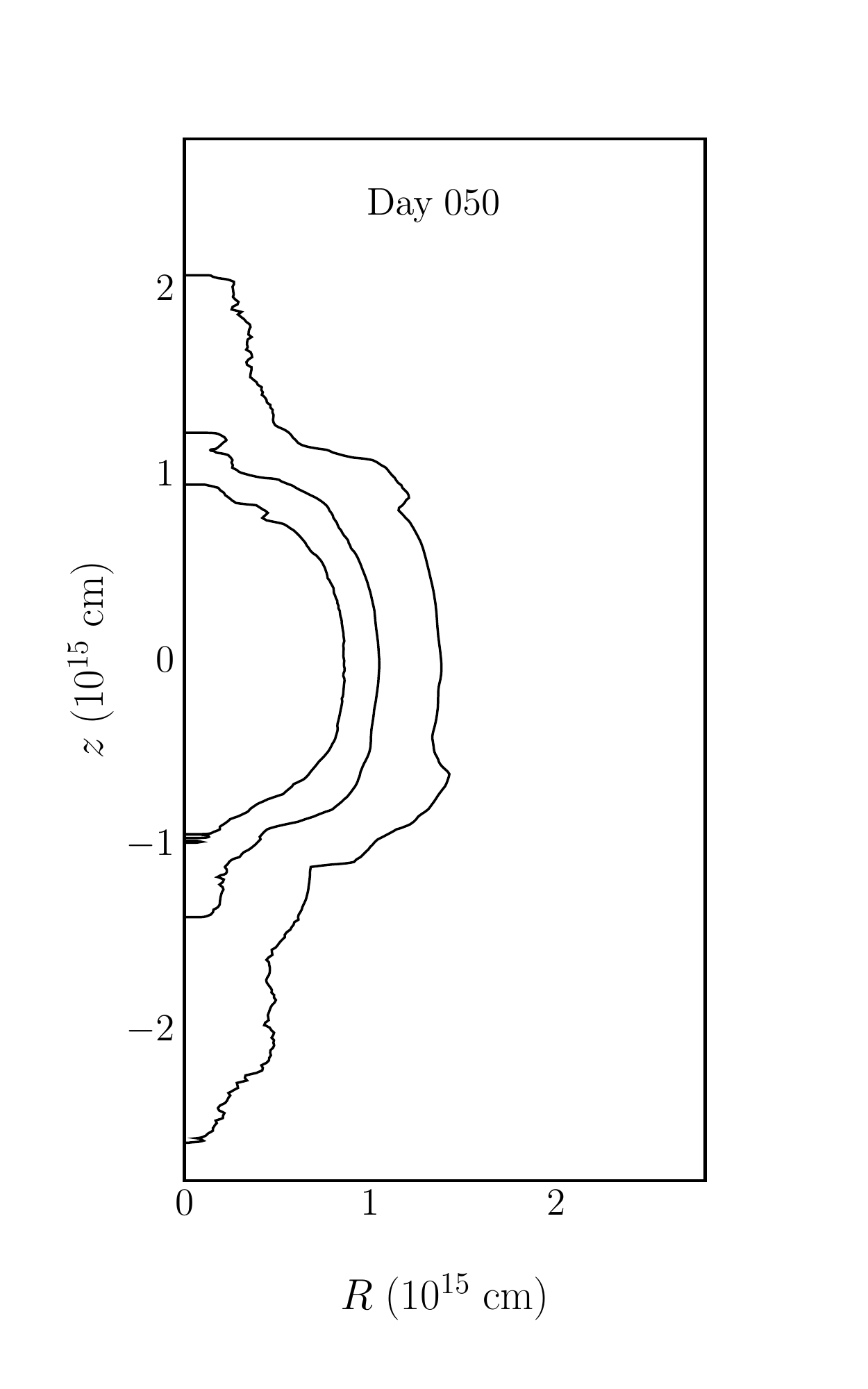} &
\includegraphics[width=1.25in, trim= .55in 0 .55in 0in, clip]{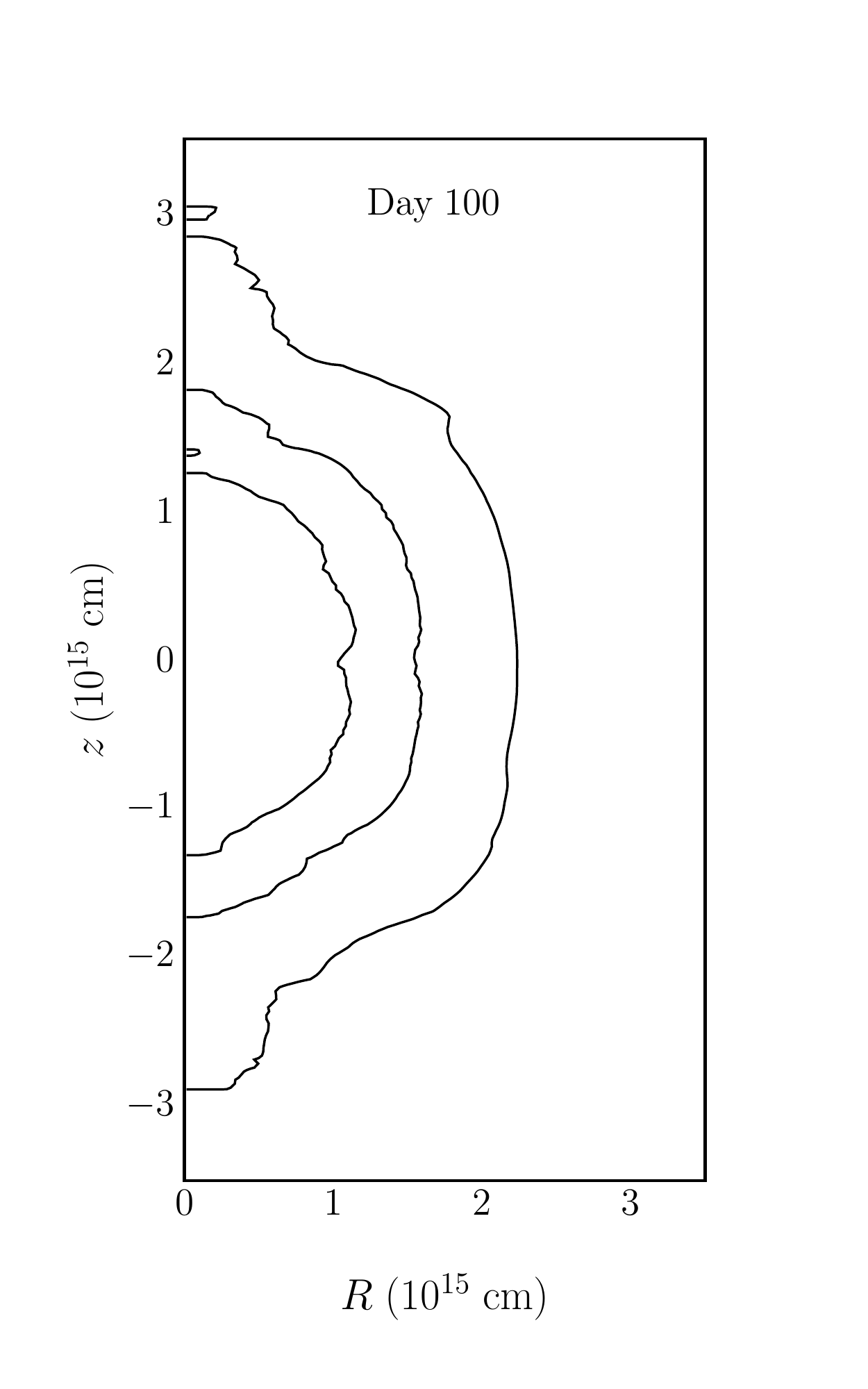} &
\includegraphics[width=1.25in, trim= .55in 0 .55in 0in, clip]{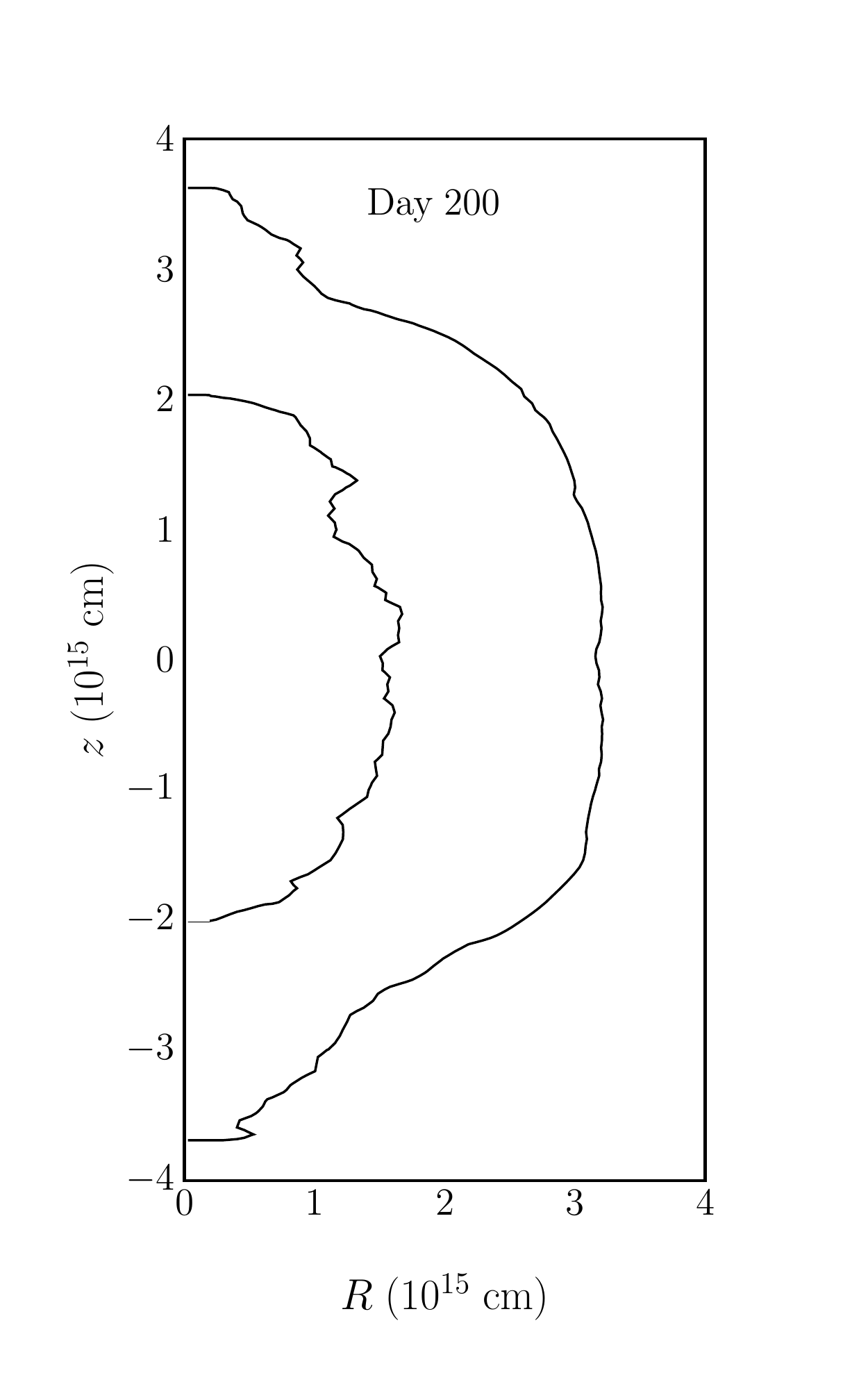}
\end{tabular}
\caption{Same as Fig. \ref{fig:tau_v1} except for kinetic energy-dominated model v3m12.  The shape of the photosphere is remarkably asymmetric at all times, in disagreement with polarimetry of SNeII-P.}
\label{fig:tau_v3}
\end{figure}

We have compared the late-time geometries of our models with observations by calculating the electron scattering optical depths under the assumption of adiabatic expansion (see Figs. \ref{fig:tau_v1} and \ref{fig:tau_v3}).  We show that the kinetic energy-dominated models (v3m12, v5m06 and v6m04) are too aspherical at early times to match polarimetry of SNeII-P.  Model v1m12 produces a photosphere that is nearly spherical for the first 50 days of expansion and increases in prolate elongation to axis ratios that are in close agreement with those derived for SN 2004dj \citep{Leonard:06}.  Our simplified calculation of optical depths neglect several important radiative effects, however a full radiative transfer calculation is beyond the scope of this work and is unlikely to qualitatively change our results.  

The development of instabilities may be important to the overall understanding of core-collapse supernovae in a number of ways.  Substantial observational evidence shows that the transport of heavier material to radii ahead of lighter material is a ubiquitous phenomenon in core-collapse supernovae, especially SNeII-P.  This ``overturn'' of the original ``onion-skin'' configuration of the progenitor has been explained in neutrino-driven explosions by the growth of fluid instabilities \citep[see][]{kifon03,kifon06}.  The substructure of certain strong lines (such as H$\alpha$) indicates large, fast-moving clumps of nickel in the ejecta.  Similar to the calculations of \citet{kifon03}, many of the instabilities that develop in our models are of very small scale and it is unlikely that such small instabilities can explain the substructure of spectral lines, however, the jets are effectively a large, single-mode asymmetry capable of producing high-velocity bipolar clumps.  The jet fluid, which we assume to be nickel, is transported ahead of lighter materials, providing significant ``overturn.''  In our models, a large amount of nickel mass is transported to radii where it can excite the H$\alpha$ and He I lines, as is required by spectral modeling of SNeII-P \citep{Fassia:98, Elmhamdi:03, Chugai:05, Chugai:06}.   \citet{kifon06} show that the mode of the instabilities that grow in neutrino-driven explosions is sensitive to the nature of the neutrino emission.  In jet-driven models, the jets create a robust, large-scale asymmetry that is not very sensitive to the detailed nature of the jets.  

We show that He clumps form due to RT instabilities in the transition layer between the H and He in the explosion ejecta in all four jet models.  These clumps may contribute the pristine helium required by \citet{Fassia:98} to fit the strength of the He I 10830 \AA\ line, though the amount of clumping we find is below that required by Fassia et al.  As we have noted, the extra nickel at high velocities in the jet models (about a factor of two greater than that assumed by Fassia et al.) may compensate for the low helium clump fraction by providing excess gamma-ray excitation.

\begin{figure}
\centering
\begin{tabular}{cc}
\includegraphics[width=3in]{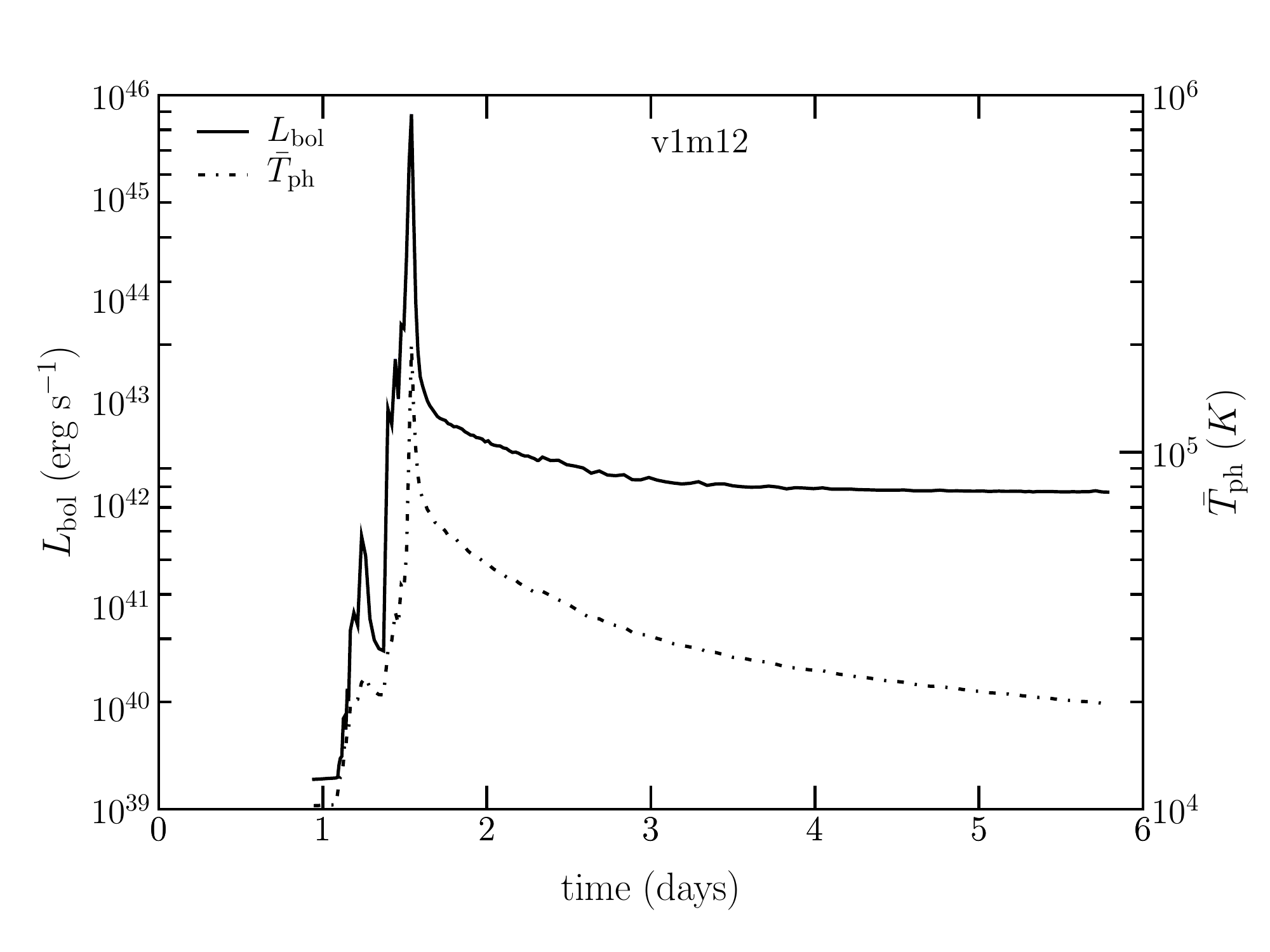} &
\includegraphics[width=3in]{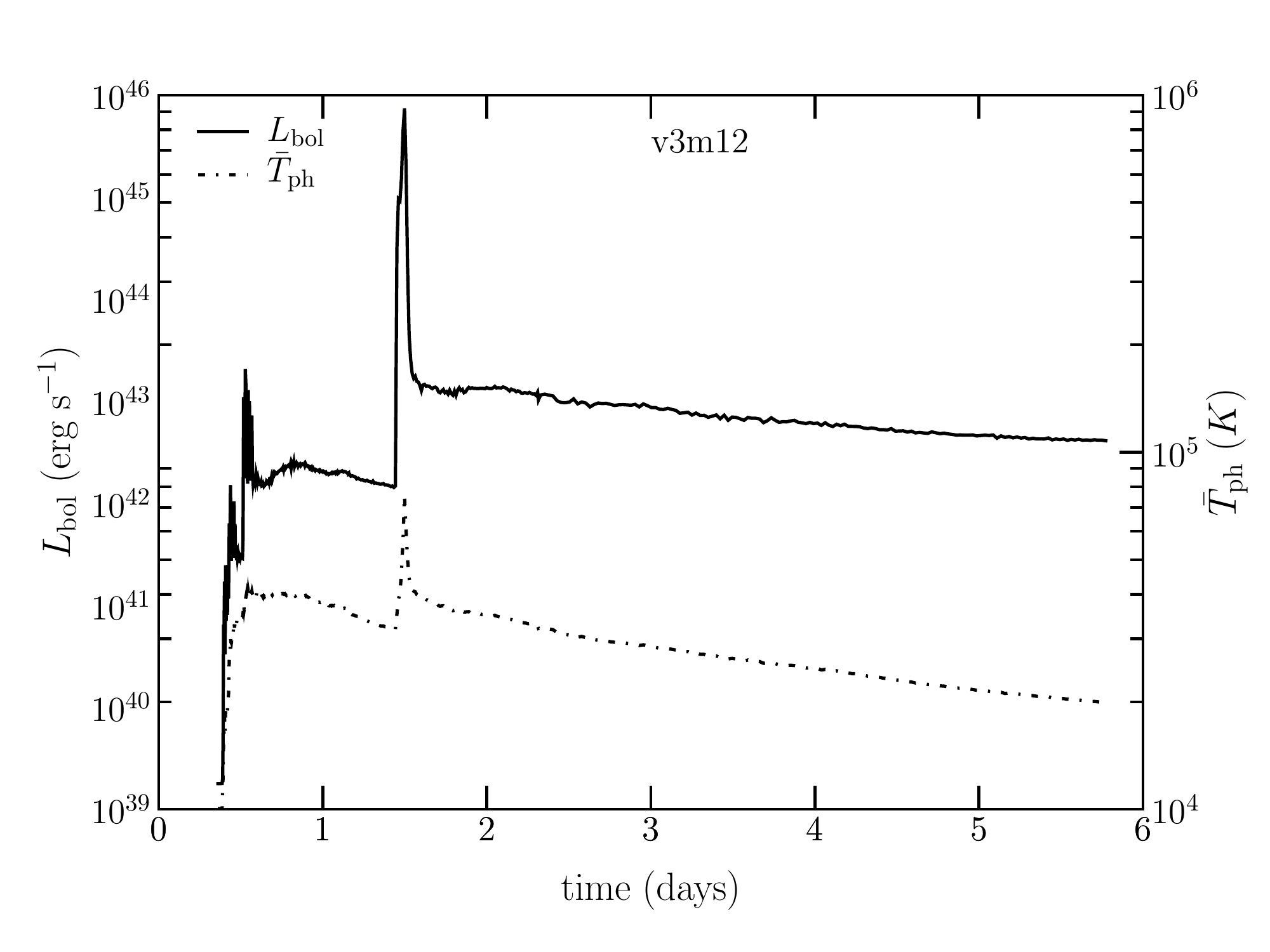}
\end{tabular}
\caption{Bolometric shock breakout light curves for models v1m12 and v3m12.  The light curve of thermal model v1m12 is similar to that for a spherically-symmetric breakout and is consistent with {\it GALEX} observations of shock breakout from SNLS 04D2dc.  The kinetic model, v3m12, has a double-peaked breakout light curve that would have been evident in the {\it GALEX} observations, but was absent.  We also plot the angle-averaged photospheric temperature.}
\label{fig:breakout}
\end{figure}

Evolving our calculations beyond shock breakout allows us to study the breakout characteristics of the explosion models.  In Figure \ref{fig:breakout} we present the bolometric shock breakout light curves for models v1m12 and v3m12.  The nearly-spherical breakout of model v1m12 produces a light curve that is very similar to spherically-symmetric model light curves.  The dramatically aspherical breakouts of the kinetic models show a first peak in the light curve associated with the breakout of the jets followed by a roughly constant luminosity period created by the expanding and cooling jet ``cocoons'' and then a large peak as the equatorial shocks break out.  The early peak and plateau in the light curves of the kinetic models would have been evident in {\it GALEX} observations of SNLS 04D2dc, if they were present.  Their absence argues against the likelihood of a significantly aspherical break out for this SNeII-P, as does the comparison to polarimetry.  The light curve of model v1m12 agrees with the observations for shock breakout for the two SNeII-P caught by {\it GALEX} \citep{Schawinski:08, Gezari:08}.

In Table \ref{tab:obs}, we summarize the observational characteristics of SNeII-P and the four jet models.  We also include the characteristics a neutrino-driven SN model based loosely on that of \citet{kifon03, kifon06} for comparison.  A multi-dimensional neutrino-driven model that includes fluid instabilities can be expected to account for transport of heavy elements to high ejecta velocities as well as the formation of pristine helium clumps in the hydrogen envelope; however, it is not clear that neutrino-driven models can consistently produce fast-moving, bipolar nickel clumps in the ejecta.  Likewise, it is unlikely that these models can account for the time dependence of SNeII-P polarization.  Significantly aspherical neutrino-driven explosions may result from the standing accretion shock instability \citep[SASI;][]{Blondin:03, Marek:07} or the acoustic mechanism \citep{Burrows:06, Burrows:07a}, however the large-scale structure of supernovae resulting from these mechanisms has not been explored.  As we have discussed, the kinetic energy-dominated jet models produce explosions that are too aspherical at early times and are, thus, incompatible with polarization and shock breakout observations of SNeII-P.   In contrast, our thermal energy-dominated model (v1m12) shows good agreement with all of the observations we have discussed here.  

The comparison of observational evidence with our simulations consistently indicates that jets capable of retaining a large fraction of their initial kinetic energy while crossing the hydrogen envelope of a red supergiant star are unlikely to be present in SNeII-P explosions.  This is evidence that if jets are present, or are the driving mechanism in core-collapse supernovae, they must be jets that are capable of transferring their kinetic energy to the hydrogen envelope of a RSG.  This may not require thermal energy-dominated jets, as we have presented here, but it does seem to rule out fast, dense jets that remain ballistically detached from the surrounding ejecta and punch out of the progenitor with a large fraction of their initial kinetic energy intact.  \citet{hof01} and \citet{kho01} have noted that if jets efficiently penetrate the core of a stripped envelope supernova, but are mostly trapped in the envelope of extended hydrogen-rich supernovae, then average stripped-envelope supernovae should have lower kinetic energies than average SNeII-P. Since this is not the case, H\"{o}flich deduced that any jet energy must already be fairly efficiently deposited in the cores of supernova progenitors by dint of any jets being rather wide and slow. The conclusions of our current study, that thermal jets are favored
over kinetic energy jets, is roughly in accord with Hoeflich's supposition. We note, however,
that even our thermal jets that ultimately deposit their energy rather uniformly in the hydrogen
envelope are less efficient in depositing energy in the helium core. The energy deposition cannot
be completely isotropized in the helium core without violating the observations from spectropolarimetry
that the inner core is substantially aspherical.  How these two constraints of energetics versus
asymmetry play off against one another needs to be studied in more detail.

\acknowledgements
The authors are grateful to Peter H\"{o}flich, Elaine Oran, and Jay Boris for helpful comments  on this work.  We would like to express our immense gratitude for the efforts of the staff at the Texas Advanced Computing Center, particularly Paul Navratil, in helping us to complete and visualize this work.  The software used in this work was in part developed by the DOE-supported ASC/Alliance Center for Astrophysical Thermonuclear Flashes at the University of Chicago.  The authors acknowledge the Texas Advanced Computing Center (TACC) at The University of Texas at Austin for providing high-performance computing, visualization, and data storage resources that have contributed to the research results reported within this paper.  This research was supported in part by NSF grant AST-0707769 and NASA grant NNG04GL00G.
 
\bibliography{RSGjets}

\clearpage

\begin{deluxetable}{lccccc}
\tablewidth{0pt}
\tablecaption{Parameters of jet models}
\tablehead{
\colhead{Jet Model} & \colhead{$\rho$ (g cm$^{-3}$)} & \colhead{$T_{\rm max}$ (K)} & 
\colhead{$P$ (erg cm$^{-3}$)} & \colhead{$v_{\rm max}$ (cm/s)} & \colhead{$M_{\rm tot}$ ($M_\odot$)}
}
\startdata
v3m12 & $6.5\times10^5$ & $2\times10^9$ & $10^{23}$ & $3.2\times10^9$ & 0.12 \\
v5m06 & $2.2\times10^5$ & $2\times10^9$ & $10^{23}$ & $5\times10^9$ & 0.06 \\
v1m12 & $2\times10^6$ & $4.4\times10^9$ & $10^{24}$ & $ 10^9 $ & 0.12 \\
v6m04 & $10^5$ & $2\times10^9$ & $ 3\times10^{22}$ & $6\times10^9$ & 0.04 \\
ambient\tablenotemark{a} & $8\times10^5$ & $ 2\times10^9$ & $1.4\times10^{23}$ & \nodata & \nodata \\
\enddata
\tablenotetext{a}{The ambient values given are those of the progenitor model and the inner radius of our simulations:  $3.8\times10^8$ cm.}
\label{tab:jets}
\end{deluxetable}

\begin{deluxetable}{lcccccc}
\tablewidth{0pt}
\tablecaption{Simulation stages and grid parameters}
\tablehead{
\colhead{Stage} & \colhead{$t_i$ (s)} & \colhead{$t_f$ (s)} & \colhead{$r_{\rm in}$ (cm)} &
\colhead{$r_{\rm out}$ (cm)} & \colhead{$N_{r,0}$}\tablenotemark{a} & \colhead{$\Delta r_{\rm min}$ (cm)}\tablenotemark{b}
}
\startdata
0: & 0 & 5 & $3.82\times10^8$ & $3.2\times10^{10}$ & 192 & $2.6\times10^6$ \\
1: & 5 & 25 & $3.82\times10^8$ & $7.5\times10^{10}$ & 192 & $6.1\times10^6$ \\
2: & 25 & 100 & $1.0\times10^9$ & $2.5\times10^{11}$ & 192 & $2.0\times10^7$ \\
3: & 100 & 300 & $2.0\times10^9$ & $9.0\times10^{11}$ & 192 & $7.3\times10^7$ \\
4: & 300 & $1\times10^3$ & $4.0\times10^9$ & $3.0\times10^{12}$ & 320 & $1.5\times10^8$ \\
5: & $1\times10^3$ & $3\times10^3$ & $1.0\times10^{10}$ & $6.0\times10^{12}$ & 320 & $2.9\times10^8$ \\
6: & $3\times10^3$ & $1\times10^4$ & $2.0\times10^{10}$ & $2.0\times10^{13}$ & 352 & $8.9\times10^8$ \\
7: & $1\times10^4$ & $3\times10^4$ & $5.0\times10^{10}$ & $6.0\times10^{13}$ & 352 & $2.7\times10^9$ \\
8: & $3\times10^4$ & $1\times10^5$ & $1.0\times10^{11}$ & $2.0\times10^{14}$ & 384 & $8.1\times10^9$ \\
9: & $1\times10^5$ & $2\times10^5$ & $2.0\times10^{11}$ & $4.0\times10^{14}$ & 384 & $1.6\times10^{10}$ \\
10: & $2\times10^5$ & $5\times10^5$ & $5.0\times10^{11}$ & $1.0\times10^{15}$ & 384 & $4.1\times10^{10}$ \\
\enddata
\tablenotetext{a}{The number of radial zones at the lowest level of refinement.}
\tablenotetext{b}{The radial resolution at the highest level of refinement.}
\label{tab:grid}
\end{deluxetable}

\begin{deluxetable}{lccccc}
\tablewidth{0pt}
\tablecaption{Observational features of supernova models}
\tablehead{
\colhead{Observed characteristic of SNeII-P} & \colhead{v1m2} & \colhead{v3m12} & \colhead{v5m06} & \colhead{v6m04} & \colhead{neutrino-driven\tablenotemark{a}}
}
\startdata
early polarization small & yes & no & no & no & yes \\
late polarization large & yes & yes & yes & yes & no \\
spectral substructure due to Ni clumps & yes & yes & yes & yes & no\tablenotemark{b} \\
pristine He clumps in H & yes & yes & yes & yes & yes \\
{\it GALEX} breakout light curve & yes & no & no & no & yes \\
\enddata
\tablenotetext{a}{Based loosely on the results of \citet{kifon03, kifon06}.}
\tablenotetext{b}{\citet{kifon06} do find large-scale nickel clumps that may account for spectral substructure, however this result seems to depend sensitively on the nature of the neutrino emission.  A better understanding of neutrino physics is needed in order to know if this result can be obtained consistently in all SNeII.}
\label{tab:obs}
\end{deluxetable}

\clearpage

\end{document}